%% file: main.tex
\documentclass[sigconf,10pt]{acmart}

\usepackage{amsmath}
\usepackage{graphicx}
\usepackage{xcolor}
\usepackage{subcaption}
\usepackage{url}
\usepackage{float}
\usepackage{algpseudocode}
\usepackage{algorithm}
\usepackage{multirow}
\usepackage{cleveref}
\usepackage{tikz}
\usepackage{wrapfig}
\usepackage{hyperref}
\usepackage{makecell}
\usepackage{xspace}

\newcommand{\jc}[1]{\textcolor{blue}{[JC: #1]}}

\newcommand{\rev}[1]{\textcolor{black}{#1}}

\newcommand{\system}{{\sf \small L3GS}\xspace}

\definecolor{darkred}{RGB}{150,0,0}
\definecolor{darkgreen}{RGB}{0,150,0}
\definecolor{darkblue}{RGB}{0,0,200}

\newcommand{\eg}{\emph{e.g.,} }
\newcommand{\ie}{\emph{i.e.,} }

\newcommand{\squishlist}{
   \begin{list}{$\bullet$}
    { \setlength{\itemsep}{0pt}      \setlength{\parsep}{3pt}
      \setlength{\topsep}{3pt}       \setlength{\partopsep}{0pt}
      \setlength{\leftmargin}{1.0em} \setlength{\labelwidth}{1em}
      \setlength{\labelsep}{0.5em} } }
\newcommand{\squishend}{
    \end{list}  }

\def\equationautorefname#1#2\null{%
  Eq.#1(#2\null)%
}

\newtheorem{theorem}{Theorem}
\newtheorem{lemma}{Lemma}
\newtheorem{problem}{Problem}

\usepackage{xr}
\begin{document}

\title[L3GS: Layered 3D Gaussian Splats for Efficient 3D Scene Delivery]{L3GS: Layered 3D Gaussian Splats\\ for Efficient 3D Scene Delivery}

\author{Yi-Zhen Tsai*}
 \thanks{*Equal contributors.}
\affiliation{%
\institution{University of California, Riverside}
\city{Riverside}
\state{California}
\country{USA}
}
\email{ytsai036@ucr.edu}

\author{Xuechen Zhang*, Zheng Li}
\affiliation{%
  \institution{University of Michigan}
  \city{Ann Arbor}
  \state{Michigan}
  \country{USA}
}
\email{{zxuechen,jimmyli}@umich.edu}

\author{Jiasi Chen}
\affiliation{%
  \institution{University of Michigan}
  \city{Ann Arbor}
  \state{Michigan}
  \country{USA}
}
\email{{jiasi@umich.edu}}

\begin{abstract}
Traditional 3D content representations include dense point clouds that consume large amounts of data and hence network bandwidth, while newer representations such as neural radiance fields suffer from poor frame rates due to their non-standard volumetric rendering pipeline. 3D Gaussian
splats (3DGS) can be seen as a generalization of point clouds that meet the best of both worlds, with high visual quality and efficient rendering for real-time frame rates. However, delivering 3DGS scenes from a hosting server to client devices is still challenging due to high network data consumption (\eg 1.5 GB for a single scene). The goal of this work is to create an efficient 3D content delivery framework that allows users to view high quality 3D scenes with 3DGS as the underlying data representation. The main contributions of the paper are: (1) Creating new layered 3DGS scenes for efficient delivery, (2) Scheduling algorithms to choose what splats to download at what time, and (3) Trace-driven experiments from users wearing virtual reality headsets to evaluate the visual quality and latency. Our system for Layered 3D Gaussian Splats delivery (\system) demonstrates high visual quality, achieving 16.9\% higher average SSIM compared to baselines, and also works with other compressed 3DGS representations. The code is available at \url{https://github.com/mavens-lab/layered_3d_gaussian_splats}.
\end{abstract}

\copyrightyear{2025}
\acmYear{2025}
\setcopyright{cc}
\setcctype{by}
\acmConference[ACM MOBICOM '25]{The 31st Annual International Conference on Mobile Computing and Networking}{November 4--8, 2025}{Hong Kong, China}
\acmBooktitle{The 31st Annual International Conference on Mobile Computing and Networking (ACM MOBICOM '25), November 4--8, 2025, Hong Kong, China}\acmDOI{10.1145/3680207.3723485}
\acmISBN{979-8-4007-1129-9/2025/11}



\maketitle

\input{intro.tex}
\input{system_design}
\input{models}

\input{formulation.tex}
\input{user_network_predictor.tex}
\input{results.tex}
\input{limitations.tex}

\input{conclusions.tex}

\newpage
\bibliographystyle{ACM-Reference-Format}
\bibliography{refs.bib}

 \input{appendix}

\end{document}

%% file: intro.tex
\section{Introduction}
\label{sec:intro}

Traditional representations of 3D content include meshes and point clouds.
Recently, new technologies to model 3D scenes have emerged that outperform traditional representations in terms of realism and modeling capability, such as neural radiance fields (NeRF~\cite{mildenhall2021nerf}) and 3D Gaussian splats (3DGS~\cite{kerbl3Dgaussians}).
NeRF requires substantial training of machine learning models to represent the 3D scene and relies on slow volumetric rendering techniques.
3D Gaussian splats, introduced in 2023, can be seen as a generalization of point clouds, where each 3D ``splat'' has position, volume, and color features.
They have gained prominence for their real-time rendering capabilities and excellent visual quality.

Typically, these 3D scenes are created and stored on a server, due to the substantial amount of computation needed to create the 3D scenes. 
Clients seeking to view the 3D scenes can download these scene models and render them locally for viewing.
This creates several network delivery challenges for viewers of 3DGS scenes:
(1) The 3DGS scenes can be very large (\eg 1.52 GB for the bicycle scene from Mip-NeRF360~\cite{barron2022mipnerf360}), and downloading the entire scene before viewing will create a long startup delay for the viewer.
(2) 3DGS scenes are comprised of a large number of splats (700k to 1M splats in standard datasets), which are of varying importance to the visual quality. 
It's not clear which splats should be prioritized for delivery to the client.
(3) 3DGS scenes immersively surround the user.
Users have the full 6 degrees of freedom (6DoF) to walk around and view different layers and parts of the scene from different angles, making it difficult to determine what parts to deliver to users.

In this work, we design an efficient and high-fidelity delivery framework for 3D scenes, \system, using 3DGS as the underlying data representation.
Our framework addresses the above challenges as follows.
(1) Leveraging the unique structure of the Gaussian splats, we design a custom training scheme that produces \emph{layered} 3DGS. 
This representation enables a ``base layer'' of splats to be displayed first, followed by additional ``enhancement layers'' on top.
This enables progressive download of different parts of the scene based on network bandwidth, while re-using previously downloaded layers.
(2) To enable fine-grained, scalable selection of visually important splats for downloading, instead of making individual decisions for each splat, we segment the 3D scene into objects, where each object is a set of splats.
This grouping also enables interactive editing of the 3D scene, so the user can interact with semantic objects instead of individual splats.
(3) We collected traces of users wearing virtual reality headsets and moving around standard 3DGS scenes. This powers an a user prediction module that determines which splats are likely to be relevant and important to the user's viewport, and hence require priority delivery.

Overall, the contributions of the paper are:
\squishlist
\item Custom training to create layered 3D Gaussian splats that represent 3D scenes. Objects are segmented in the scene to provide fine-grained control for downloading and editing.
\item Scheduling algorithms to choose the right sets of splats to download maximize visual quality, based on predictions of the user's future viewport and network bandwidth. The scheduler accommodates other  compressed 3DGS representations too.
\item Experiments to measure visual quality and latency, driven by traces that we collected of users exploring 3DGS scenes using VR headsets (Meta Quest 3).
\squishend

The paper is organized as follows.
\Cref{sec:related} discusses the background and related work.
\Cref{sec:system_design} describes the overall system design and the individual modules.
\Cref{sec:results} shows the experimental results and we conclude in \Cref{sec:conclusions}.
The code and technical report are available at \url{https://github.com/mavens-lab/layered_3d_gaussian_splats}.

\vspace{-10pt}
\section{Background and Related Work}
\label{sec:related}

\textbf{3D representations.}
Traditional 3D representations such as point clouds and meshes usually represent 3D scenes explicitly. \emph{\textbf{Point clouds}} represent 3D scenes using a set of points in 3D space. It can also include other features like colors to better represent the scene. \emph{\textbf{3D meshes}} use vertices, edges and faces to represent 3D objects. These vertices can also have feature vectors associated with them.
Despite many efforts to efficiently stream such representations~\cite{han2020vivo,guan2023metastream,liu2024muv2}, they do not leverage the latest advances in 3D representations.

Emerging 3D representations such as NeRF and 3DGS can more accurately represent 3D scenes. \emph{\textbf{NeRF}} \cite{mildenhall2021nerf} represents 3D scenes as a continuous volumetric field using a multi-layer perceptron (MLP) \cite{barron2021mip}. 
\emph{\textbf{3DGS}}~\cite{kerbl3Dgaussians} represents 3D scenes as a set of 3D Gaussian splats. Each splat is represented by a set of attributes including its position, opacity, a covariance matrix containing size and rotation information, and Spherical Harmonic coefficients representing view-dependent color. When rendered, these 3D Gaussian splats will be projected into 2D camera coordinates and a tile-based rasterizer will be used for color computation.

\textbf{3DGS training}. 
To train a set of splats to represent a 3D scene ~\cite{kerbl3Dgaussians}, 
the splats will be initialized with a sparse point cloud produced by Structure-from-Motion (SfM) methods.
The training proceeds in iterations, where in each iteration adjusts the values of the splat attributes (color, position, radius, etc.) to create rendered images that match the ground truth.
At the same time, densification and pruning processes are used to improve the overall quality of the 3DGS model and ensure that splats are created in the right places to accurately represent the 3D scene.
Specifically, to cover the geometry in under-reconstructed regions (regions with too few splats), the training algorithm will  \emph{\textbf{clone}} the splats in the region by simply creating a copy of the splats.
Also, because larger splats in visually complex regions may not be able to capture all the details, the training will \emph{\textbf{split}} a splat into smaller splats (replace a splat with two new ones). The training will also \emph{\textbf{prune}} unimportant splats (\eg splats with too low opacity) to keep the scene a reasonable size.



\textbf{NeRF and 3DGS optimizations for efficient delivery.}
Although NeRF and 3DGS have high visual quality, they still face problems that impede their efficient delivery.
NeRF suffers from large model size and slow rendering~\cite{muller2022instant}, while 3DGS faces the challenges of large model size \cite{shi2024lapisgslayeredprogressive3d, papantonakis2024reducing}. 
Compression techniques are needed to decrease model size for efficient network delivery and rendering. 
For NeRF compression, efforts usually focus on decreasing the size of the MLP~\cite{chen2023mobilenerf,muller2022instant,chen2024nerfhub}.
Standard compression techniques such as quantization, pruning and knowledge distillation can also be used for NeRF compression \cite{chen2024far} and 3DGS compression \cite{papantonakis2024reducing}. 
For 3DGS, LightGaussians~\cite{fan2023lightgaussian} uses knowledge distillation, pseudo-view augmentation and global significance scores to compress 3D scenes.
To eliminate structural redundancies and further compress 3DGS, Scaffold-GS ~\cite{lu2024scaffold} used anchors to cluster 3DGS.
HAC~\cite{chen2024hachashgridassistedcontext} further introduced a hash grid to make 3DGS representation more compact.
Based on octrees, Octree-GS ~\cite{ren2024octree} anchors Gaussians with different level-of-details to improve the rendering performance of 3DGS. 
Mip-Splatting~\cite{yu2024mip} uses a 3D smoothing filter and a 2D Mip filter to improve rendering quality. 
These works mainly focus on non-layered representations and do not consider their network delivery.
Further, we will demonstrate that \system can work alongside versions of \cite{shuai2024LoG,kerbl2024hierarchical,chen2024hachashgridassistedcontext}. 

Recently, LapisGS~\cite{shi2024lapisgslayeredprogressive3d} also proposed layered 3DGS, but lack the ability to control the target number of splats, and hence the scene size. Based on LapisGS, LTS~\cite{sun2025lts} adapts layered representations for dynamic 3DGS scene streaming but does not segment the scene into objects for fine-grained splat scheduling, as we do. Other recent work studies 3DGS volumetric video streaming and integrates traditional video delivery optimization into their system design \cite{sun20243dgstream, sun2025lts,liu2024swings,wang2024v}, whereas we focus on scene delivery.



\textbf{Multimedia delivery.}
Layering and viewport prediction are two important techniques that can be used to improve the efficiency of multimedia streaming. 
Layering can improve streaming efficiency by making the streaming more adaptive.
Scalable Video Coding (SVC)~\cite{schwarz2007overview} encodes video stream in layered structure, where a base layer provides a minimum quality level, and enhancement layers improve the resolution, frame rate, and quality.
\rev{Multiple Description Coding (MDC) differs from layered codecs by encoding multiple independent descriptions that can be sent over separate network paths, with any additional received streams enhancing quality. 
\system has more similarities to SVC but for 3D scenes, not 2D videos.}

Viewport prediction makes streaming more efficient by only fetching parts of the scene that include what a user is about to view.
Regression, machine learning, and video saliency features are commonly used techniques for viewport prediction~\cite{10.1145/3241539.3241565,10.1145/3359989.3365413,li2023viewport}.
In our work, we focus on lightweight 6 DoF viewport prediction where we have to predict not only orientation but position in a 3D scene, particularly for the standard 3DGS scenes for which user traces and viewport prediction has not been well studied.

%% file: system_design.tex
\section{System Design}
\label{sec:system_design}

\begin{figure}[htbp]
    \centering
    \includegraphics[width=0.45\textwidth]{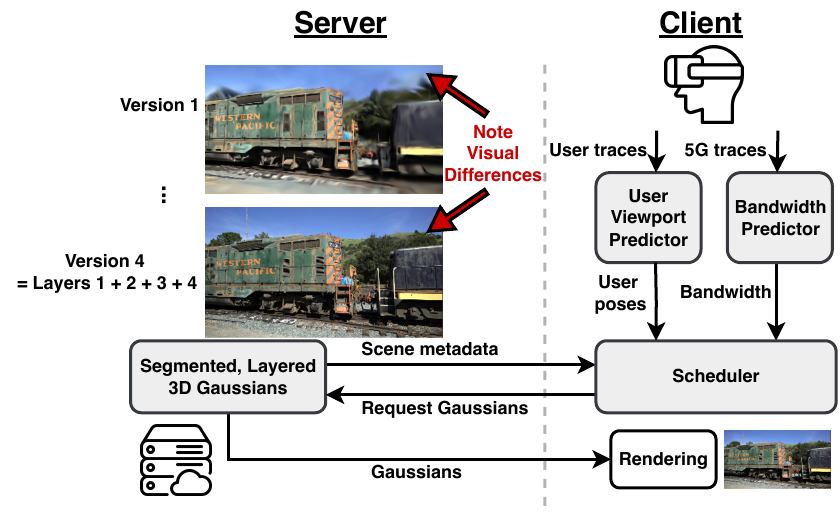}
    \vspace{-5pt}
    \caption{System architecture. Given a set of layered and segmented 3D Gaussian splats, \system retrieves the most useful splats within the user's predicted viewport and network bandwidth. 
    }
    \label{fig:framework}
\end{figure}

The architecture of \system is presented in \Cref{fig:framework}. Given a 3D scene comprised of 3DGS, the system decides what are the best splats to retrieve in order to render the content in the user's viewport while respecting the estimated network bandwidth. To accomplish this, there are four components:

\squishlist
\item \textbf{Segmented, layered 3D Gaussian splats (\Cref{sec:layered_splats})}. To provide users with progressively improving quality, we create 3DGS scenes with layers, including a base layer and several enhancement layers. Further, we also create more complex layered scenes that are automatically segmented into semantically meaningful object, enabling user interactions and scene editing.
\item \textbf{Splat download scheduler (\Cref{sec:scheduler}).} Given a 3DGS scene, our scheduler determines what splats to download for each object in each layer,
based on the utility values of each segmented object in each layer, plus the available network bandwidth. 
We formally define the optimization problems for various cases (layered, non-layered, segmented, non-segmented), prove the NP-hardness of the main cases, and design optimal algorithms to solve them.
\item \textbf{User viewport predictor (\Cref{sec:viewport}).} We collect our own traces of users  wearing VR headsets (Meta Quest 3) and their 6-DoF movements around standard 3DGS scenes.
To predict the user's future viewport based on past history, we use linear regression due to its simplicity and success.
\item \textbf{Bandwidth predictor (\Cref{sec:bandwidth}).} We use outdoor 5G users' walking traces~\cite{narayanan2020lumos5g} to simulate variable 5G network bandwidth. To predict the available network bandwidth, 
we borrow from existing methods~\cite{yin2015mpc}.
\squishend

%% file: models.tex
\subsection{Segmented, Layered 3D Gaussian Splats}
\label{sec:layered_splats}


\begin{figure}[htbp]
\includegraphics[scale=0.35]{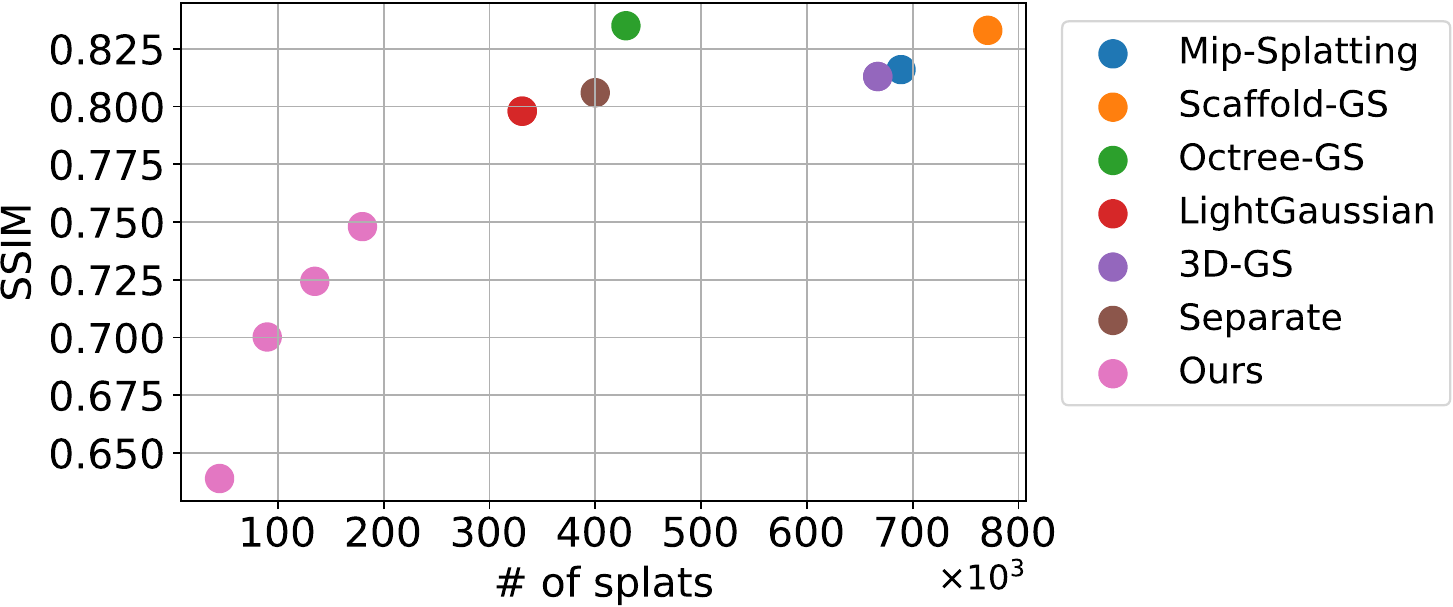}
    \vspace{-20pt}
    \caption{
    Our approaches (``Ours'' and ``Separate'') can gracefully trade off visual quality for the number of splats. Data from the ``Train'' scene \cite{knapitsch2017tanksntemple}.
    } \label{fig:layered_baseline_simple}
\end{figure}

A key component of \system is the layered 3DGS representation.
When creating layered splats, we need to balance between the number of splats (which is directly proportional to model size) and the visual quality (in terms of SSIM).
\Cref{fig:layered_baseline_simple} presents this tradeoff for existing methods, alongside our proposed methods (to be discussed below).
With default pre-trained 3DGS models~\cite{kerbl3Dgaussians}, although they have high visual quality, each scene can contain anywhere from 650K to more than 5M Gaussian splats, which is challenging for network delivery due to the large size (\eg 1.52 GB for the bicycle scene from Mip-NeRF360~\cite{barron2022mipnerf360}).
This is represented by ``3D-GS'' for a specific 3D scene in \Cref{fig:layered_baseline_simple}.
Existing 3DGS methods (Mip-Splatting~\cite{yu2024mip}, Scaffold-GS~\cite{lu2024scaffold}, Octree-GS~\cite{ren2024octree}, LightGaussian~\cite{fan2023lightgaussian}) can achieve similarly high visual quality (SSIM $\geq 0.8$), but still require at least 300k splats.
In contrast, our goal is to create splats that achieve a graceful tradeoff between visual quality and number of splats, represented by ``Ours'' and ``Separate'' in the figure.

\begin{algorithm}
\small
\caption{The overall pipeline of Layered Model Training}\label{algo:layered}
\begin{algorithmic}[1]
\Require Pretrained 3D-GS: $\mathcal{G}$, Target size $\mathcal{D} = \{d_l\}_{l=1}^{L}$
\Ensure Layered 3D-GS $\mathcal{LG}_{1},...,\mathcal{LG}_{L}$
\State $\mathcal{G}_{L} = \{G_i\}_{i=1}^{d_{l}}$ $\leftarrow$ \textsc{Prune2TargetSize}($\mathcal{G}$, $d_{L}$) \Comment{\Cref{algo:Prune}}
\State $\mathcal{LG}_{1},...,\mathcal{LG}_{L}$ $\leftarrow$ \textsc{ProgressiveTraining}($\mathcal{G}_{L},\mathcal{D}$) \Comment{\Cref{algo:layered_train}}
\end{algorithmic}
\end{algorithm}

\begin{figure}
    \centering
        \includegraphics[scale=0.13]{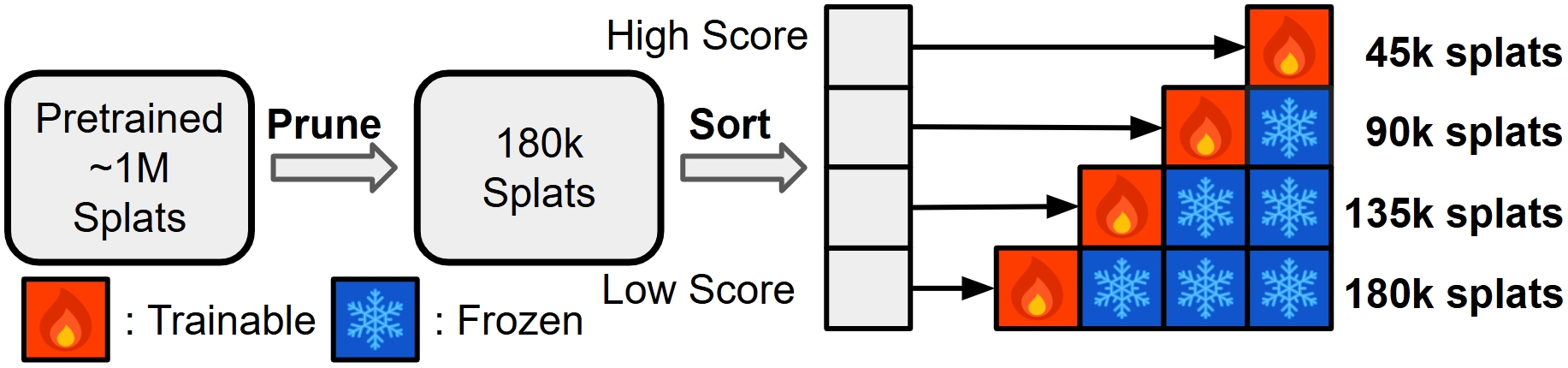}
        \vspace{-10pt}
    \caption{Training framework for layered 3DGS. The layers are iteratively trained, with subsequent layers relying on frozen splats from preceding layers.}\label{fig:layered}
\end{figure}

The overall procedure of training these layered 3DGS is summarized in \Cref{algo:layered} and \Cref{fig:layered}.
There are two main steps in the example shown, where we want to create a scene with 180k total splats, split into layers of 45k splats each.
First, we create an initial scene with a controlled number of splats by pruning from the default pretrained model (1M splats) down to the total target number (180k splats on the left side of \Cref{fig:layered}).
Second, we take the 180k splats and split them into layers, 45k splats each in the example, and train them in an iterative fashion (right side of \Cref{fig:layered}).
Below, we delve into these steps in more detail.
Throughout this work, we call the different qualities of a 3D scene as separate ``versions'', and the delta between versions as ``layers''.
In the above example, the versions contain 45k, 90k, 135k, or 180k splats, while each layer has 45k splats.




\textbf{(1) Creating an initial scene with a controlled total number of splats.} 
To create splats that can trade off between visual quality and size, we need to be able to control the total number of splats in the initial 180k model.
In the original 3DGS training pipeline, the number of splats is indirectly controlled by a set of hyperparameters, such as $\tau_p$ for cloning and splitting decisions, or the densification interval.
However, since these hyperparameters interact with each other in complex ways to produce the final output, the mapping between the hyperparameter settings and total number of splats is unclear.
Furthermore, the mapping could change for different 3D scenes being created.

Therefore, to effectively control the model size, we design \Cref{algo:Prune} to obtain a layer of precisely the desired size while achieving good visual quality. 
We start with the default pre-trained model from \cite{kerbl3Dgaussians} (line~\ref{line:init}), and prune down to the target number of splats from there.
This is a top-down approach; the basic idea is to repeatedly prune the number of splats (which decreases the visual quality), and then grow (which improves the visual quality) multiple times.
Note that we also experimented with a bottom-up approach where we grew the number of splats to the target size, but found top-down approach worked better in practice.

Specifically, if the number of splats is much larger than the target $d$, we \textsc{Prune} the splats down to remove the worst $r$ fraction of splats, according to their global significance score~\cite{fan2023lightgaussian} (line \ref{line:prune}).
After pruning, we run \textsc{Recovery} to recover the visual quality by training the splat parameters, following a regular training iteration that allows cloning, splitting, and pruning (line~\ref{line:recovery}).
This might cause the number of splats to increase again, so we repeat the pruning and recovery process multiple times.
For the final pruning, to ensure the number of splats in the final model exactly matches the target, we \textsc{Finetune} the splat attributes according to a regular training iteration but do not allow cloning, splitting, or pruning (line~\ref{line:finetune}).
After the desired number of splats are obtained, \system sorts the splats by their score to prepare for the next step, layered model training (line~\ref{line:final_sort}). 
\begin{algorithm}
\caption{\textsc{Prune2TargetSize}}\label{algo:Prune}
\begin{algorithmic}[1]
\Require Pretrained 3DGS: $\mathcal{G}$, Target size $d$, Max pruning ratio $r$, Training view images $\mathcal{I}$, and their associated camera poses $\mathcal{P}$
\Ensure pruned and sorted 3DGS $\mathcal{G}_{L} = \{G_i\}_{i=1}^{d}$ 
\State Initialize 3D-GS $\mathcal{G}^{*}$ $\leftarrow$ $\mathcal{G}$ \Comment{$\mathcal{G}$ pretrained by \cite{kerbl3Dgaussians} } \label{line:init}
\State $n$ = \textsc{NumOfGS}($\mathcal{G}^{*}$) \Comment{Compute the number of  splats}
\While{$n$ $>$ $d$} 
\State $\mathcal{S}$ $\leftarrow$ \textsc{CalGS}($\mathcal{G}^{*}, \mathcal{I}, \mathcal{P}$) \Comment{Compute global significance score~\cite{fan2023lightgaussian}}
\If{$n*(1-r)$ $>$ $d$ \label{line:splat_number_condition}}
\State	$\mathcal{G}^{*}$ $\leftarrow$ \textsc{Prune}($\mathcal{G}^{*},\mathcal{GS},n*r$) \Comment{Prune $r$ faction of splats with lowest $\mathcal{S}$} \label{line:prune}
\State $\mathcal{G}^{*}$ $\leftarrow$ \textsc{Recovery}($\mathcal{G}^{*}, \mathcal{I}, \mathcal{P}$, IsRefinementIteration = True)\Comment{Training method \cite{kerbl3Dgaussians} with prune, split, and clone} \label{line:recovery}
\State $n$ = \textsc{NumOfGS}($\mathcal{G}^{*}$) 
\Else \State $\mathcal{G}^{*}$ $\leftarrow$ \textsc{Prune}($\mathcal{G}^{*},\mathcal{GS},n-d$)
\State $\mathcal{G}^{*}$ $\leftarrow$ \textsc{Finetune}($\mathcal{G}^{*}, \mathcal{I}, \mathcal{P}$, IsRefinementIteration=False) \Comment{Training method \cite{kerbl3Dgaussians} without prune, split and clone.} \label{line:finetune}
\EndIf
\EndWhile
\State $\mathcal{S}$ $\leftarrow$ \textsc{CalGS}($\mathcal{G}^{*}, \mathcal{I}, \mathcal{P}$)
\State $\mathcal{G}_L$ $\leftarrow$ \textsc{Sort}($\mathcal{G}^{*}$, $\mathcal{S}$ ) \Comment{Sort splats by their global significance score in descending order.}  \label{line:final_sort}
\end{algorithmic}
\end{algorithm}

\textbf{(2) Layered model training.} In order to support efficient 3DGS delivery, we need to create layered splats that can overlay on top of one another to increase visual quality. 
A naive strategy to create a layered structure is to create a combined loss function and jointly train splats in the $L$ layers \emph{simultaneously}.
For example, if there are $L=2$ layers, the first model would be trained with loss function $\ell_1$ (corresponding to splats in layer 1), and the second model trained simultaneously with loss function $\ell_1 + \ell_2$ (corresponding to splats in layer 1 and layer 2).
However, we found that with this technique, the splats in layer 1 would effectively get weighted twice in every training iteration compared to layer 2, leading to unstable layer 1 splats and poor performance.

To overcome this, we had to design a custom training procedure instead of simply modifying the training loss function, given in \Cref{algo:layered_train}.
The main idea, as shown in the right side of \Cref{fig:layered}, is to incrementally train the models from the lower to higher layers.
To ensure the higher layers can overlay on the preceding lower layers, when training each layer, we freeze the lower layers and only train on the added splats. 
In the example in \Cref{fig:layered}, we first train layer 1 (45k splats in the figure), freeze layer 1, add on another 45k splats for layer 2 and train them, and so on.

Specifically, we first obtain the splats output by \Cref{algo:Prune} and partition its splats into $L$ layers, denoted as $\Delta \mathcal{LG}_l$ (line~\ref{line:split_layers}).
We train the smallest layer $\mathcal{LG}_1$ using the \textsc{Finetune} function, which doesn't change the number of splats (line~\ref{line:finetune_smallest}).
For subsequent layers, we freeze the splats from the preceding layers (line~\ref{line:freeze_gradient}) and only allow training for the current version's newly added splats (line~\ref{line:train_gradient}), training them using \textsc{Finetune}, which again doesn't change the number of splats.
This continues until all the enhancement layers have been trained.
Note that in each training iteration, the loss function is computed based on rendering all splats in a given layer and its preceding layers ($\mathcal{LG}_l$), but only the latest enhancement layer's splats ($\Delta \mathcal{LG}_l$) are allowed to be updated to reduce the loss.
This enables the layered effect.
\begin{algorithm}
\caption{\textsc{ProgressiveTraining}}\label{algo:layered_train}
\begin{algorithmic}[1]
\small
\Require Pruned and sorted 3DGS: $\mathcal{G}_{L} = \{G_i\}_{i=1}^{d_{L}}$, cumulative target \# of splats $\mathcal{D} = \{d_1, d_2, \ldots d_L\}$
\Ensure Layered 3D-GS $\mathcal{LG}_1,...,\mathcal{LG}_{L}$
\State $\mathcal{LG}_1,\Delta \mathcal{LG}_{2},...,\Delta \mathcal{LG}_{L}$ $\leftarrow$ $\{G_i\}_{1}^{d_{i=1}},\{G_i\}_{i=d_{1}+1}^{d_{2}}, \ldots ,\{G_i\}_{i=d_{L-1}+1}^{d_{L}}$ \Comment $\Delta \mathcal{LG}_{l}$ means  $\mathcal{LG}_{l} \setminus \mathcal{LG}_{l-1}$ \label{line:split_layers}
\State $\mathcal{LG}_1$ $\leftarrow$ \textsc{Finetune}($\mathcal{LG}_1, \mathcal{I}, \mathcal{P}$,IsRefinementIteration=False) \Comment{Train base layer} \label{line:finetune_smallest}
\For {$l \leftarrow 2$ to $L$} 
\State $\mathcal{LG}_l =  \{G_i\}_{i=1}^{d_{l}} $ $\leftarrow$ \textsc{Concatenate}($\mathcal{LG}_{l-1}$, $\Delta \mathcal{LG}_{l}$)
\State $\{G_i\}_{1}^{d_{l-1}}$.requires\_grad\_(false) \Comment{Freeze splats from previous layers} \label{line:freeze_gradient}
\State $\{G_i\}_{d_{l-1}+1}^{d_{l}}$.requires\_grad\_(true) \Comment{Newly added splats are trainable} \label{line:train_gradient}
\State $\mathcal{LG}_l$ $\leftarrow$ \textsc{Finetune}($\mathcal{LG}_l, \mathcal{I}, \mathcal{P},IsRefinementIteration=False)$ \Comment{Train} \label{line:train_layer}
\EndFor
\end{algorithmic}
\end{algorithm}

\textbf{Segmentation.}
Segmenting the 3D scene into objects is useful in order to group splats together into semantically meaningful objects and do more efficient scheduling later in \Cref{sec:scheduler}.
To implement segmentation in our 3DGS models, we add the object ID as an additional feature to each splat in \Cref{algo:layered}, assigning the initial object IDs according to \cite{ye2024gaussiangroupingsegmentedit}.
These object IDs are refined throughout the training process described above.
Although the total number of splats in each layer $d_{l}$ is fixed, we did not constrain the object ID during training, so the number of splats assigned to each object is non-uniform in a given layer.
In other words, more/less splats can be automatically allocated to different objects by the training process to achieve the best visual quality. 

%% file: formulation.tex
\subsection{Splat Download Scheduler}
\label{sec:scheduler}

Given the layered splats produced by the algorithms in \Cref{sec:layered_splats}, the scheduler needs to determine what splats to download, in what order.
This section describes the problem setup, including the splat utility definition, problem formulation, and scheduling algorithms.
Due to space constraints, proofs and some problem definitions are provided in the technical report~\cite{tech-report}.


\begin{table}
    \small
    \begin{center}
    \vspace{0.1in}
    \begin{tabular}{ |c| p{6cm} | } 
     \hline
     $B[t]$ & predicted bandwidth in time slot $t$\\ \hline
         $c_{jl}$ & for separate splat representations, the cost of object $j$ of version $l$. \\ \hline
    $\Delta c_{jl}$ & for layered splat representations, the cost of  object $j$ of layer $l$. $\Delta c_{jl}[t] = c_{jl}[t] - c_{j,l-1}[t]$\\ \hline
    $d_l$ & Target number of splats for layer $l$\\ \hline
     $\mathcal{G} = \{G_i\}$ & entire set of splats in scene \\ \hline 
     $hw$ & length of history time window (s) \\ \hline
     $pw$ & length of prediction time window (s)\\ \hline    
     $N$ & number of layers per scene\\ \hline
     $M_{jl}$ & number of splats for object $j$ in layer $l$ \\ \hline
     $T$ & duration of time slot \\ \hline
     $U_{jl}[t]$ & for separate splat representations, the utility of object $j$ of version $l$ at time $t$. \\ \hline
     $\Delta U_{jl}[t]$ & for layered splat representations, the utility of object $j$ from layer $l$ at time $t$. See (\ref{eqn:utility_diff}). \\ \hline
     $\tilde{U}_{jl}[t]$ & utility of object $j$ from layer $l$ from time $t$ onwards. See (\ref{eqn:cum_utility}). \\ \hline
     $x_{jl}[t]$ & decision variable of whether to download object $j$ from layer $l$ at time $t$\\ \hline
     $y_{jl}[t]$ & binary indicator of whether object $j$ from layer $l$ is stored at time $t$\\ \hline
    \end{tabular}
    \end{center}
    \caption{Table of Notation.}
    \label{table:notation}
    \vspace{-10pt}
\end{table}

\subsubsection{Problem Setup}
\label{sec:problem_setup}

As described in \Cref{sec:layered_splats}, a scene is comprised of splats encoded into different layers.
An enhancement layer of an object is only useful if the base layer and all lower enhancement layers were previously downloaded.
Optionally, the scene can be segmented into objects, and multiple splats belong to a single object.
Notation-wise, this means each splat has a layer ID $l$ and optionally an object ID $j$.
Time is divided into slots of duration $T$ indexed by $t$.
Given the bandwidth $B[t]$ over time, the goal is to select which splats to download that maximize the total utility while staying under the available bandwidth.
We denote the main decision variable $x_{jl}[t]$ as whether to download the splats comprising object $j$ at layer $l$ at time slot $t$. 
There are several scenario variations, depending on whether the splats at layered (or not) and whether the scene is segmented into objects (or not).
Our scheduler is designed to work with all these scenarios, which are summarized in \Cref{table:prob_taxonomy}.
The mathematical notation is summarized in \Cref{table:notation}.

\begin{table}[]
\small
\begin{tabular}{|p{1cm}|p{3cm}|p{3cm}|}
\hline
 & \thead{\textbf{segmented objects}} & \thead{\textbf{not segmented}} \\ \hline
\thead{\textbf{layered}} & \makecell{\textbf{case I} (\Cref{sec:case1});\\ NP-hard (\Cref{thm:layered_np_hard});\\ Solved by knapsack.} & \makecell{\textbf{case III} (\Cref{sec:case3_4});\\ NP-hard (\Cref{thm:layered_np_hard});\\ Solved by knapsack.} \\ \hline
\thead{\textbf{separate}}
& \makecell{\textbf{case II} (\Cref{sec:case2});\\ NP-hard (\Cref{thm:separate_np_hard});\\ Solved by progressive\\ loading.}
& \makecell{\textbf{case IV} (\Cref{sec:case3_4});\\ NP-hard (\Cref{thm:separate_np_hard});\\ Solved by progressive\\loading.} \\ \hline
\end{tabular}
\caption{Taxonomy of problems and algorithms for different 3D Gaussian splat representations.}
\label{table:prob_taxonomy}
\end{table}

\paragraph{Splat utility definition.}
We must assign a numerical utility value to each splat to determine its importance to the scene and whether its download should be prioritized by the scheduler.
However, it is difficult to assign a utility to an individual splat because it is just one component of the rendered view; typical methods of estimating visual quality like PSNR or SSIM require an entire image (similarly, calculating the SSIM of a 3D point in a point cloud is ill-defined).
Therefore, we design a splat utility function for use by the scheduler.
The utility of a splat $i$ in object $j$ in version $l$ at time $t$ is:
\begin{align} 
    U_{ijl}[t] = \text{closeness}_{ijl}[t] \times \text{overlap}_{ijl}[t] \times \text{opacity}_{ijl}
    \label{eqn:utility}
\end{align}
The ``closeness'' measures the 3D Euclidean distance between the center of the splat to the center of the user's viewport.
The ``overlap'' is the 2D area of the 3D splat when projected onto the user's 2D viewport.
The ``opacity'' is a standard feature stored within the splat data structure.
Note that closeness and overlap depend on the user's viewport at time $t$, while opacity is a static property.
The intuition is that large, opaque splats that are near the center of the viewport should have higher utility.
\rev{The total utility of an object is the sum of utilities of splats that comprise that object, \ie $U_{jl} = \sum_i U_{ijl}$.}

\rev{These splat utilities are pre-computed offline, and retrieved online for the relevant splats and objects in the viewport as needed.}
Note that this utility function differs from the global significance score (GSG)~\cite{fan2023lightgaussian}, which is used to train the layered splats in \Cref{algo:Prune}.
\rev{The GSG does not depend on the viewport; our utility function incorporates similar factors as the GSG, such as splat opacity, but adds viewport dependence because we seek to optimize the visual quality of the user's current viewport, rather than a general set of splats.}
We experiment with a version of \rev{the global significance score} as a baseline in \Cref{sec:results}.

\begin{figure}
\centering
\includegraphics[width=0.4\textwidth]{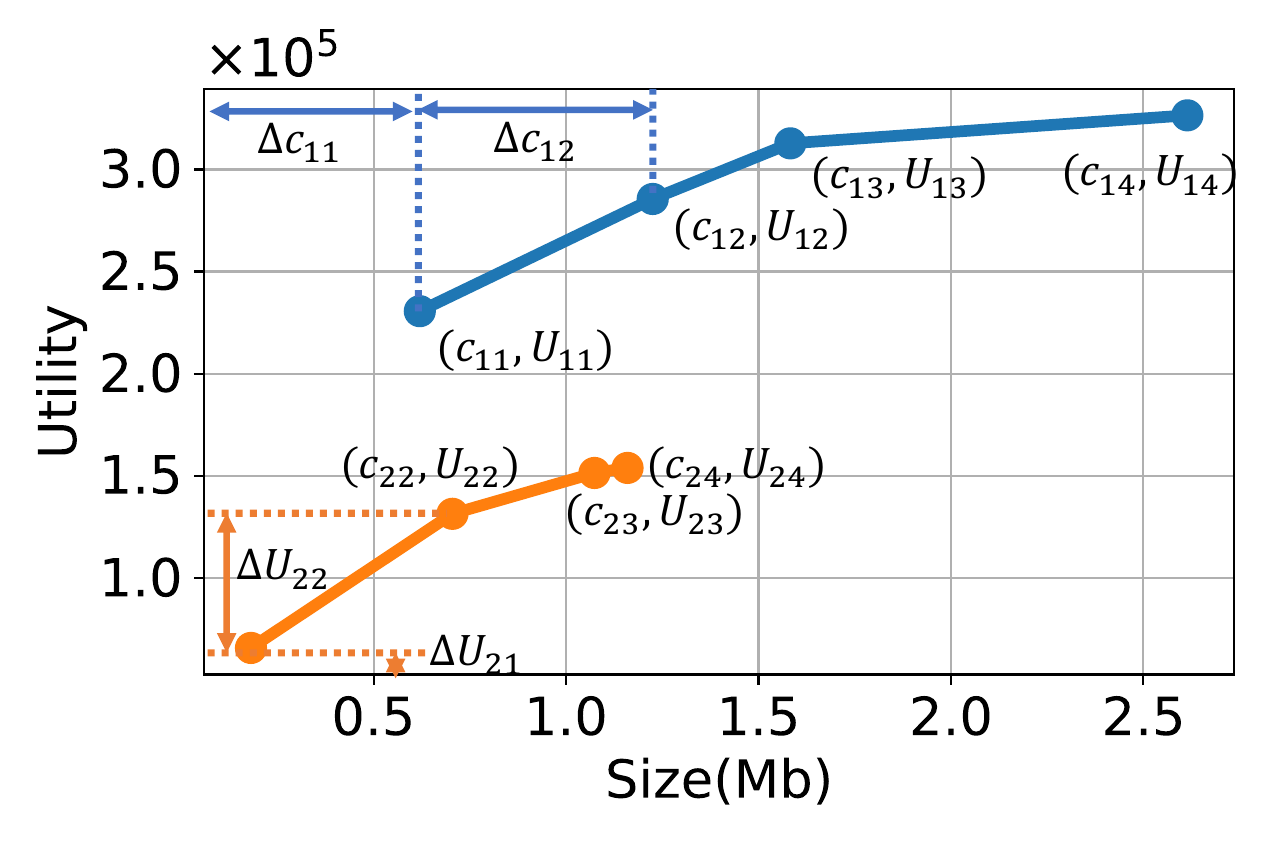}
    \vspace{-10pt}
  \caption{Illustration of the utility function (\ref{eqn:utility}) for two objects (egg and pork belly) from the ``Ramen'' scene~\cite{ye2024gaussiangroupingsegmentedit}.
  }\label{fig:utility}
\end{figure}

It's also useful to define the utility of an object in a layer (rather than a version in (\ref{eqn:utility})) as the difference from the previous version:
\begin{align} 
    \Delta U_{jl}[t] = U_{jl}[t] - U_{j,l-1}[t]
    \label{eqn:utility_diff}
\end{align}
Intuitively, this represents the additional utility provided by layer $l$.An empirical example of the utility function for a single splat at time $t$ is shown in \Cref{fig:utility}.
Note that $U_{jl}$ and $\Delta U_{jl}$ are used only by the scheduler to determine the utility of individual splats; the evaluations in \Cref{sec:results} are done using overall scene SSIM according to standard practice.


\subsubsection{Case I: Layered splats with segmented objects}
\label{sec:case1}

This is the main scenario we consider in \system, because it incorporates both layers
semantic segmentation into objects, thus enabling the most fine-grained control.
The scheduling problem is defined as follows.

\begin{problem}{Layered splats with segmented objects} 
\begin{align} 
    \max_{x,y} \sum_j \sum_l \sum_t \Delta U_{jl}[t] y_{jl}[t] \label{eqn:obj}\\
    \text{s.t. }  
    \sum_{j} \sum_l \Delta c_{jl}x_{jl}[t] \leq B[t] \; \; \forall t \label{eqn:bandwidth_constraint} \\
    \sum_t x_{jl}[t] \leq 1 \; \; \forall j,l \label{eqn:select_one_constraint}\\
    y_{jl} = \sum_{t^\prime < t} x_{jl}[t^\prime] \; \; \forall j,t,l \label{eqn:store_splats_constraint}\\
    x_{jl}[t] \leq y_{j,l-1}[t] \; \; \forall j,l,t \label{eqn:layer_constraint} \\
    x_{jl}[t], y_{jl}[t] \in \{0, 1\} \; \; \forall j,t
\end{align}
\label{prob:full}
\end{problem}

The main variable $x_{jl}[t]$ is 1 if layer $l$ of object $j$ is downloaded in time slot $t$, and 0 otherwise.
The objective (\ref{eqn:obj}) is to maximize the total utility across all splats in the viewport across all time.
Constraint (\ref{eqn:bandwidth_constraint}) states that the total size of the downloaded splats in each time slot must not exceed the available predicted bandwidth.
Constraint (\ref{eqn:select_one_constraint}) states that each layer of each object can be downloaded only once over the entire duration of the user trace.
Note that all objects in a given layer do not have be downloaded before proceeding to the next layer, \ie layers can be partially downloaded.
Constraint (\ref{eqn:store_splats_constraint}) keeps track of previously downloaded splats in the helper variable $y_{jl}[t]$, which is 1 if layer $l$ of object $j$ has been downloaded before time slot $t$, and 0 otherwise. In other words, $x_{jl}[t]$ is 1 when the splat is downloaded, and the corresponding $y_{jl}[t]$ is 1 thereafter.
Constraint (\ref{eqn:layer_constraint}) states that layer $l$ can only be downloaded if the preceding layer $l-1$ has previously been downloaded.

To understand the difficulty of \Cref{prob:full}, we first transform it into an equivalent \Cref{prob:gap} according to \Cref{lemma:gap_equiv}.

\begin{lemma} 
{Problem~\ref{prob:gap} is equivalent to Problem~\ref{prob:full}.}
\label{lemma:gap_equiv}
\end{lemma}

\noindent The transformation is done by defining a new cumulative utility  $\Delta \tilde{U}_{jl}[t]$ as 
\begin{align} 
\Delta \tilde{U}_{jl}[t] \equiv \sum_{t^\prime \geq t} \Delta U_{jl}[t], \label{eqn:cum_utility}
\end{align}
\noindent summing up the utility of layer $l$ of object $j$ for a user from time $t$ onward.
Intuitively, it aggregates the total future utility, from time $t$ onwards, of a user for an object based on the user's movements around the scene.
The new \Cref{prob:gap} (see technical report~\cite{tech-report})
This adds a complication to the already NP-hard generalized assignment problem.
However, one subtlety is that adding another constraint to the generalized assignment problem does not necessarily mean that it is also NP-hard; constraining the feasible set could potentially make it easier to solve the problem.
Therefore, we require \Cref{thm:layered_np_hard} to state that our precedence-constrained generalized assignment is NP-hard.

\begin{theorem}
Problem~\ref{prob:full} is NP-Hard.
\label{thm:layered_np_hard}
\end{theorem}

Because \Cref{prob:full} is NP-hard, we turn our attention to heuristic solutions.
In the simple case where $B[t]$ and $U_{jl}[t]$ are static over time, a greedy algorithm (sort the utility functions by their slopes, and pick the best slope each time) is optimal; however, static bandwidth and user viewport assumptions are unrealistic.
In practice, bandwidth and user prediction far into the future is difficult.
We therefore focus on one individual time slot and optimize within that.
This turns out to be a multiple choice knapsack problem, which can be solved in pseudo-polynomial time via dynamic programming.
\Cref{lemma:knapsack} shows the equivalency of \Cref{prob:full} to a knapsack problem for a single time slot.
\begin{lemma} 
    Within a single time slot, Problem~\ref{prob:full} is a multiple choice knapsack problem.
    \label{lemma:knapsack}
\end{lemma}

\subsubsection{Case II: Separate splats with segmented objects}
\label{sec:case2}

Next, we consider the case where the splats are still segmented into different objects but not layered.
We need to choose the right version for each object, and subsequent better quality versions replace prior low quality versions.
We consider this non-layered scenario because recent work in the computer vision community design new compression schemes for non-layered splats, and we would like our framework to be able to work with them (see \Cref{sec:type}).
The main difference from \Cref{prob:full} is we now consider the \emph{maximum} utility across layers for a given object (rather than summation) in the objective function (\ref{eqn:jiasi_obj2}), because subsequent versions replace earlier ones rather than adding onto them as in the layered approach.
The other difference is the lack of precedence constraints, as each layer is downloaded independently, so constraint (\ref{eqn:layer_constraint}) is not needed in \Cref{prob:case2} below.

\begin{problem}{Case II: Separate splats with segmented objects}
\begin{align} 
    \max \sum_j \sum_t \max_{l} U_{jl}[t] y_{jl}[t] \label{eqn:jiasi_obj2}\\
    \text{s.t. }  
    \sum_{j} \sum_l c_{jl} x_{jl}[t] \leq B[t] \; \; \forall t \label{eqn:jiasi_bandwidth2} \\
    \sum_t x_{jl}[t] \leq 1 \; \; \forall j,l \label{eqn:jiasi_select_one2}\\
    y_{jl} = \sum_{t^\prime < t} x_{jl}[t^\prime] \; \; \forall j,t,l \label{eqn:jiasi_y_def2}\\
    x_{jl}[t] \in \{0, 1\} \; \; \forall j,t
\end{align}
\label{prob:case2}
\end{problem}
\noindent It turns out that this problem is also NP-hard according to \Cref{thm:separate_np_hard}, so in practice we solve it by simply loading the versions in sequence, starting version $0, \ldots, L$, similar to a progressive JPEG image.

\begin{theorem} 
Problem~\ref{prob:case2} is NP-hard.
\label{thm:separate_np_hard}
\end{theorem}



\subsubsection{Cases III and IV}
\label{sec:case3_4}
Case III, layered splats without object segmentation, is similar to Case I and is covered by \Cref{prob:full} and hence its knapsack algorithm.
This is because without segmented \emph{objects} in the scene to choose from, the decision is what \emph{layer} of the entire scene to download.
Thus the decision variable $x_{jl}[t]$ represents the choice of whether to download layer $l$ of the entire scene, and the index $j$ is moot.
Once the layer is chosen, the splats within the layer are downloaded in order according to their score (\ref{eqn:utility}).

Case IV, without object segmentation and with separate versions of the scene, can be covered by \Cref{prob:case2} following similar arguments as the previous paragraph. 
The representation is similar to using existing splat compression methods~\cite{fan2023lightgaussian,chen2024hachashgridassistedcontext} and tuning their parameters to create different versions of the same scene.
We consider this as a baseline in the evaluation.

%% file: user_network_predictor.tex
\subsection{User Viewport Predictor}
\label{sec:viewport}
\begin{figure}[htbp]
\centering
\begin{subfigure}[b]{0.15\textwidth}
\includegraphics[scale=0.15]{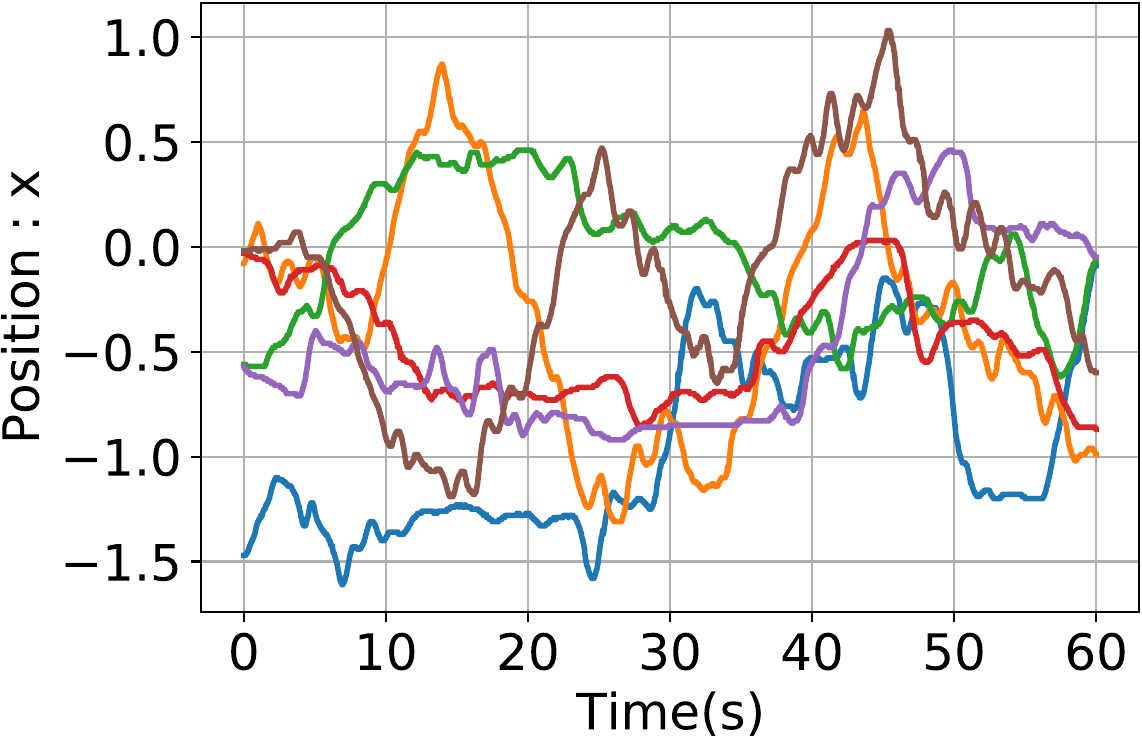}
\end{subfigure}
\begin{subfigure}[b]{0.15\textwidth}
\includegraphics[scale=0.15]{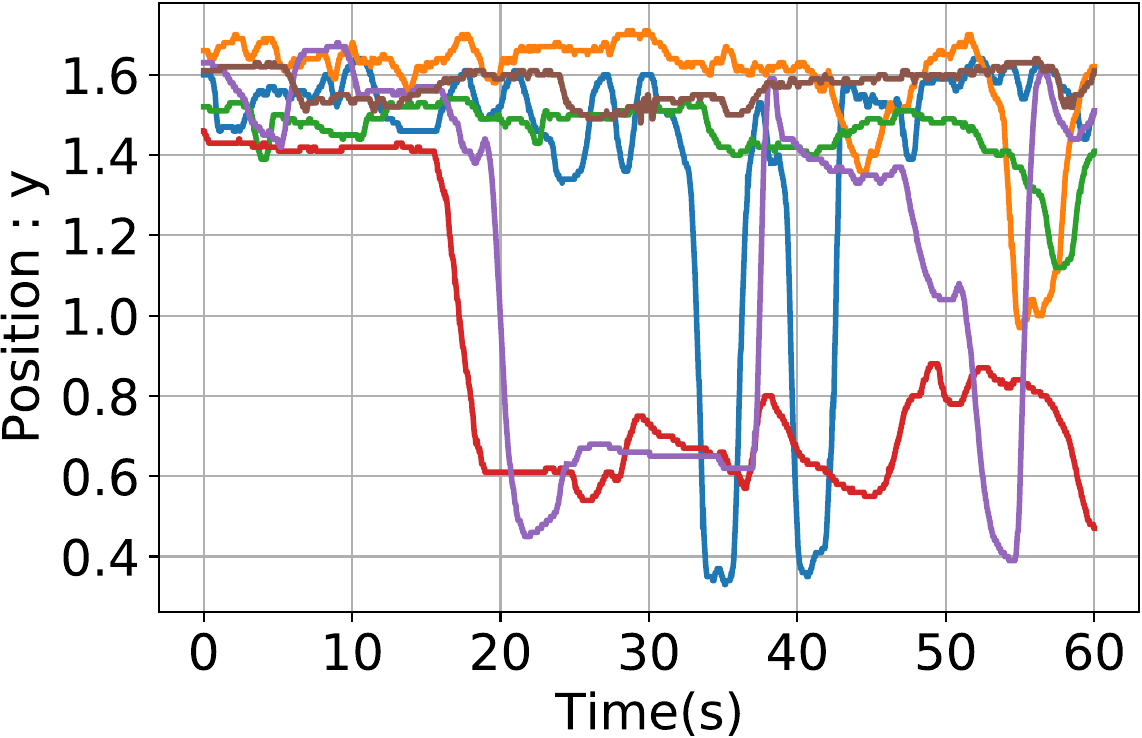}
\end{subfigure}
\begin{subfigure}[b]{0.15\textwidth}
\includegraphics[scale=0.15]{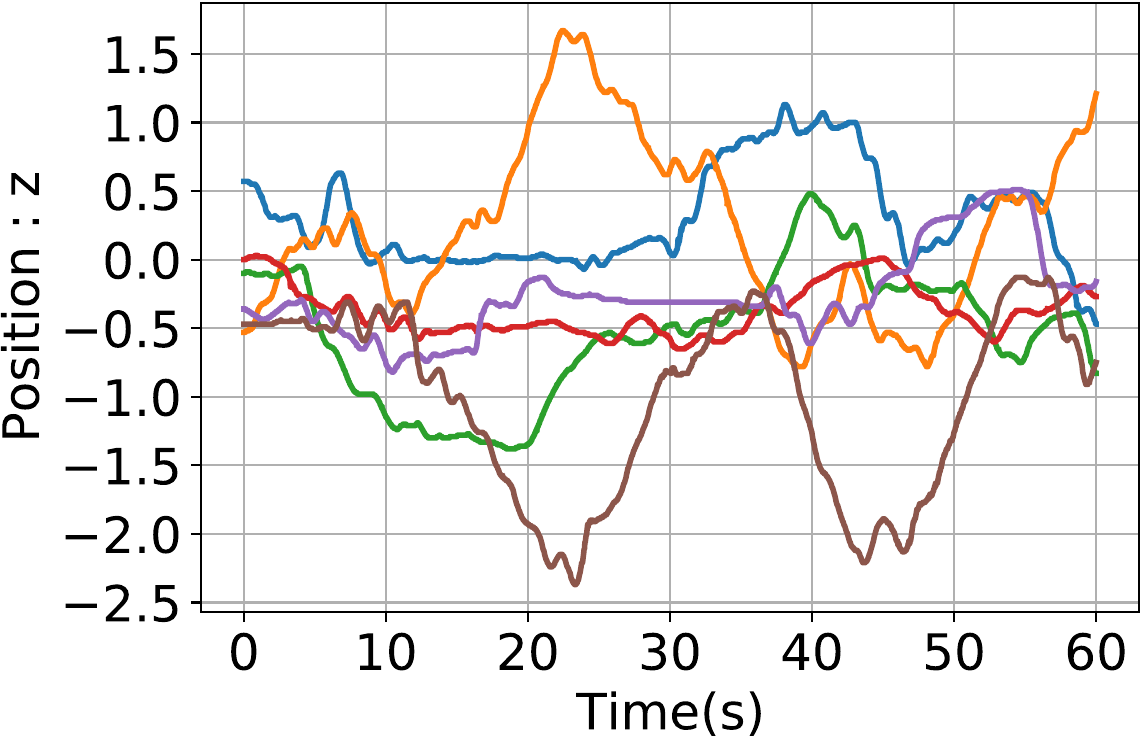}
\end{subfigure}\\
\begin{subfigure}[b]{0.15\textwidth}
\includegraphics[scale=0.15]{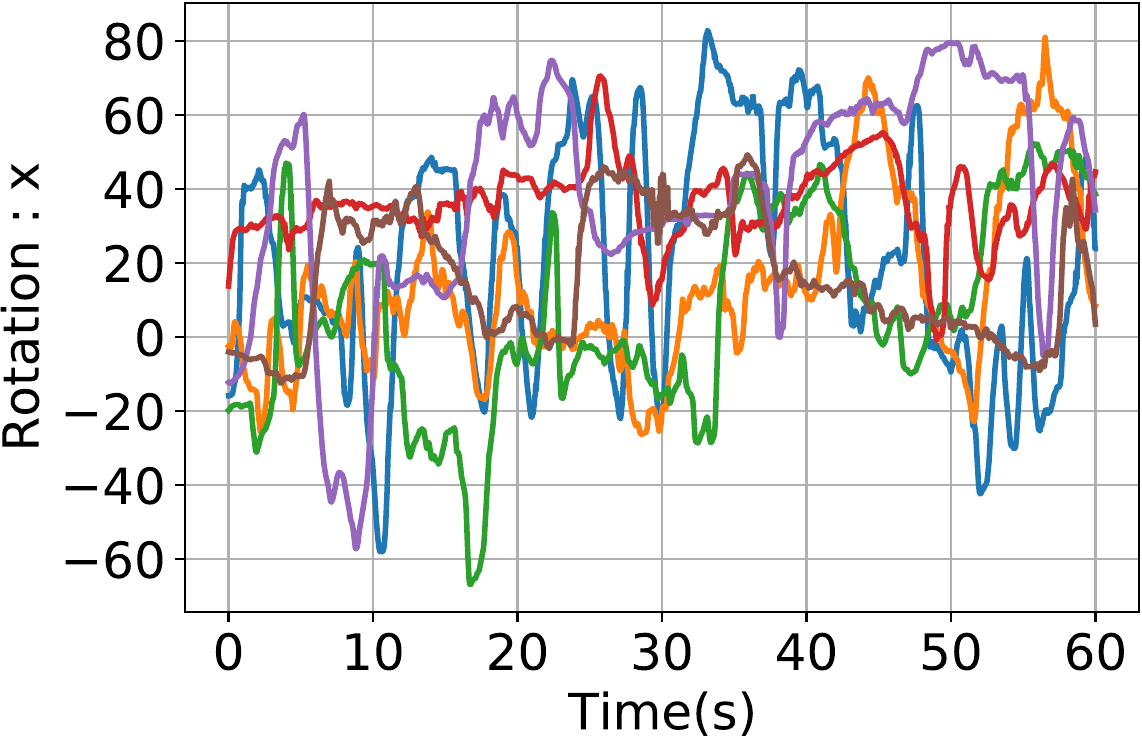}
\end{subfigure}
\begin{subfigure}[b]{0.15\textwidth}
\includegraphics[scale=0.15]{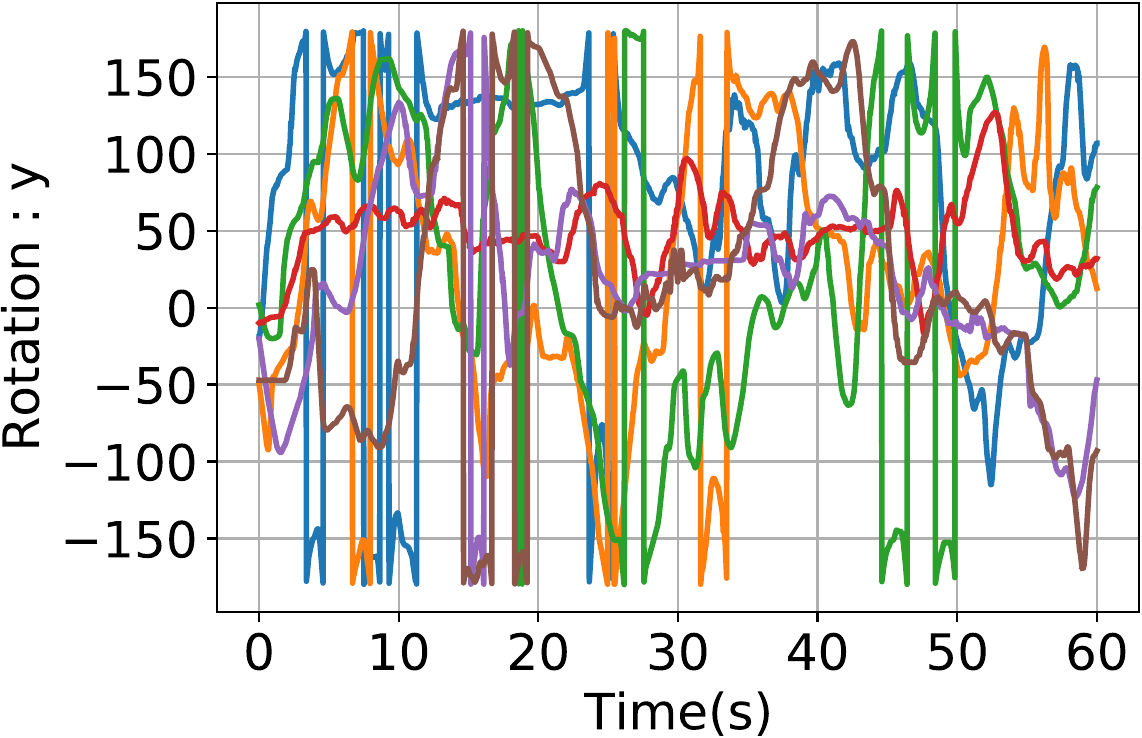}
\end{subfigure}
\begin{subfigure}[b]{0.15\textwidth}
\includegraphics[scale=0.15]{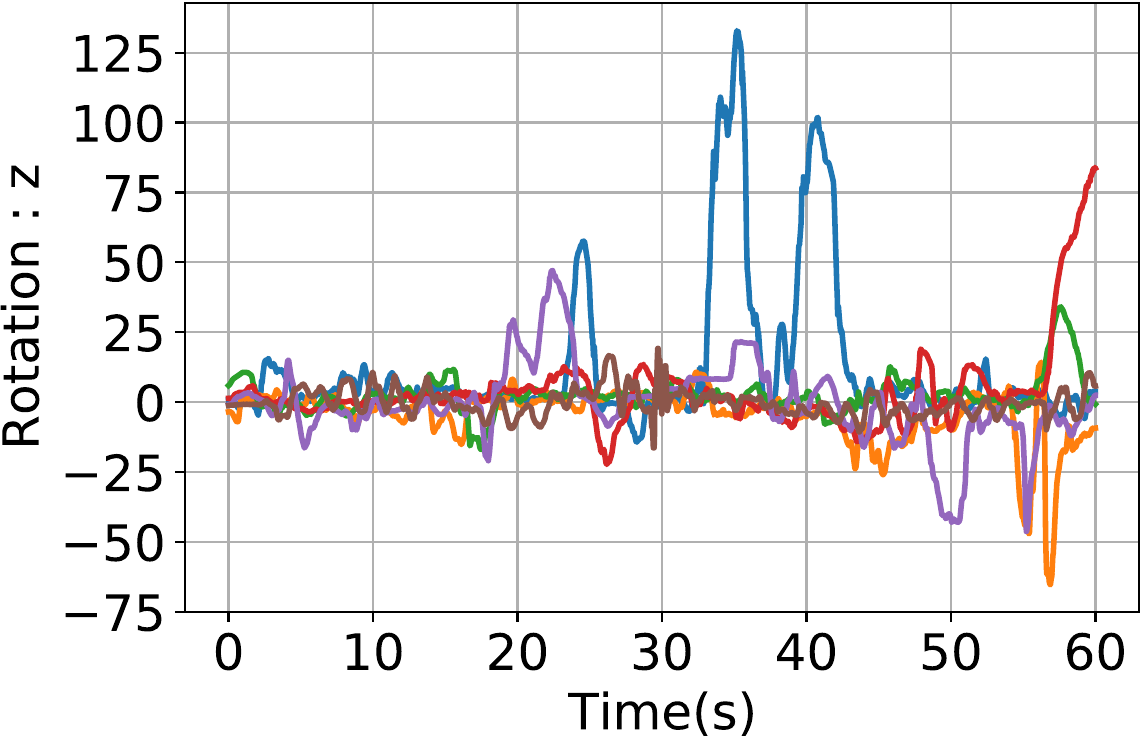}
\end{subfigure}
    \vspace{-10pt}
\caption{Real viewport trajectory data from the bicycle scene, each color representing a different user.}
\label{fig:realtrace}
\end{figure}

\textbf{Trace collection and generation.}
We collected viewport traces from 6 users for 8 scenes each. The viewport trajectory data is recorded as a time series using the Meta Quest 3 VR headset's inside-out tracking with 6DoF. This 6DoF tracking system can track rotational and positional movements along the x, y, and z axes. The sampling rate is set to 36Hz. Examples of the viewport trajectories are shown in \Cref{fig:realtrace}. Additionally, 6 synthetic traces were generated, including elliptical, circular, and spiral paths, and three other paths derived from the original test sets of the 3DGS datasets.

\textbf{Prediction model.}
A linear regression model is used in several viewport-adaptive systems due to its simplicity, speed, and reasonable accuracy in short-term forecasting~\cite{10.1145/3359989.3365413,10.1145/3241539.3241565,10.1145/3210240.3210323}. 
Given the recent viewports within a history window ($hw$), we apply the model to predict the subsequent viewports within a prediction window ($pw$). During runtime, the model is continuously fitted to the data points from the recent history window ($hw$) and then used to forecast future positions in the prediction window ($pw$). The six features of 6DoF are predicted separately. We treat the value of each feature as the dependent variable and its relative order in the sequence, compared to the first point in $hw$, as the independent variable. A challenge arises from the cyclical nature of angles, where 0$^\circ$ is equivalent to 360$^\circ$. To address this, we re-center the data before each prediction to ensure the model accurately adapts to this angular continuity. We set the first point in $hw$ as 0. Using it as the reference, angles increase in the clockwise direction and decrease in the counterclockwise direction, resulting in a final range of [-180, 180]. In practice, we set $hw = 0.5 \text{ s}$ and $pw = 1 \text{ s}$.
\rev{The modular design of \system allows for more sophisticated viewport prediction modules, such as deep learning for panoramic videos~\cite{10.1145/3359989.3365413}, to be slotted in in the future.}


\subsection{Bandwidth Predictor}
\label{sec:bandwidth}

\system makes downloading decisions based on the available network bandwidth.
Specifically, we used a harmonic mean bandwidth predictor with a history window ($hw$) of size 0.5 seconds to predict the next second into the future ($pw$).
\rev{The harmonic mean is defined as $\frac{n}{\sum_{i=1}^n \frac{1}{a_i}}$, where $n$ is the number of bandwidth samples in $hw$, and $a_i$ are the sample values.
We further add linear interpolation to the $hw$ to make the bandwidth sample frequency the same as that of the viewport trace, 36 FPS.}
\rev{We chose the harmonic mean predictor due to its simplicity and successful use in prior multimedia systems~\cite{jiang2012improving,yin2015mpc}.
More sophisticated bandwidth predictor modules could easily be slotted into our framework (Section \ref{sec:limitations}). 
}
To simulate an outdoor 5G network, we sampled outdoor user walking traces from~\cite{narayanan2020lumos5g}, scaling down the bandwidth to adapt to our use case as has been done in other work~\cite{ye2024dissecting}. 
The scaled average network bandwidth is 11.8 Mbps. 

%% file: results.tex
\section{Experimental Results}
\label{sec:results}
\subsection{Setup}
We conduct evaluations on 8 scenes from public datasets, including 3 scenes from Mip-NeRF360~\cite{barron2022mipnerf360}, 2 scenes from Tanks\&Temples~\cite{knapitsch2017tanksntemple}, and 3 segmented scenes from Gaussian Grouping~\cite{ye2024gaussiangroupingsegmentedit}. We create each scene with 4 layers, 
each layer having 45k splats. 
To evaluate our delivery framework, we report the visual quality (SSIM) of the user viewport every 1 second within the 60-second user traces, averaging over all scenes and user traces. To decrease the influence of the difficult position, we randomly sample 3 different starting points for each trace and show the average SSIM. Since we do not have the ground truth images for all viewports in our real user traces (the standard datasets only provide ground truth images for a handful of viewports), we calculate the SSIM with reference to renderings generated by the original pre-trained 3DGS models with the full number of splats. 

\subsection{Baselines} \label{sec:baseline}
We evaluate our 3DGS delivery pipeline with our layered representations, marked as ``Ours'', along with several other baselines described below. 
\squishlist
    \item \textbf{Splat sorting (``Sort'').} We sort the Gaussian splats of the officially pre-trained 3DGS models by their global significance score proposed by LightGaussians\cite{fan2023lightgaussian}.
    Recall that this score is viewport-independent, so it does not adjust to where the user is currently looking.
    When downloading, splats are progressively retrieved with higher scores first. 
   \item \textbf{Non-layered models (``Separate'').} We use \Cref{algo:Prune} to create independent models with different target number of splats, without the layered structure. When downloading, a complete stand-alone version of the scene is downloaded sequentially at increasing quality. Once a lower version is retrieved and displayed, a higher version is subsequently downloaded. This baseline 
   corresponds to cases II and IV from \Cref{table:prob_taxonomy}.
    \item \textbf{Optimal (``Pre-load'').} We compare to the ideal performance that could be achieved by pre-loading our largest separate model (180k splats, generated after \Cref{algo:Prune}), onto the client in advance, instead of downloading it. 
\squishend
\noindent Finally, we evaluate \system with various 3DGS models \cite{fan2023lightgaussian,ren2024octree,yu2024mip,lu2024scaffold}, particularly \cite{fan2023lightgaussian} and \cite{ren2024octree} that are specifically designed for model compression.


\subsection{Visual quality of our layered splats}
\begin{figure}
\includegraphics[scale=0.33]{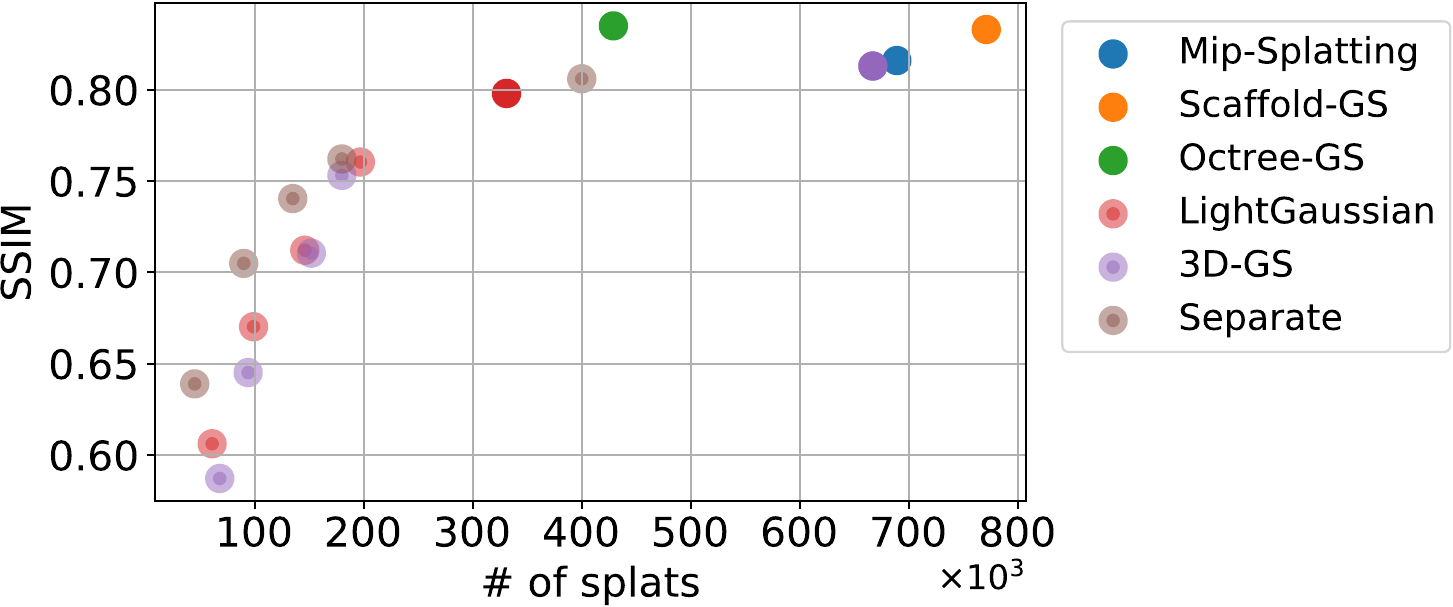}
\vspace{-10pt}
    \caption{Visual quality vs number of splats for the Train scene~\cite{knapitsch2017tanksntemple}. \system provides viable options to deliver 3D scenes with fewer splats. 
    }\label{fig:layered_baseline}
\end{figure}

We first examine the visual quality of the splats produced by \system without the network delivery pipeline.
\textbf{Our methods are able to precisely control the number of splats to achieve a good tradeoff between visual quality and size, compared to 3DGS models from the literature.} 
The results, presented in \Cref{fig:layered_baseline} and \Cref{fig:layered_baseline_simple}, utilize ground truth images and positions from the test set to compute the SSIM. The performance values of models from other works (dots without transparency) are directly taken from \cite{ren2024octree,fan2023lightgaussian}.
Our models  (``Ours'' and ``Separate'') significantly reduce the number of splats (and hence size) and provide viable options to deliver 3D scenes with fewer splats.
We created additional versions of original 3DGS model~\cite{kerbl3Dgaussians} and LightGaussians~\cite{fan2023lightgaussian} (dots with transparency) by optimizing their hyperparameters using grid search to vary the model size. 
Note that to ensure a fair comparison with other models from the literature, we exclude feature distillation and quantization strategies (such strategies can be applied to all splat models to reduce file size, although not splat numbers).
Overall, \system not only precisely controls the number of splats but also achieves higher SSIM than other compressed variants. 

\begin{figure}
\centering
\begin{subfigure}[b]{0.23\textwidth}
\includegraphics[width=\textwidth]{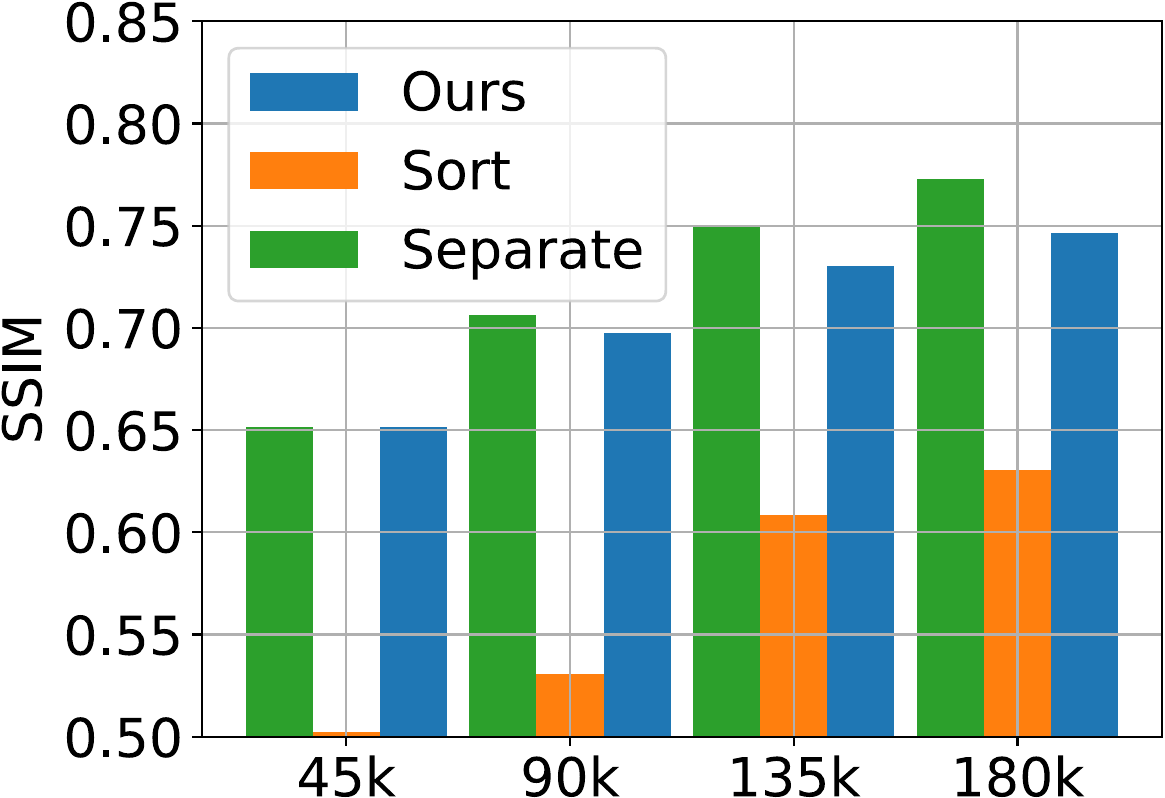}
\vspace{-15pt}
  \caption{\small{With segmentation}}
\end{subfigure}
\begin{subfigure}[b]{0.23\textwidth}
\includegraphics[width=\textwidth]{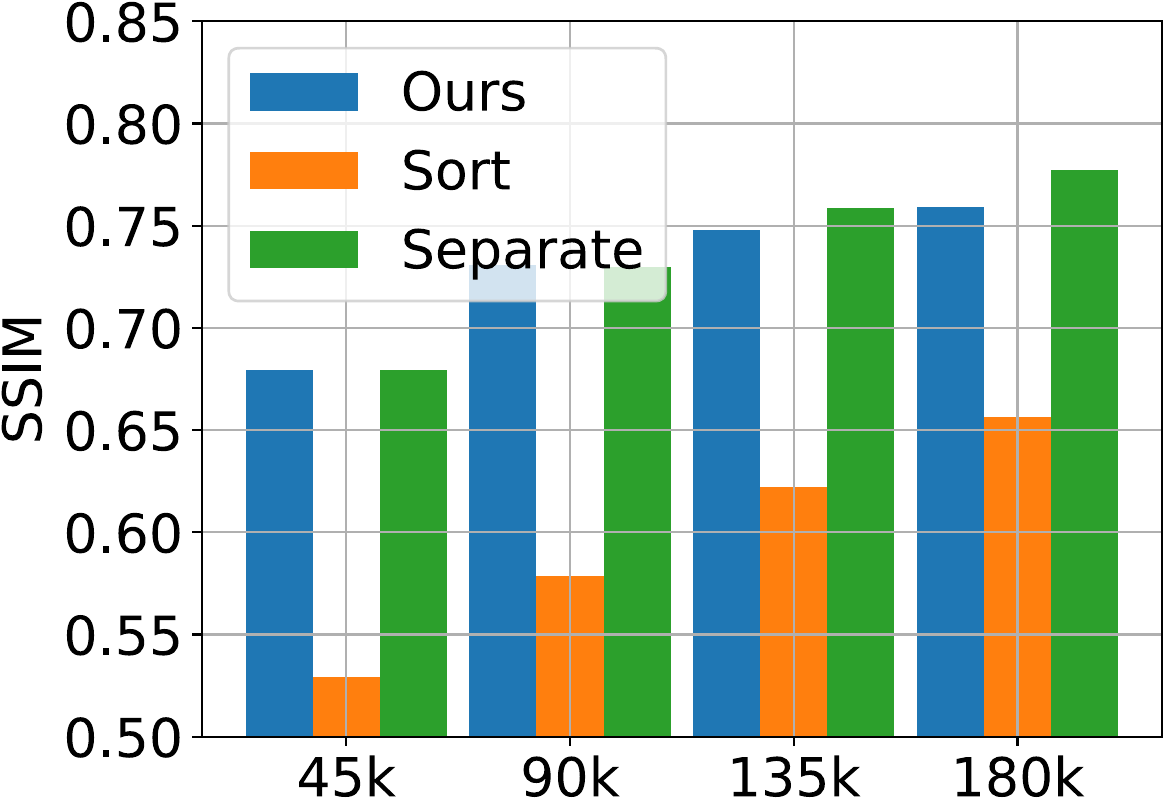}
\vspace{-15pt}
  \caption{\small{Without segmentation}}
\end{subfigure}
\vspace{-10pt}
\caption{Visual quality of our layered splats (Ours) compared to baselines. The layering achieves high visual quality, even close to the non-layered versions (Separate).}
\label{fig:diff}
\end{figure}

\textbf{\system constructs layered scenes without significantly compromising performance.} We also evaluate \Cref{algo:layered_train} by comparing against the splat-sorting and non-layered baselines across all 8 scenes with varying model sizes. In \Cref{fig:diff}, we showed that our layered 3DGS (``Ours'') outperforms the ``Sort'' baseline significantly. Further, although \system constructs layered splats that could potentially constrain the optimization space and result in worse visual quality, \Cref{algo:layered_train} still achieves comparable performance with the non-layered model (``Separate''). 

\subsection{Main results of our delivery pipeline}
\begin{figure}
\centering
\begin{subfigure}[b]{0.38\textwidth}
\includegraphics[width=\textwidth]{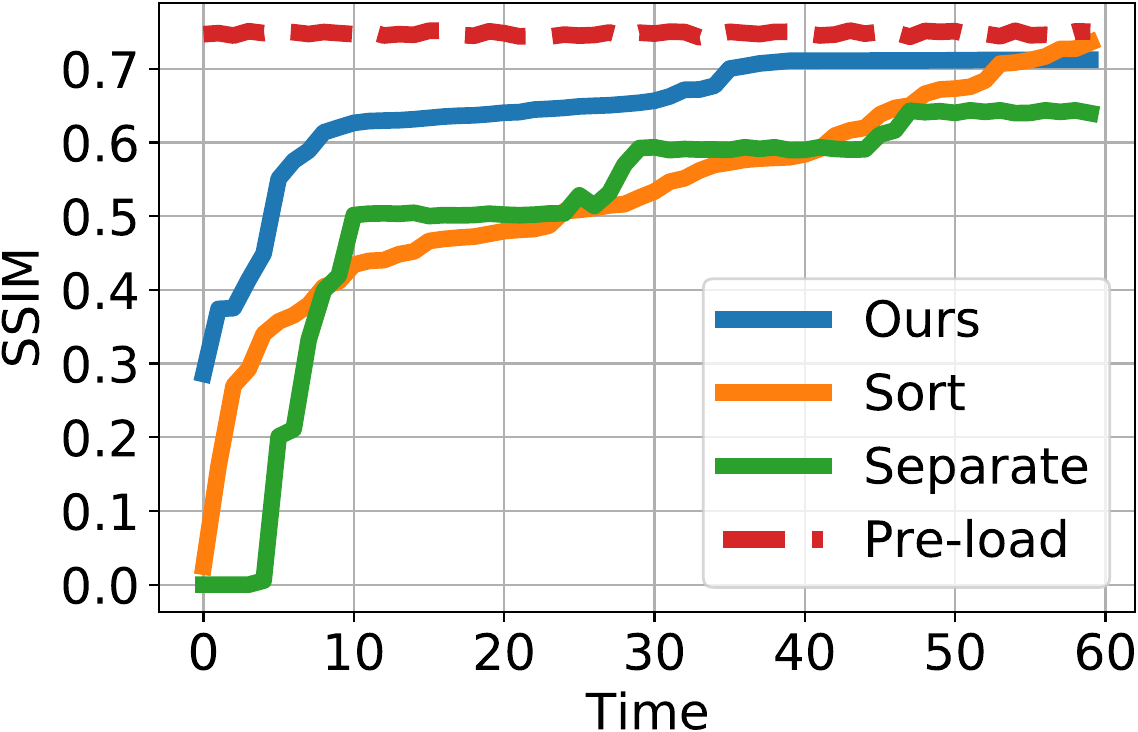}
\vspace{-15pt}
  \caption{\small{SSIM of synthetic traces.}}\label{fig:real}
\end{subfigure}
\begin{subfigure}[b]{0.38\textwidth}
\includegraphics[width=\textwidth]{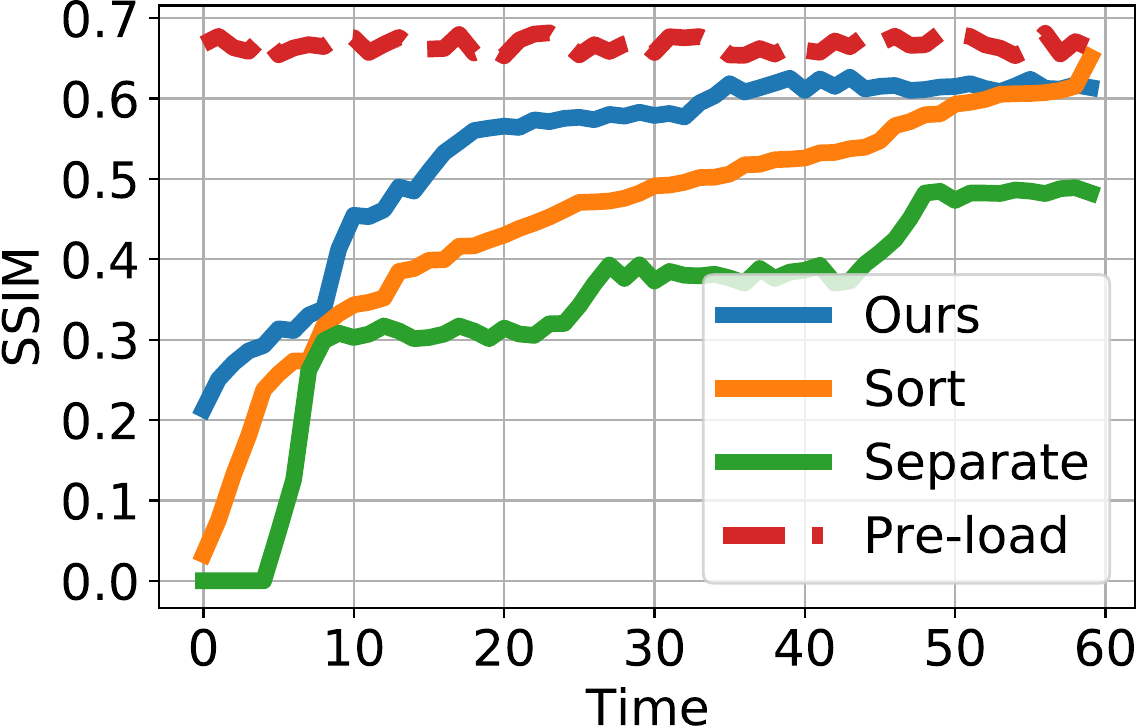}
\vspace{-15pt}
  \caption{\small{SSIM of real traces.}}\label{fig:syn}
\end{subfigure}
    \caption{We evaluate \system with both synthetic and real traces for the segmented and layered 3DGS we developed. The SSIM is significantly higher than the baselines and not far from the optimal (Pre-load).
    }\label{fig:main}
\end{figure}
\textbf{\system outperforms other baselines because its scheduler can effectively retrieve splats within the user's viewport.} The main performance results are illustrated in \Cref{fig:main} for the synthetic and real user traces.
For the first 5 seconds, ``Ours'' significantly outperforms other baselines. For ``Sort'', not only the visual quality of scenes with limited number of splats is poor, but their selection is only determined by a global significance score, which is not view-dependent. Worst of all, ``Separate'' cannot finish downloading even the lowest quality version, resulting in blank rendered images and an initial SSIM of 0. In contrast, \system efficiently selects the most critical splats for the current viewport, thereby achieving superior performance. Even when the base 45k-splat ``Separate'' model is fully loaded (at around 8 seconds on average), our method achieves better performance by prioritizing important splats based on predicted future viewports, such as those closer to the user.

With the layered approach of \system, performance stabilizes after downloading all 180k splats, which takes approximately 30 seconds. The performance is comparable to the ``Pre-load'' baseline, with only minor losses due to the layered structure. However, ``Sort'' can achieve better performance  towards the end of the traces by loading more splats (360k splats at time 60s), because it has access to a larger model.
Regarding ``Separate'', although the model's visual quality is slightly better than ours for the same number of splats, the lack of progressively overlapping splats in different versions necessitates loading the entire larger model and discarding the previous one. This process results in substantial bandwidth wastage, leading to lower SSIM compared to \system under limited bandwidth conditions.

\begin{figure}
\includegraphics[scale=0.45]{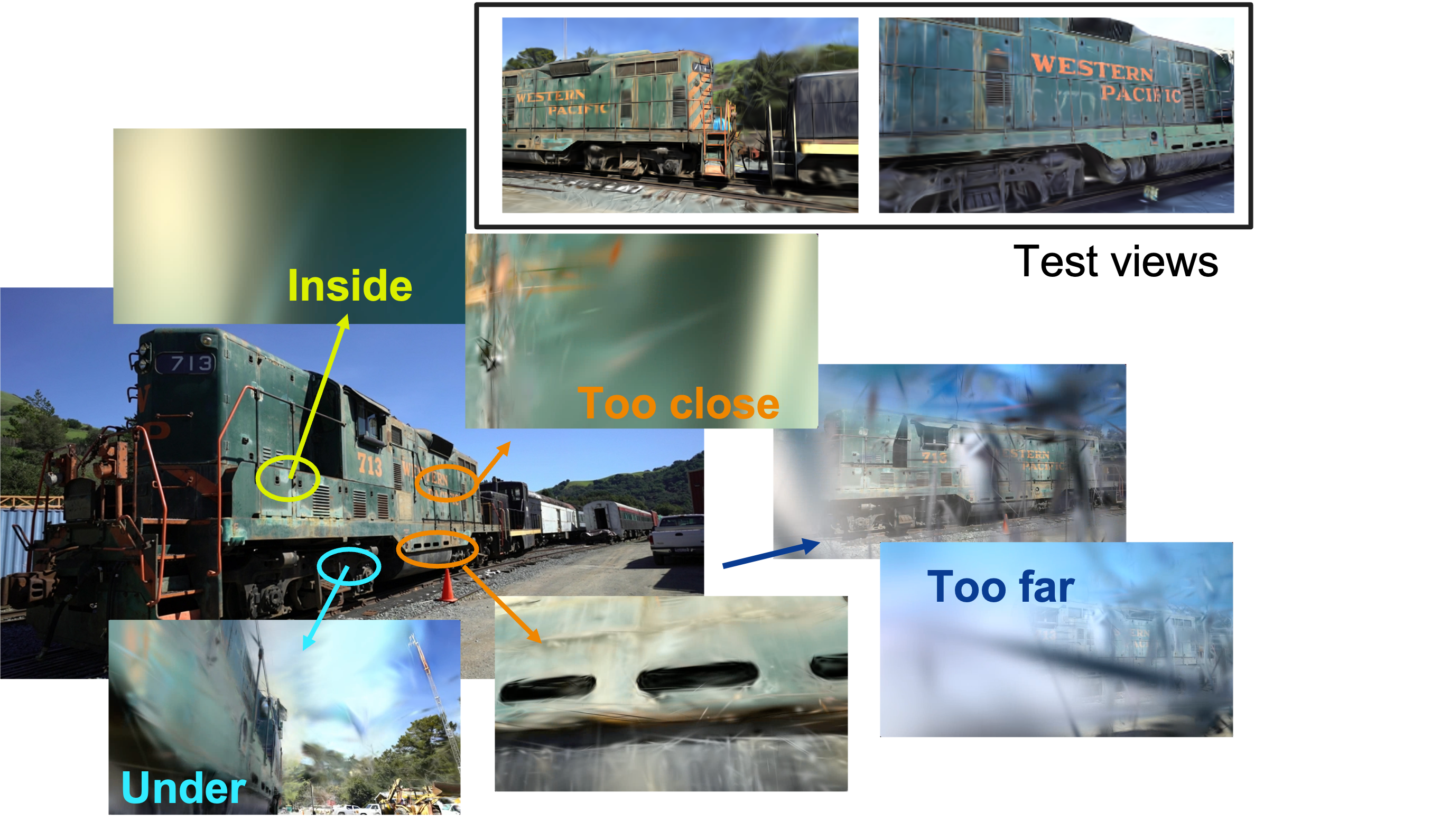}
    \caption{Screenshots capturing various viewpoints. Views from the test set demonstrate higher visual quality. However, some views derived from real-world traces display uncommon characteristics resulting in low visual quality, which affects all methods. 
    }\label{fig:minority_screenshot}
\end{figure}

The best SSIM \system can achieve is lower than the ideal values shown in \Cref{fig:diff}, particularly for real traces shown in  \Cref{fig:real}. This is because users may explore some strange positions in the scene, such as walking too close or inside the train, or attempting to walk beyond the boundaries of the scene. Our data collection revealed that users are often curious about the holes and low-quality parts, resulting in out-of-scene viewports. Even the original pre-trained 3DGS model~\cite{kerbl3Dgaussians} shows low quality in such edge cases due to lack of ground truth images. We provide some example screenshots in \Cref{fig:minority_screenshot}.
The main take-away is that such real user behavior hurts our SSIM, because the SSIM of these strange positions tends to be lower, but this issue affects all methods, not just ours.

\subsection{Adapting to other splat representations} \label{sec:type}
\begin{figure}
\centering
\includegraphics[width=0.45\textwidth]{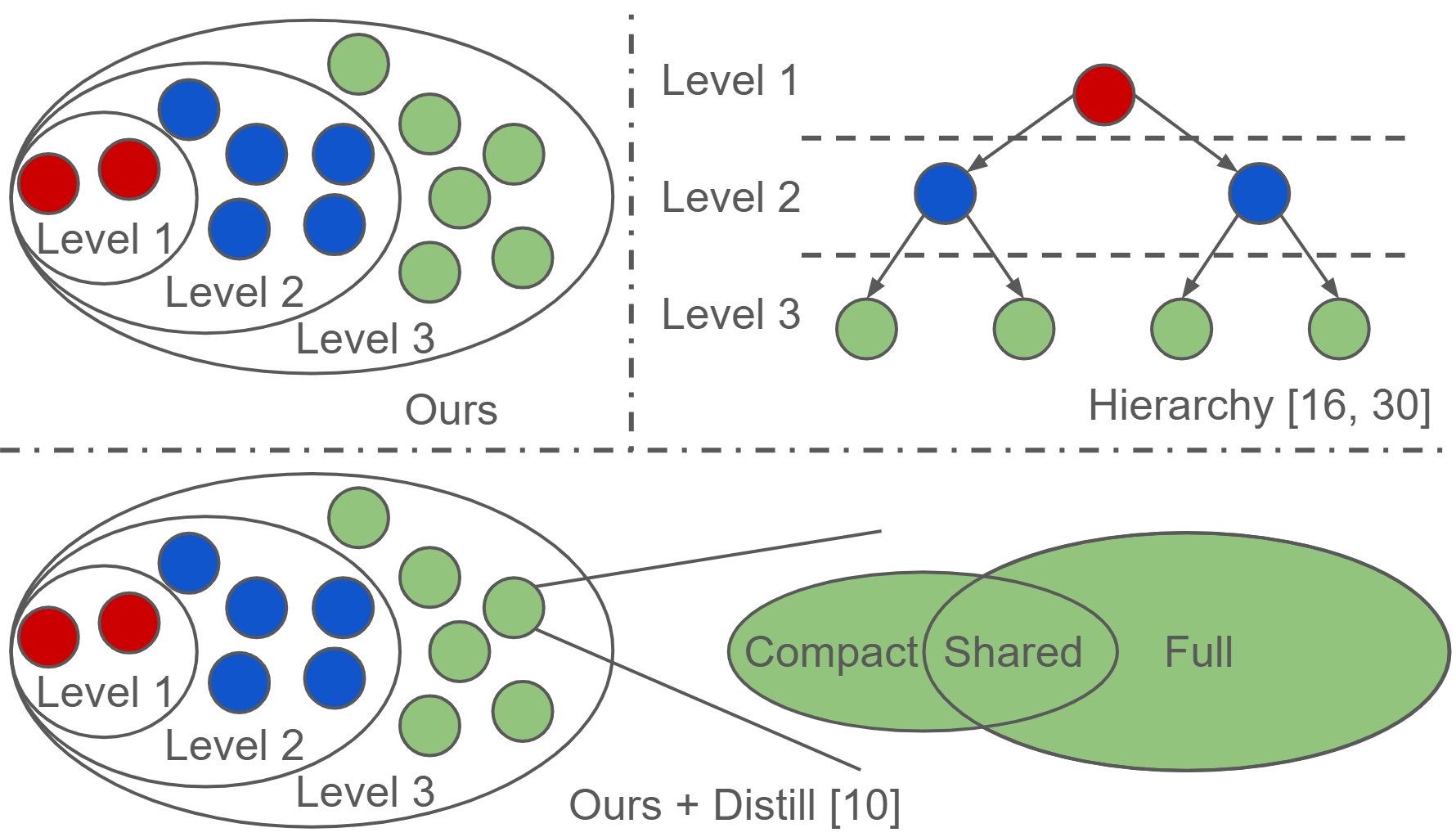}
  \caption{Other representations of 3DGS that \system can work with. Both ``Ours'' and ``Ours+Distill'' utilize layered architectures. Splats in the ``Ours+Distill'' model have two forms: shared features with compact or full SH. Splats can also be organized in a hierarchy.
  }\label{fig:type_splats}
\end{figure}

3DGS development is rapid, with new versions continuously emerging despite the original paper only appearing in 2023.
Thus the ability of \system to adapt to such improvements is an important consideration. Two popular strategies are being explored: compressing the feature size of each splat through distillation or quantization, and building hierarchical models for more efficient rendering. We explore the performance of our delivery framework under these two alternative 3DGS representations, to demonstrate the flexibility of our framework.
The alternative methods are:
\squishlist
    \item \textbf{Ours + spherical harmonics distillation~\cite{fan2023lightgaussian}.} 
    We incorporate the spherical harmonics (SH) distillation module from \cite{fan2023lightgaussian} into our layered model to create two versions of splats. A schematic of the model is illustrated at the bottom of \Cref{fig:type_splats} as ``Ours+Distill''. 
    The main idea is that splats are still organized into layers as in our model, but each splat contains both a compact and full version of the SH representation.
    Meanwhile, other shared features such as opacity and location remain consistent with the original model. When downloading the splats, we add the dimension (compact or full) of the SH as a component to our utility function, and thus the scheduler can choose from layered splats with the full or compact SH.

    \item \textbf{Hierarchy~\cite{shuai2024LoG,kerbl2024hierarchical}} Inspired by the classical idea of Levels of Detail (LoD), some works store the splats in a tree structure. Each node in this tree can be represented in greater detail by several of its children. To showcase the hierarchy model in our framework, we use the non-layered models (``Separate'') with different target sizes to form the tree structure, with lower-version splats set as root nodes and the higher-version splats as their child nodes, sharing the same object ID and highest similarity score. An illustration of this method is shown in the top half of \Cref{fig:type_splats}.
\squishend

\begin{figure}
\centering
\includegraphics[width=0.35\textwidth]{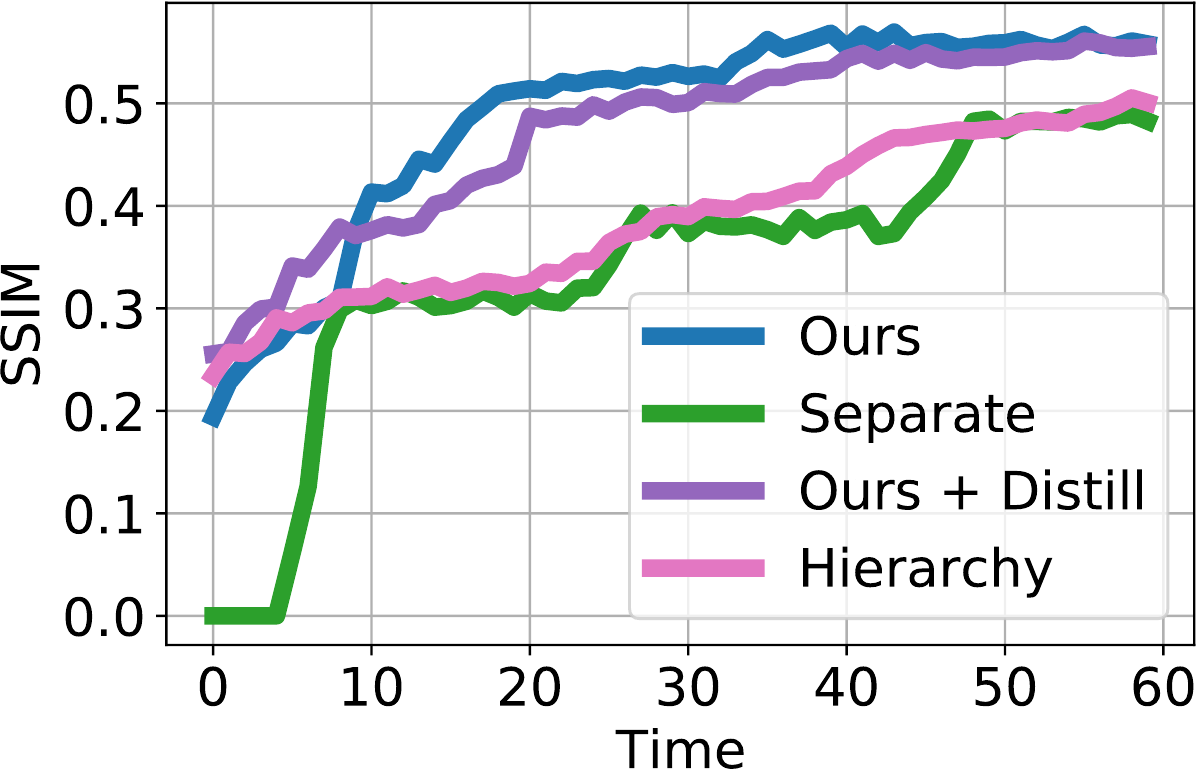}
\vspace{-10pt}
  \caption{\system works with other compressed splat representations.}\label{fig:model_type}
\end{figure}

\begin{figure}
\centering
\includegraphics[width=0.35\textwidth]{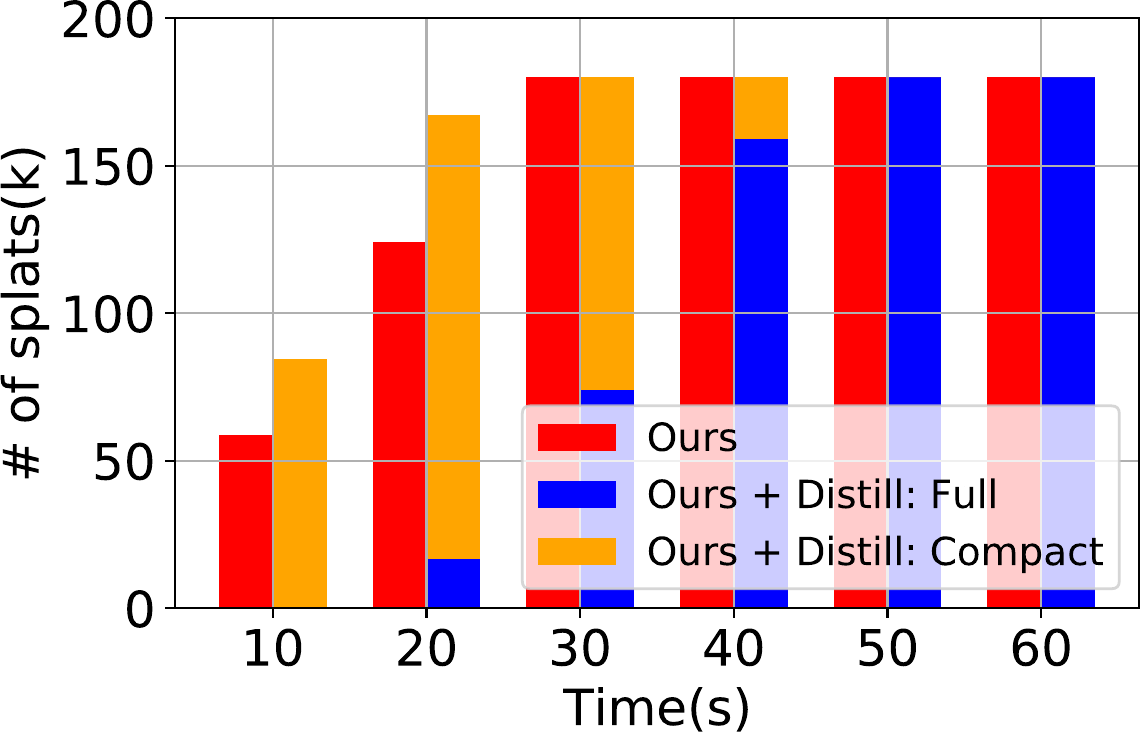}
\vspace{-10pt}
  \caption{\rev{\system works with 3DGS distilled spherical harmonics (SH). Over time, a greater fraction of splats are upgraded from compact to full SH.}}\label{fig:distill_splats}
\end{figure}

The results with these various 3DGS compression methods are shown in \Cref{fig:model_type}.
The ``Ours + Distill'' variant demonstrates that beyond the layered approach of our model, when there are more variants of individual splats available by compressing the spherical harmonics, \system can achieve good performance.
\rev{To further understand these results, we look at the proportion of splats with full or compact SH.
As shown in \Cref{fig:distill_splats},}
initially, the compact SH version of the splats tends to be loaded (orange bars), allowing more splats to be downloaded within the same bandwidth and achieving higher SSIM. Over time, as the total number of bits transferred increases, the compact SH will be replaced by the full SH (blue bars) to enhance visual quality. 
Since the other splat features are shared, only the SH features need to be upgraded, enabling ``Ours + Distill'' to quickly reach performance levels comparable to our original model.

The ``Hierarchy'' variant in \Cref{fig:model_type} demonstrates our framework's adaptability to hierarchical splat organization, similar to level of details (LoD). As with the ``Separate'' baseline, no splats are shared across levels of the tree, requiring the replacement of multiple splats, leading to bandwidth inefficiencies compared to our layered approach. However, since parent nodes can be replaced by child nodes, ``Hierarchy'' allows fine-grained progressive splat replacement, achieving a smoother performance enhancement compared to the original ``Separate'' baseline that can only do coarse-grained replacement at the per-scene level, rather than per-splat.

\subsection{3D scenes with and without object segmentation}
\begin{figure}
\centering
\includegraphics[width=0.33\textwidth]{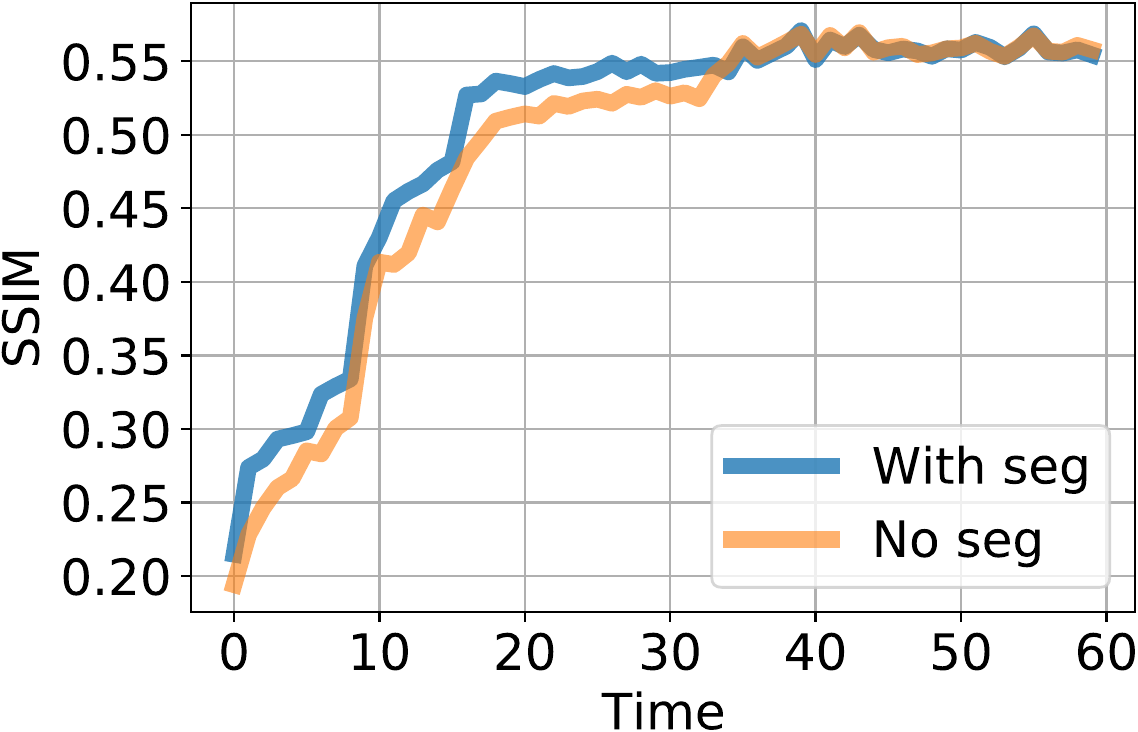}
\vspace{-10pt}
  \caption{\system with and without object segmentation in the scene.}\label{fig:seg}
\end{figure}

We compare the performance of the scheduler with and without segmenting the 3D scene into objects. With segmentation, referred to as \textbf{Case I} in \Cref{table:prob_taxonomy}, the scheduler has flexibilty to choose both objects and layers, resulting in a larger selection space. In contrast, in \textbf{Case II}, which uses layered splats without object segmentation, we only have coarse-grained control over which layer to download.
The results, shown in \Cref{fig:seg}, indicate that the scheduler with segmentation has better performance. This suggests that incorporating object-level decisions enhances \system. 

\begin{figure}
\centering
\begin{subfigure}[b]{0.15\textwidth}
\includegraphics[scale=0.25]{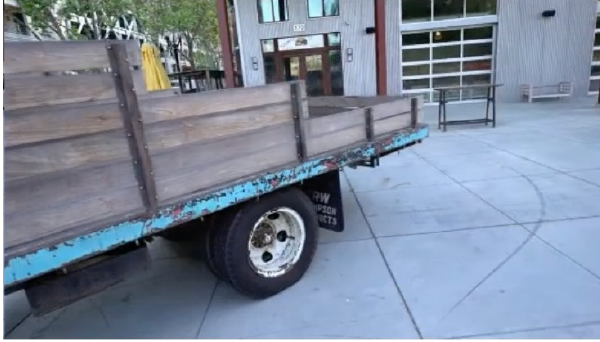}
  \caption{\small{GroundTruth}}
\end{subfigure}
\begin{subfigure}[b]{0.15\textwidth}
\includegraphics[scale=0.25]{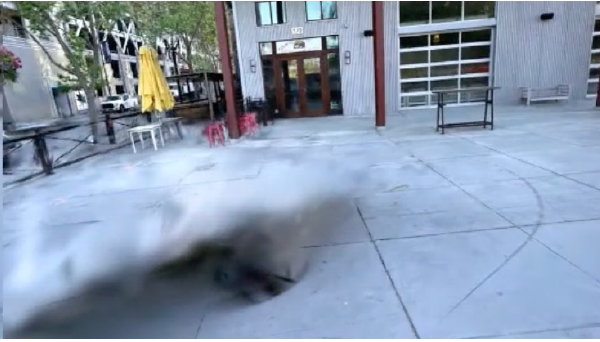}
  \caption{\small{Remove truck}}\label{fig:removal}
\end{subfigure}
\begin{subfigure}[b]{0.15\textwidth}
\includegraphics[scale=0.25]{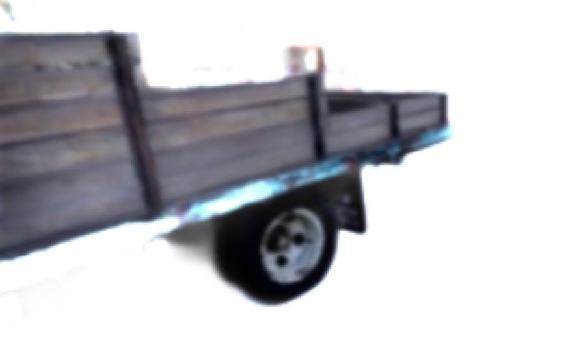}
  \caption{\small{Truck Only}}\label{fig:truck_only}
\end{subfigure}
\caption{With segmentation, we can render objects independently.
To visualize the truck clearly (right), the background is set to white.}
\label{fig:seg_screenshot}
\end{figure}

With segmentation, we can also selectively render specific objects. For instance, in \Cref{fig:removal}, we can exclude the truck and render only the background. Conversely, in \Cref{fig:truck_only}, we only render with the splats associated with the truck, which takes only 25.9\% of the total splats. This approach enables more efficient rendering based on user preferences. For example, users may prefer to prioritize the rendering of the main foreground object before the background ~\cite{li2023viewport,wang2018revisiting,9729212}. 
 
\subsection{Latency overhead}
\begin{figure}
\centering
\includegraphics[width=0.45\textwidth]{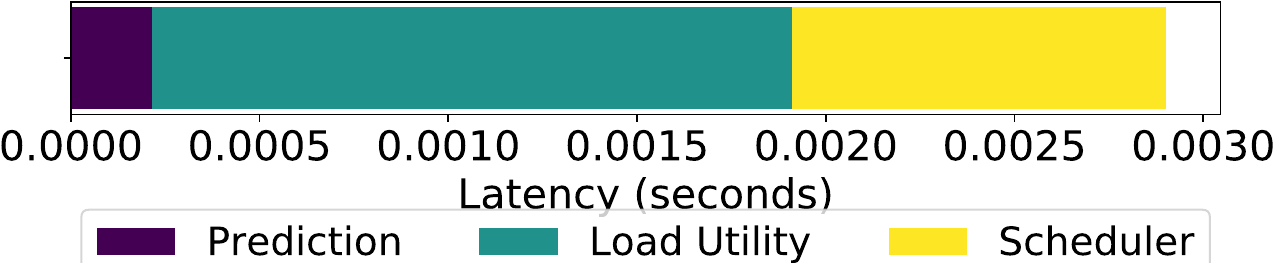}
\vspace{-5pt}
  \caption{Latency breakdown per scheduling period.}\label{fig:latency}
\end{figure}

We also measure the latency every scheduler period (every 1 second), as shown in \Cref{fig:latency}. To perform the rendering, we first initialize the camera parameters, a process that takes approximately 1.30 seconds but occurs only once. Following this, the viewport and bandwidth predictor modules output the viewport and bandwidth in parallel; the maximum latency across these two predictions is labeled as ``prediction'' in the figure. Once the predictions are completed, we use the predicted viewport for utility calculation and the predicted bandwidth to determine which splats to download with the scheduler. Calculating the utility of all splats for a single viewport takes around 10.29 seconds, which is infeasible for real-time rendering. To address this, we pre-compute the utility values by uniformly sampling 1,000 viewport positions and 1,728 viewport orientations, then load the values from disk at runtime. This pre-computation step is labeled as ``Load Utility'' in \Cref{fig:latency}. 
Finally, we run the scheduling algorithm to decide which splats to download (``Scheduler'').
The rendering time of our largest model (180k splats) is 3.69 ms.
All latency values are measured with 2vCPU @ 2.2GHz.
\rev{In summary, the latency overhead of \system is negligible, consuming less than 3 ms.}





\subsection{Ablation studies}

\begin{figure}[htbp]
\centering
\begin{subfigure}[b]{0.22\textwidth}
\includegraphics[scale=0.2]{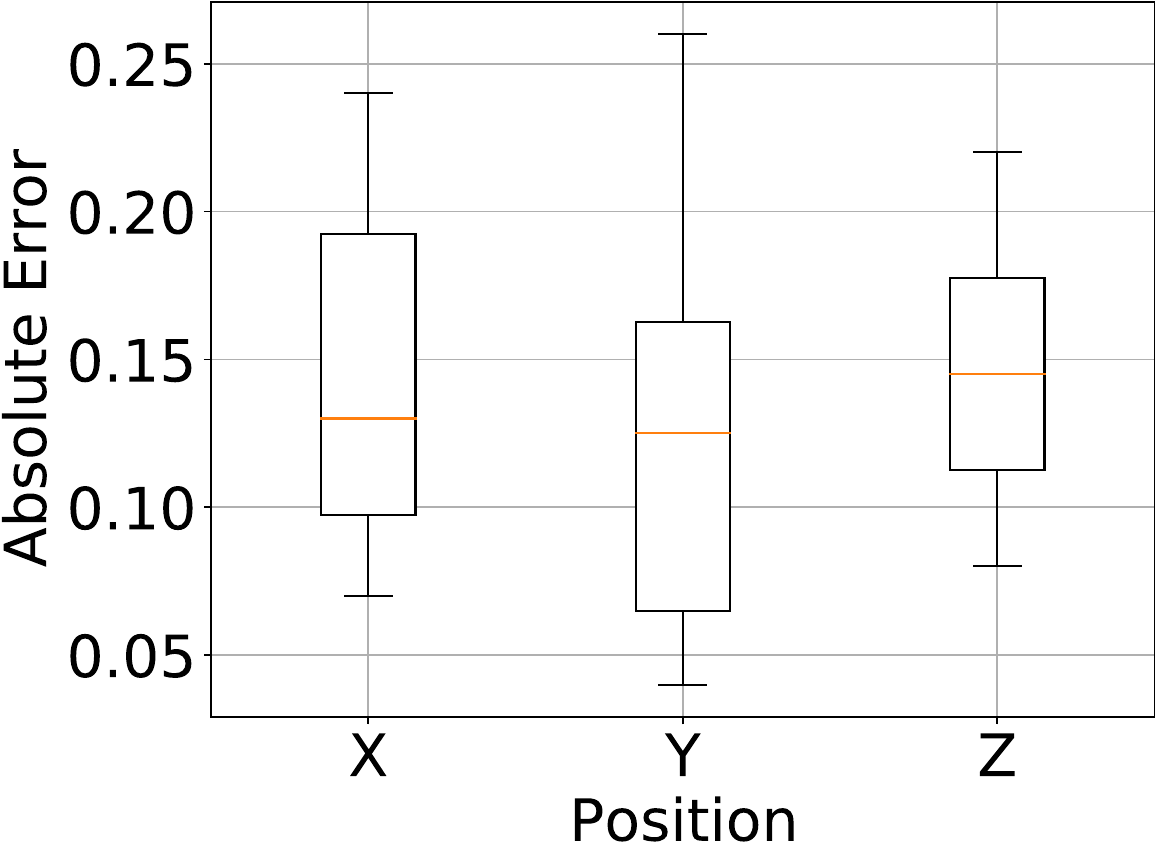}
\end{subfigure}
\begin{subfigure}[b]{0.22\textwidth}
\includegraphics[scale=0.2]{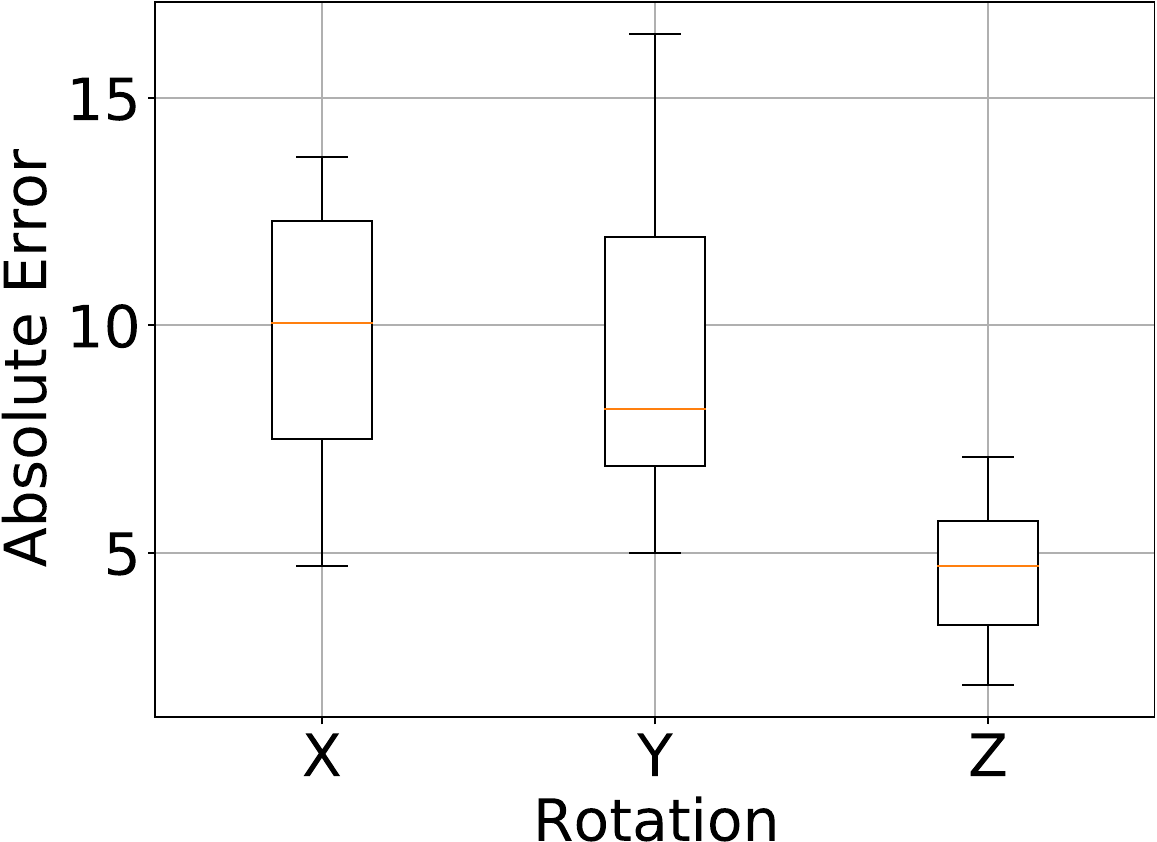}
\end{subfigure}
\caption{Absolute error of viewport prediction.}
\label{fig:viewport_error}
\end{figure}

\begin{figure}[htbp]
\centering
\includegraphics[scale=0.4]{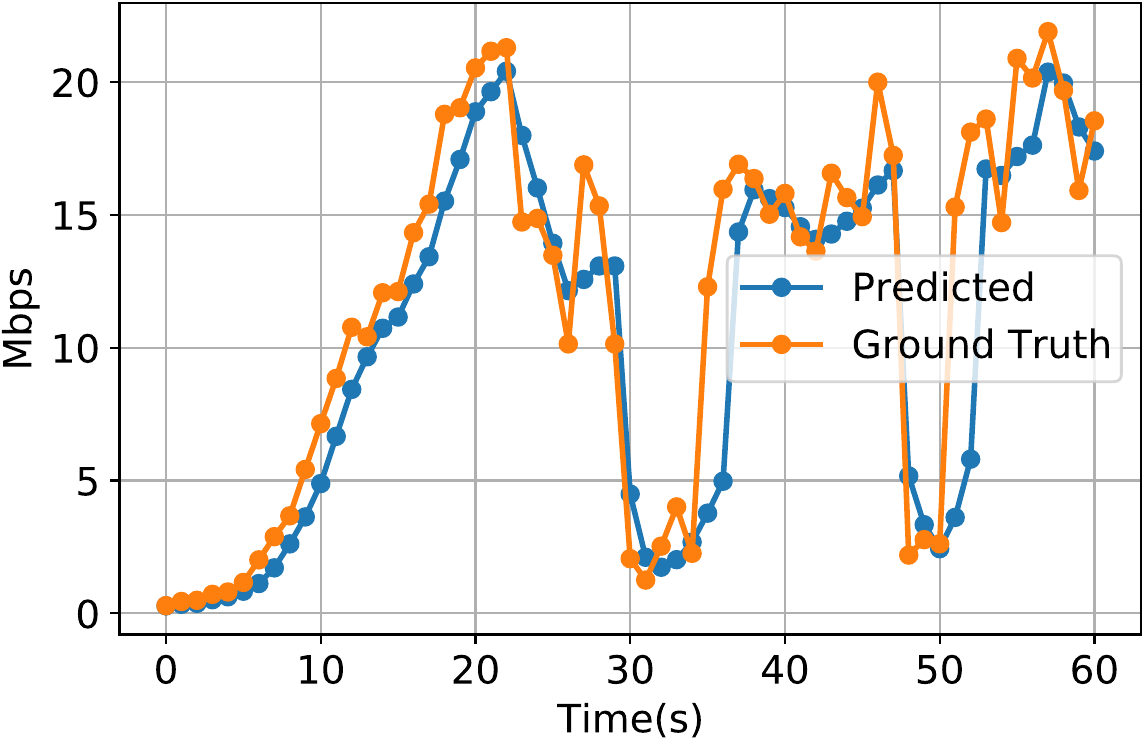}
\caption{Sample bandwidth prediction trace. The mean absolute error across all traces is 3.11 Mbps, with a standard deviation of 1.43.}
\label{fig:throughput_error}
\end{figure}

User viewport prediction is inherently imperfect due to the unpredictable nature of human behavior and attention. 
Similarly, bandwidth prediction can be difficult due to fluctuations in network conditions. The average error of our viewport predictor is shown in \Cref{fig:viewport_error}. In~\Cref{fig:throughput_error}, we show an example of the predicted network bandwidth with the ground truth network trace. The performance of our method is reasonable,
but the question is still how any remaining inaccuracies affect overall system performance?

\begin{figure}
\centering
\includegraphics[width=0.5\textwidth]{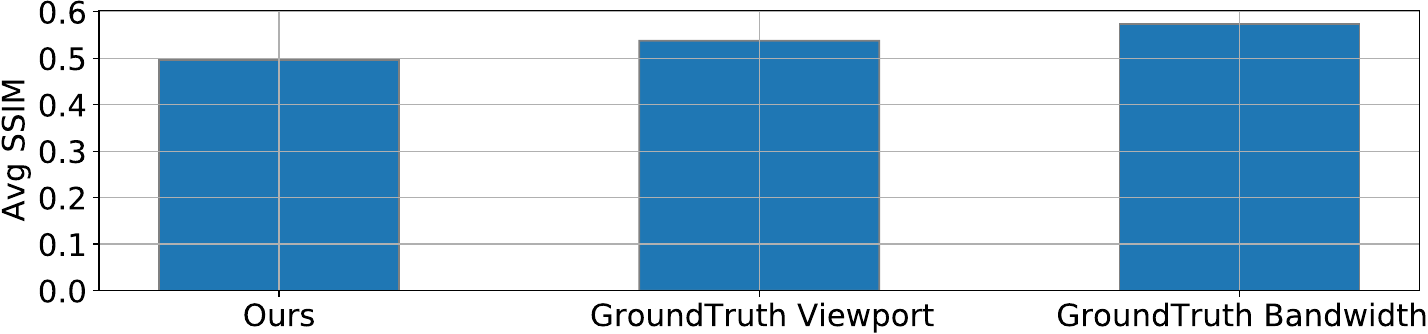}
  \caption{Ablation study compared to perfect user viewport or throughput prediction.}\label{fig:gt}
\end{figure}

To investigate this, we conducted an ablation study, evaluating \system with perfect prediction of viewport or bandwidth, with the results presented in \Cref{fig:gt}. Replacing both predictors with ground truth values (a hypothetical scenario) improves average visual quality, as imperfect predictions can lead to suboptimal splat downloads and potential bandwidth wastage; however, the observed gaps are relatively modest, with average SSIM differences of 0.077 for bandwidth and 0.041 for viewport over a 60-second period. This suggests that the \rev{prediction mechanisms work} effectively.

\begin{figure}
\centering
\includegraphics[width=0.5\textwidth]{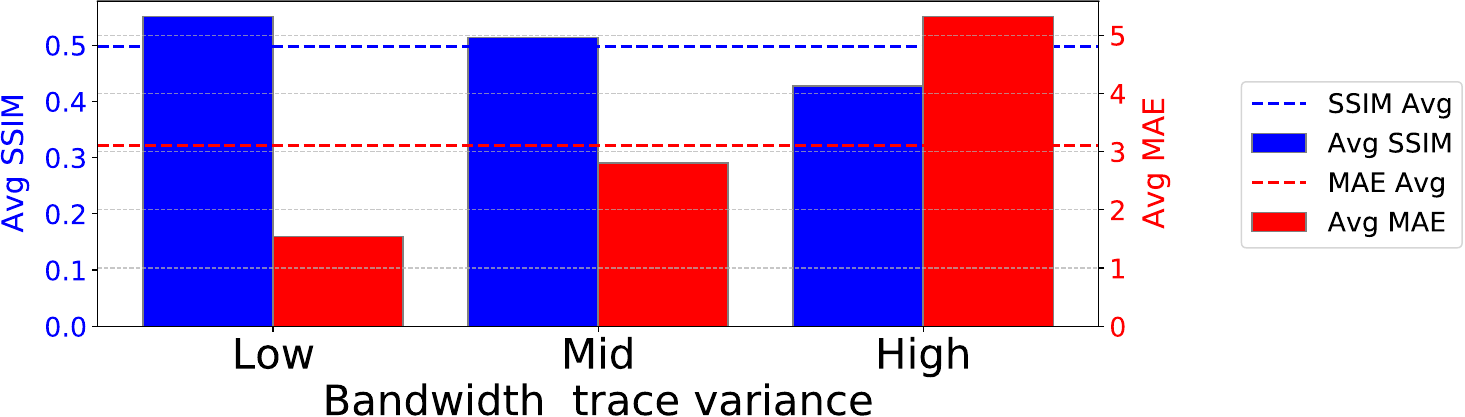}
  \caption{\rev{SSIM improves as the bandwidth prediction accuracy increases, for bandwidth traces with different levels of variance.}
  }\label{fig:bw_var_combined}
\end{figure}

\rev{To further understand the system’s robustness to fluctuating network conditions, we partition the bandwidth traces into three levels of variance (low/medium/high, corresponding to 0-29, 29-45, 45+) and plot the bandwidth prediction accuracy (in terms of mean absolute error) and visual quality (in terms of SSIM) across these categories. The results are shown in \Cref{fig:bw_var_combined} and indicate that higher network bandwidth variance  negatively impacts bandwidth prediction accuracy, and hence results in some decrease in SSIM.  However, even for the bandwidth traces with the highest variance, the performance remains comparable to the overall average SSIM across all traces.}

%% file: limitations.tex
\section{Limitations and Future Work}
\label{sec:limitations}


\rev{Limitations of \system include the difficulty of accuracy viewport and bandwidth prediction.
Viewport prediction is challenging in 3DGS scenes because users have 6 degrees of freedom, both translation and rotation (different from, for example, rotation-only user movements in panoramic videos).
Beyond harmonic mean or linear predictors, machine learning models that could be used, but they require larger datasets of user movements in 3DGS scenes that are not currently available.
These machine learning models could incorporate additional features such as visual saliency estimation or eye tracking data to aid prediction accuracy, as has been found useful in the related domains of VR viewport prediction~\cite{sitzmann2018saliency,stein2022eye}.}

\rev{The bandwidth prediction module currently used by \system is relatively simple, although it has been effective in previous multimedia adaptation works~\cite{jiang2012improving,yin2015mpc}.
\system could incorporate more sophisticated predicton modules employing adaptive filtering~\cite{bentaleb2019bandwidth} or LSTM~\cite{narayanan2020lumos5g}.
The system uses application-layer bandwidth to make decisions, abstracting away from fluctuations like packet loss or handovers.
To adapt in real time if the actual bandwidth differs from the predicted bandwidth, we can adopt heuristic rules to change the splat download schedule accordingly.
For example, if the actual bandwidth is higher than predicted and the splats finish downloading earlier, the framework can re-run the scheduler sooner and start fetching the next set of splats.
}

\rev{
Future work includes integrating new compressed 3DGS representations that are constantly being developed into our framework, deploying a prototype on live networks, and user studies to evaluate perceptual visual quality.
Finally, it would be interesting to explore scenarios of multiple clients sharing a bottleneck link, and how splats from 3DGS scenes could be scheduled fairly and efficiently across users.
}

%% file: conclusions.tex
\section{Conclusions}
\label{sec:conclusions}
In this paper, we created an efficient delivery framework for 3D scenes using layered 3D Gaussian splats.
We developed a training pipeline to create layered 3DGS models, where the scenes can further be segmented into objects to provide fine-grained control for downloading and editing. 
By layering the 3DGS models and creating splat scheduling algorithms based on user viewport and network bandwidth, our system can adapt to varying network bandwidths while maintaining high visual quality.  
In addition, the scheduler also works with various types of 3DGS representations.
The experimental results show higher performance compared to the baseline, achieving higher average SSIM scores and low overhead.


\section*{Acknowledgments}
Thank you to our shepherd, Robert LiKamWa, and the anonymous reviewers for their valuable suggestions. Thank you also to the volunteers who participated in our data collection. This work was supported by NSF CAREER 1942700 and Meta research funds.

%% file: appendix.tex
\onecolumn
\appendix

\section{Proofs and Problem Definitions}
\label{sec:appendix}

\begin{problem}{Generalized assignment problem with precedence constraints}
\begin{align}\small
    \max \sum_j \sum_l \sum_{t} \Delta \tilde{U}_{jl}[t] x_{jl}[t] \label{eqn:gap_obj}\\
    \text{s.t. }  
    \sum_{j} \sum_l \Delta c_{jl} x_{jl}[t] \leq B[t] \; \; \forall t  \label{eqn:gap_bandwidth}\\
    \sum_t x_{jl}[t] \leq 1 \; \; \forall j,l \label{eqn:gap_select_one}\\
    x_{jl}[t] \leq \sum_{t^\prime < t} x_{jl}[t^\prime] \; \; \forall j,l,t \label{eqn:gap_layers}\\
    x_{jl}[t] \in \{0, 1\} \; \; \forall j,t
\end{align}
\label{prob:gap}
\end{problem}

\textbf{Lemma~\ref{lemma:gap_equiv}}
\begin{proof}
\small
Substituting (\ref{eqn:store_splats_constraint}) into the objective function (\ref{eqn:obj}), we can re-write the objective as
\begin{align}
\sum_{j,l} \sum_{t=1}^T \Delta U_{jl}[t] y_{jl}[t] &= \sum_{j,l} \sum_{t=1}^T \Delta U_{jl}[t] \sum_{t^\prime=1}^t  x_{jl}[t^\prime]\\ 
&= \sum_{j,l} \sum_{t^\prime=1}^T \sum_{t=t^\prime}^T \Delta U_{jl}[t] x_{jl}[t^\prime]\\
&= \sum_{j,l} \sum_{t^\prime=1}^T \Delta \tilde{U}_{jl}[t^\prime] x_{jl}[t^\prime] \label{eqn:lemma_lastline}
\end{align}
\noindent where (\ref{eqn:lemma_lastline}) is just (\ref{eqn:gap_obj}), the objective of Problem~\ref{prob:gap}.
The other constraints are equivalent: (\ref{eqn:bandwidth_constraint}) and (\ref{eqn:gap_bandwidth}), (\ref{eqn:select_one_constraint}) and (\ref{eqn:gap_select_one}), and (\ref{eqn:layer_constraint}) and (\ref{eqn:gap_layers}).
\end{proof}

\textbf{Theorem~\ref{thm:layered_np_hard}}
\begin{proof} \small The first step is a reduction from the generalized assignment problem (GAP) to an instance of Problem~\ref{prob:gap}. The GAP can be stated as: given $T$ bins (with capacity $c_t$) and $n$ items (with profit $p_{jt}$ and weight $w_{jt}$), find the assignment of items to bins to maximize the profit without exceeding each bin's capacity.
Consider the following simplified instance of our Problem~\ref{prob:gap} where there is only one layer ($L=1$).
Map each item in the GAP to a splat in Problem~\ref{prob:gap} by setting $\Delta \tilde{U}_{j}[t] = p_{jt}$, $c_{j} = w_{jt} \; \forall t$, and $B[t] = c_t$.
Hence the reduction from GAP to Problem~\ref{prob:gap} is shown, so Problem~\ref{prob:gap} is NP-hard.
Second, we note that Problem~\ref{prob:gap} is equivalent to Problem~\ref{prob:full} by Lemma~\ref{lemma:gap_equiv}, so Problem~\ref{prob:full} is also NP-hard.
\end{proof}

\textbf{Theorem~\ref{thm:separate_np_hard}}

\begin{proof}\small 
The proof involves a reduction from the generalized assignment problem (GAP) to an instance of Problem~\ref{prob:case2}. The GAP can be stated as: given $T$ bins (with capacity $c_t$) and $n$ items (with profit $p_{jt}$ and weight $w_{jt}$), find the assignment of items to bins to maximize the profit without exceeding each bin's capacity.
Consider the following simplified instance of our Problem~\ref{prob:case2} where there is only one layer ($L=1$).
Then the objective function (\ref{eqn:jiasi_obj2})
can be written as
\begin{align}\small
    \sum_j \sum_{t=1}^T \tilde{U}_j[t]x_j[t]
\end{align}
\noindent using (\ref{eqn:cum_utility}) and following the same derivation as Lemma~\ref{lemma:gap_equiv} with $\tilde{U}_j = \sum_{t^\prime \geq t} U_{j}[t]$, similar to (\ref{eqn:cum_utility}).
Map each item in the GAP to a splat in Problem~\ref{prob:case2} by setting $\tilde{U}_{j}[t] = p_{jt}$, $c_{j} = w_{jt} \; \forall t$, and $B[t] = c_t$.
Or to be more precise, set $U_j[t] = p_{jt}-p_{j(t+1)}$ (reverse transformation of (\ref{eqn:cum_utility})).
Hence the reduction from GAP to Problem~\ref{prob:case2} is shown.
\end{proof}

\textbf{Lemma~\ref{lemma:knapsack}}

\begin{proof} \small
The multiple choice knapsack problem is defined as:
\begin{align}
    \max \sum_j \sum_l U_{jl}[t] z_{jl}[t] \label{eqn:mckp_obj}\\
    \text{s.t. }  
    \sum_{j} \sum_l c_{jl} x_{jl}[t] \leq B[t] \; \; \forall t  \label{eqn:mckp_bandwidth}\\
    \sum_l z_{jl}[t] \leq 1 \; \; \forall j \label{eqn:mckp_select_one}\\
    z_{jl}[t] \in \{0, 1\} \; \; \forall j,l
\end{align}
We will show that Problem~\ref{prob:gap}, which is equivalent to Problem~\ref{prob:full} by Lemma~\ref{lemma:gap_equiv}, can be transformed into this knapsack problem.

Let the decision variable from Problem~\ref{prob:gap} be defined as $x_{jl} \equiv \sum_{l^\prime \geq l} z_{jl^\prime}$.
For a given time slot, the objective (\ref{eqn:gap_obj}) from Problem~\ref{prob:gap} can be re-written as:
\begin{align} \small 
    \sum_j \sum_l \Delta U_{jl} x_{jl}
    &= \sum_j \sum_l (U_{jl}-U_{j,l-1}) x_{jl}\\
    &= \sum_j \sum_l U_{jl}(x_{jl} - x_{j,l+1})\\
    &= \sum_j \sum_l U_{jl} \left( \sum_{l^\prime \geq l} z_{jl^\prime} - \sum_{l^\prime \geq l+1} z_{jl^\prime} \right)\\
    &= \sum_j \sum_l U_{jl} z_{jl}
\end{align}
A similar transformation can be done on the cost constraint (\ref{eqn:gap_bandwidth}) from Problem~\ref{prob:gap} to turn it into the knapsack cost constraint (\ref{eqn:mckp_bandwidth}).
Finally, we see that (\ref{eqn:gap_select_one}) from Problem~\ref{prob:gap} can be written as $x_{jl} = \sum_{l^\prime \geq l} z_{jl^\prime} \leq 1, \forall j,l$, which implies $\sum_l z_{jl} \leq 1, \forall j$, which is constraint (\ref{eqn:mckp_select_one}) from the knapsack problem.
\end{proof}

\section{Additional visual quality evaluation metrics}
\label{sec:appendixB}

We further evaluate the models using additional metrics of PSNR, SSIM, and LPIPS, as presented in \Cref{tab:psnr}. Our layered model consistently demonstrates comparable performance to the ``Separate'' while consistently outperforming the ``Sort''. However, due to variations in scene size, model performance differs significantly across scenes. For instance, the ``Bicycle'' scene, which is notably large with an original model size of 1.52 GB, exhibits lower visual quality.

\begin{table*}[]
\caption{Evaluation was conducted using visual quality metrics, including PSNR (Peak Signal-to-Noise Ratio), SSIM (Structural Similarity Index), and LPIPS (Learned Perceptual Image Patch Similarity).} \label{tab:psnr}
\resizebox{0.9\textwidth}{!}{%
\begin{tabular}{|l|l|lll|lll|lll|}
\hline
\multirow{2}{*}{Scene}        & \multirow{2}{*}{\# of splats} & \multicolumn{3}{c|}{PSNR} & \multicolumn{3}{c|}{SSIM} & \multicolumn{3}{c|}{LPIPS} \\ \cline{3-11} 
                  &                               & \multicolumn{1}{l|}{Separate} & \multicolumn{1}{l|}{Sort} & Ours & \multicolumn{1}{l|}{Separate} & \multicolumn{1}{l|}{Sort} & Ours      & \multicolumn{1}{l|}{Separate} & \multicolumn{1}{l|}{Sort} & Ours       \\ \hline
\multirow{4}{*}{Bicycle} & 45,000        &\multicolumn{1}{l|}{21.315} & \multicolumn{1}{l|}{14.180} &  21.315 & \multicolumn{1}{l|}{0.464} & \multicolumn{1}{l|}{0.313} &  0.464  &\multicolumn{1}{l|}{0.571} & \multicolumn{1}{l|}{0.612} &   0.571  \\ \cline{2-11} 
                         & 90,000        &\multicolumn{1}{l|}{21.771} & \multicolumn{1}{l|}{14.663} &  21.817 &  \multicolumn{1}{l|}{0.490} & \multicolumn{1}{l|}{0.342} & 0.493  &\multicolumn{1}{l|}{0.531} & \multicolumn{1}{l|}{0.572} & 0.523 \\ \cline{2-11} 
                         & 135,000       &\multicolumn{1}{l|}{22.082} & \multicolumn{1}{l|}{15.020} & 21.949  &  \multicolumn{1}{l|}{0.510} & \multicolumn{1}{l|}{0.363} & 0.497 &\multicolumn{1}{l|}{0.499} & \multicolumn{1}{l|}{0.545} & 0.502\\ \cline{2-11} 
                         & 180,000       &\multicolumn{1}{l|}{22.301} & \multicolumn{1}{l|}{15.317} & 22.028& \multicolumn{1}{l|}{0.527} & \multicolumn{1}{l|}{0.379} & 0.516   &\multicolumn{1}{l|}{0.474} & \multicolumn{1}{l|}{0.525} &  0.490 \\ \hline
\multirow{4}{*}{Bonsai} & 45,000        &\multicolumn{1}{l|}{25.930} & \multicolumn{1}{l|}{18.193} &  25.930 & \multicolumn{1}{l|}{0.797} & \multicolumn{1}{l|}{0.639} &   0.797 &\multicolumn{1}{l|}{0.371} & \multicolumn{1}{l|}{0.458} & 0.371    \\ \cline{2-11} 
                         & 90,000        &\multicolumn{1}{l|}{27.993} & \multicolumn{1}{l|}{20.279} & 27.685  & \multicolumn{1}{l|}{0.856} & \multicolumn{1}{l|}{0.707} &  0.853 & \multicolumn{1}{l|}{0.306} & \multicolumn{1}{l|}{0.399} & 0.315 \\ \cline{2-11} 
                         & 135,000       &\multicolumn{1}{l|}{29.159} & \multicolumn{1}{l|}{21.848} & 28.304  &  \multicolumn{1}{l|}{0.887} & \multicolumn{1}{l|}{0.758} & 0.875  &\multicolumn{1}{l|}{0.269} & \multicolumn{1}{l|}{0.357} & 0.287 \\ \cline{2-11} 
                         & 180,000       &\multicolumn{1}{l|}{30.045} & \multicolumn{1}{l|}{23.181} & 28.600& \multicolumn{1}{l|}{0.9065} & \multicolumn{1}{l|}{0.799} &   0.885 &\multicolumn{1}{l|}{0.244} & \multicolumn{1}{l|}{0.3240} & 0.2731  \\ \hline
\multirow{4}{*}{Counter} & 45,000        &\multicolumn{1}{l|}{24.874} & \multicolumn{1}{l|}{18.214} & 24.874  & \multicolumn{1}{l|}{0.789} & \multicolumn{1}{l|}{0.643} &    0.789 &\multicolumn{1}{l|}{0.376} & \multicolumn{1}{l|}{0.446} &    0.376 \\ \cline{2-11} 
                         & 90,000        &\multicolumn{1}{l|}{26.412} & \multicolumn{1}{l|}{20.422} & 26.410  & \multicolumn{1}{l|}{0.831} & \multicolumn{1}{l|}{0.706} &  0.834 &\multicolumn{1}{l|}{0.319} & \multicolumn{1}{l|}{0.388} & 0.315 \\ \cline{2-11} 
                         & 135,000       &\multicolumn{1}{l|}{27.270} & \multicolumn{1}{l|}{21.957} & 26.847  & \multicolumn{1}{l|}{0.854} & \multicolumn{1}{l|}{0.748} & 0.850  &\multicolumn{1}{l|}{0.286} & \multicolumn{1}{l|}{0.3497} & 0.291 \\ \cline{2-11} 
                         & 180,000       &\multicolumn{1}{l|}{27.799} & \multicolumn{1}{l|}{23.179} & 27.076 & \multicolumn{1}{l|}{0.870} & \multicolumn{1}{l|}{0.781} &  0.860  &\multicolumn{1}{l|}{0.263} & \multicolumn{1}{l|}{0.319} & 0.276  \\ \hline
\multirow{4}{*}{Train} & 45,000        &\multicolumn{1}{l|}{19.603} & \multicolumn{1}{l|}{14.694} & 19.603  &  \multicolumn{1}{l|}{0.638} & \multicolumn{1}{l|}{0.490} &    0.638 &\multicolumn{1}{l|}{0.427} & \multicolumn{1}{l|}{0.508} &  0.427   \\ \cline{2-11} 
                         & 90,000        &\multicolumn{1}{l|}{20.712} & \multicolumn{1}{l|}{16.286} & 20.610  &\multicolumn{1}{l|}{0.704} & \multicolumn{1}{l|}{0.561} & 0.700  &\multicolumn{1}{l|}{0.358} & \multicolumn{1}{l|}{0.439} & 0.364 \\ \cline{2-11} 
                         & 135,000       &\multicolumn{1}{l|}{21.288} & \multicolumn{1}{l|}{17.399} & 20.937  &\multicolumn{1}{l|}{0.740} & \multicolumn{1}{l|}{0.615} &  0.724 &\multicolumn{1}{l|}{0.319} & \multicolumn{1}{l|}{0.390} & 0.335 \\ \cline{2-11} 
                         & 180,000       &\multicolumn{1}{l|}{21.608} & \multicolumn{1}{l|}{18.355} & 21.047 & \multicolumn{1}{l|}{0.762} & \multicolumn{1}{l|}{0.658} &  0.734  &\multicolumn{1}{l|}{0.293} & \multicolumn{1}{l|}{0.352} &  0.322 \\ \hline
\multirow{4}{*}{Truck} & 45,000        &\multicolumn{1}{l|}{21.106} & \multicolumn{1}{l|}{13.870} & 21.106  &  \multicolumn{1}{l|}{0.707} & \multicolumn{1}{l|}{0.559} &    0.707 &\multicolumn{1}{l|}{0.380} & \multicolumn{1}{l|}{0.460} &  0.380   \\ \cline{2-11} 
                         & 90,000        &\multicolumn{1}{l|}{22.527} & \multicolumn{1}{l|}{14.872} & 22.5652  & \multicolumn{1}{l|}{0.767} & \multicolumn{1}{l|}{0.575} &  0.771 &\multicolumn{1}{l|}{0.315} & \multicolumn{1}{l|}{0.4352} & 0.310 \\ \cline{2-11} 
                         & 135,000       &\multicolumn{1}{l|}{23.239} & \multicolumn{1}{l|}{16.104} & 22.902 & \multicolumn{1}{l|}{0.799} & \multicolumn{1}{l|}{0.625} & 0.790  &\multicolumn{1}{l|}{0.275} & \multicolumn{1}{l|}{0.384} & 0.284 \\ \cline{2-11} 
                         & 180,000       &\multicolumn{1}{l|}{23.676} & \multicolumn{1}{l|}{17.104} & 23.071 & \multicolumn{1}{l|}{0.819} & \multicolumn{1}{l|}{0.664} &  0.800  &\multicolumn{1}{l|}{0.248} & \multicolumn{1}{l|}{0.346} &  0.268 \\ \hline
\multirow{4}{*}{Figurines} & 45,000        &\multicolumn{1}{l|}{21.228} & \multicolumn{1}{l|}{14.075} & 21.228  & \multicolumn{1}{l|}{0.461} & \multicolumn{1}{l|}{0.439} &  0.461  &\multicolumn{1}{l|}{0.541} & \multicolumn{1}{l|}{0.603} &  0.541   \\ \cline{2-11} 
                         & 90,000        &\multicolumn{1}{l|}{22.281} & \multicolumn{1}{l|}{14.872} & 22.279  & \multicolumn{1}{l|}{0.493} & \multicolumn{1}{l|}{0.451} &  0.490 & \multicolumn{1}{l|}{0.507} & \multicolumn{1}{l|}{0.587} &  0.522\\ \cline{2-11} 
                         & 135,000       &\multicolumn{1}{l|}{23.938} & \multicolumn{1}{l|}{20.986} & 23.610  &  \multicolumn{1}{l|}{0.509} & \multicolumn{1}{l|}{0.458} & 0.498  &\multicolumn{1}{l|}{0.500} & \multicolumn{1}{l|}{0.551} &  0.516\\ \cline{2-11} 
                         & 180,000       &\multicolumn{1}{l|}{24.657} & \multicolumn{1}{l|}{22.985} & 24.176& \multicolumn{1}{l|}{0.525} & \multicolumn{1}{l|}{0.573} &  0.514  &\multicolumn{1}{l|}{0.482} & \multicolumn{1}{l|}{0.535} &  0.516 \\ \hline
\multirow{4}{*}{Ramen} & 45,000        &\multicolumn{1}{l|}{25.236} & \multicolumn{1}{l|}{21.014} & 25.236  & \multicolumn{1}{l|}{0.798} & \multicolumn{1}{l|}{0.674} &  0.798  &\multicolumn{1}{l|}{0.397} & \multicolumn{1}{l|}{0.448} &  0.397   \\ \cline{2-11} 
                         & 90,000        &\multicolumn{1}{l|}{27.099} & \multicolumn{1}{l|}{22.154} & 26.804  & \multicolumn{1}{l|}{0.836} & \multicolumn{1}{l|}{0.678} &  0.814 & \multicolumn{1}{l|}{0.302} & \multicolumn{1}{l|}{0.416} & 0.338 \\ \cline{2-11} 
                         & 135,000       &\multicolumn{1}{l|}{27.499} & \multicolumn{1}{l|}{23.891} & 27.021  &  \multicolumn{1}{l|}{0.859} & \multicolumn{1}{l|}{0.701} & 0.850  &\multicolumn{1}{l|}{0.264} & \multicolumn{1}{l|}{0.411} & 0.289 \\ \cline{2-11} 
                         & 180,000       &\multicolumn{1}{l|}{27.787} & \multicolumn{1}{l|}{23.904} &27.283 & \multicolumn{1}{l|}{0.884} & \multicolumn{1}{l|}{0.704} &   0.862 &\multicolumn{1}{l|}{0.240} & \multicolumn{1}{l|}{0.407} &  0.253 \\ \hline
\multirow{4}{*}{Teatime} & 45,000        &\multicolumn{1}{l|}{27.588} & \multicolumn{1}{l|}{21.891} &27.588   & \multicolumn{1}{l|}{0.746} & \multicolumn{1}{l|}{0.685} &  0.746   &\multicolumn{1}{l|}{0.395} & \multicolumn{1}{l|}{0.403} &  0.395   \\ \cline{2-11} 
                         & 90,000        &\multicolumn{1}{l|}{28.016} & \multicolumn{1}{l|}{23.076} & 27.819  & \multicolumn{1}{l|}{0.770} & \multicolumn{1}{l|}{0.694} &  0.762 &\multicolumn{1}{l|}{0.341} & \multicolumn{1}{l|}{0.402} & 0.348 \\ \cline{2-11} 
                         & 135,000       &\multicolumn{1}{l|}{28.476} & \multicolumn{1}{l|}{23.098} &  28.051 & \multicolumn{1}{l|}{0.845} & \multicolumn{1}{l|}{0.700} &  0.816 &\multicolumn{1}{l|}{0.276} & \multicolumn{1}{l|}{0390} &  0.300\\ \cline{2-11} 
                         & 180,000       &\multicolumn{1}{l|}{28.912} & \multicolumn{1}{l|}{23.371} & 28.517 & \multicolumn{1}{l|}{0.856} & \multicolumn{1}{l|}{0.703} &  0.833  &\multicolumn{1}{l|}{0.255} & \multicolumn{1}{l|}{0.387} &   0.296\\ \hline
\end{tabular}
}
\end{table*}
\section{Synthetic traces generation}
\begin{figure}
\centering
\begin{subfigure}[b]{0.3\textwidth}
\includegraphics[width=\textwidth]{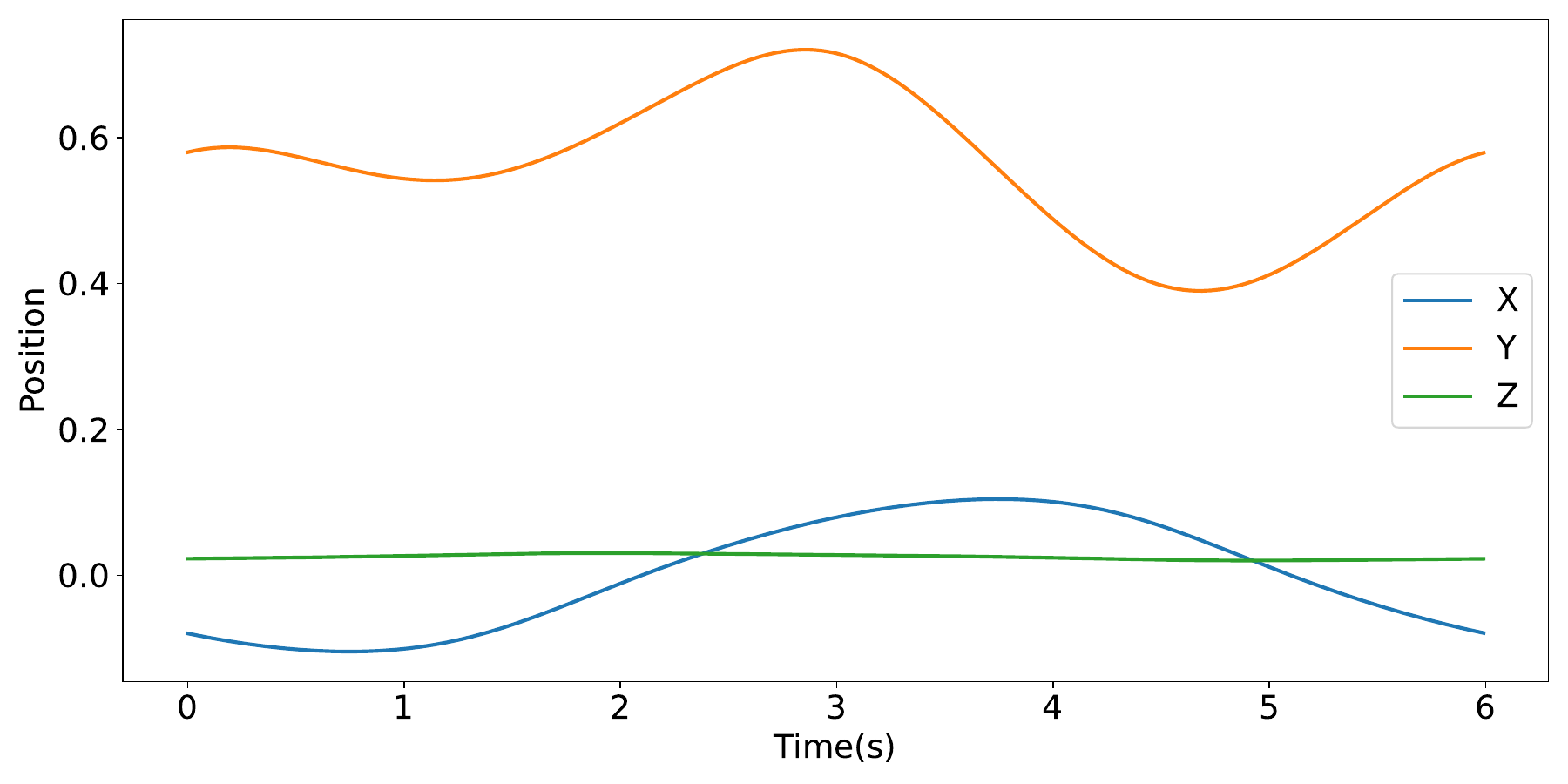}
\vspace{-15pt}
\end{subfigure}
\begin{subfigure}[b]{0.3\textwidth}
\includegraphics[width=\textwidth]{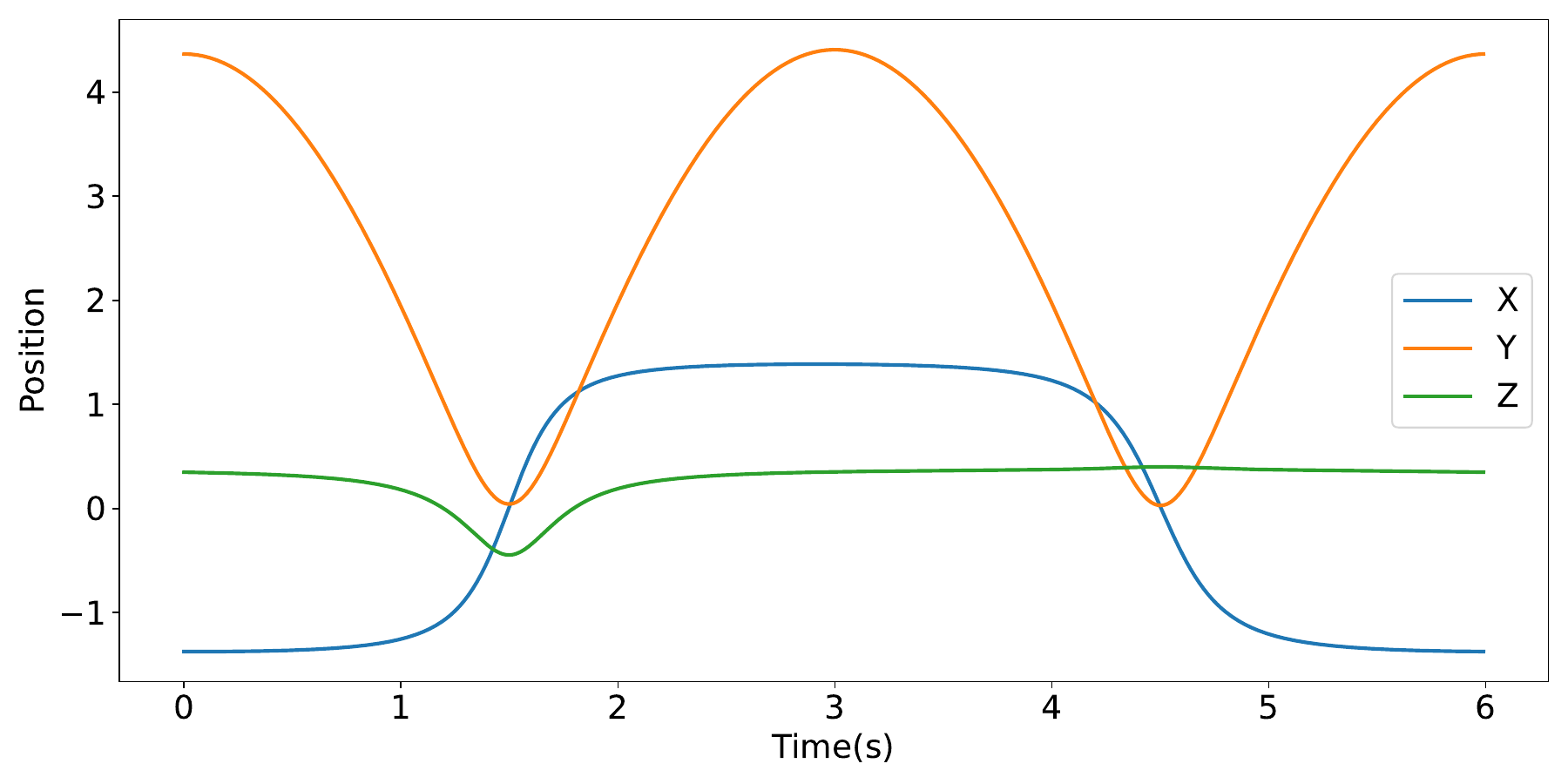}
\vspace{-15pt}
\end{subfigure}
\begin{subfigure}[b]{0.3\textwidth}
\includegraphics[width=\textwidth]{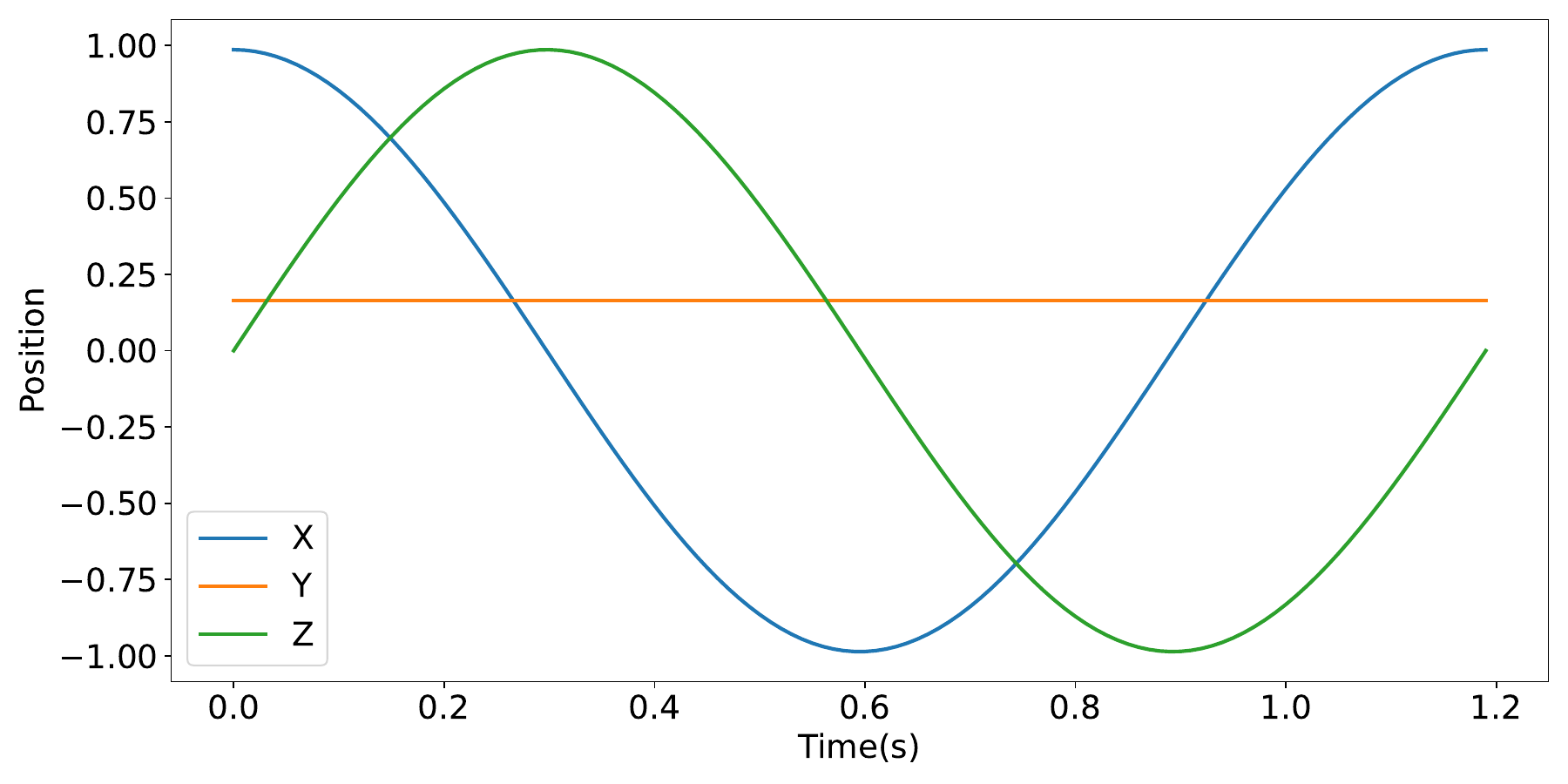}
\vspace{-15pt}
\end{subfigure}
\begin{subfigure}[b]{0.3\textwidth}
\includegraphics[width=\textwidth]{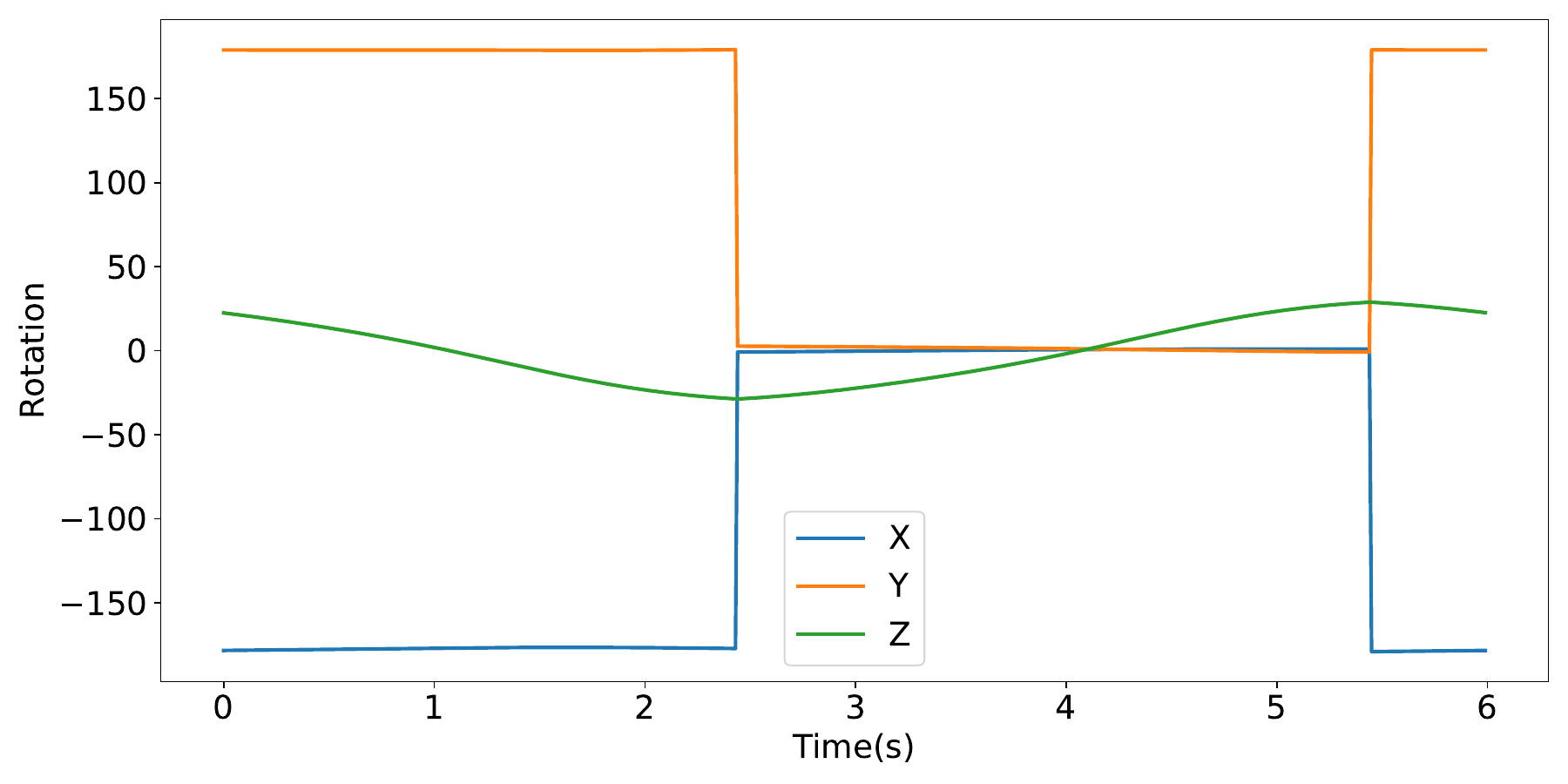}
\vspace{-15pt}
\end{subfigure}
\begin{subfigure}[b]{0.3\textwidth}
\includegraphics[width=\textwidth]{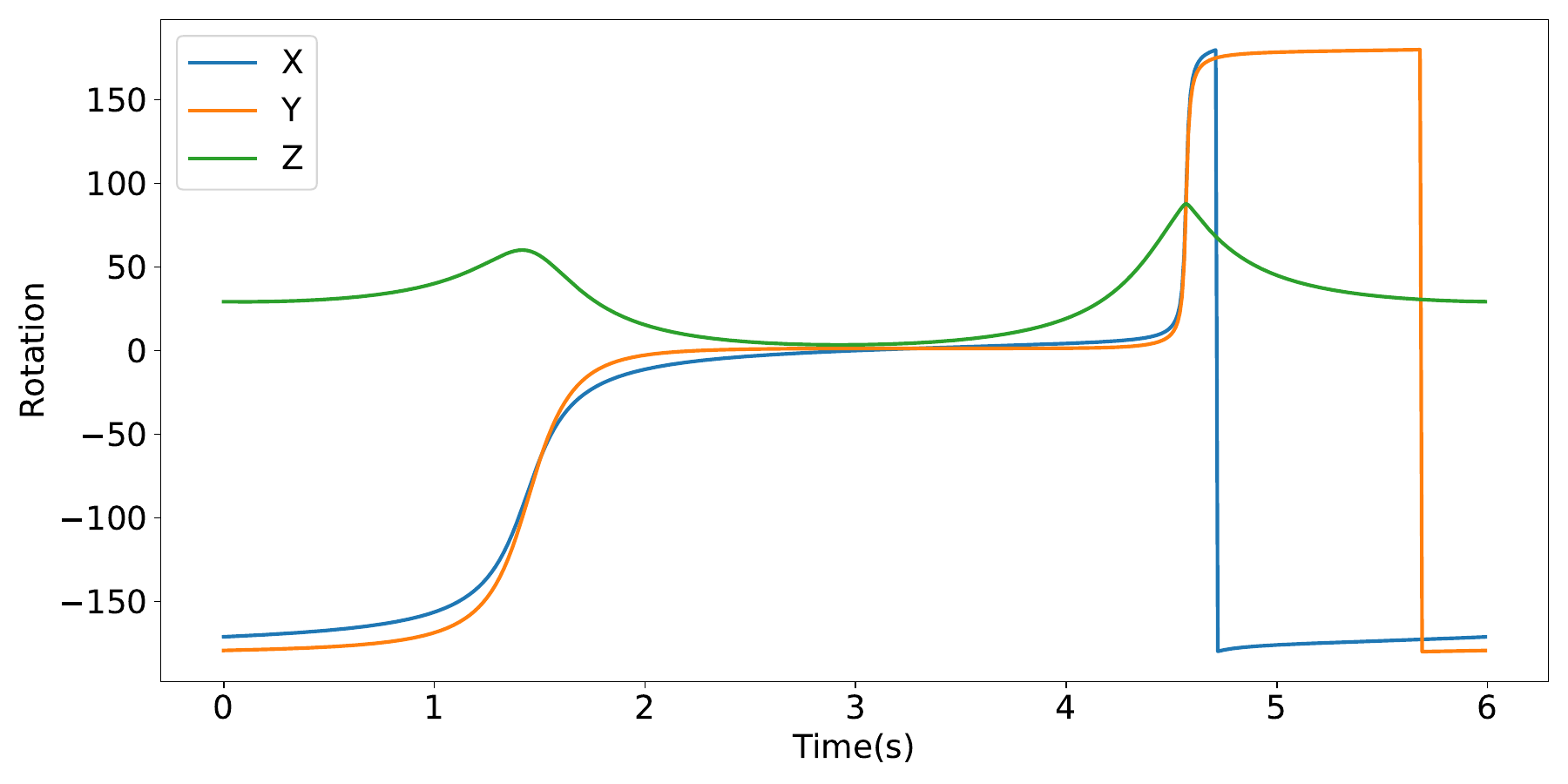}
\vspace{-15pt}
\end{subfigure}
\begin{subfigure}[b]{0.3\textwidth}
\includegraphics[width=\textwidth]{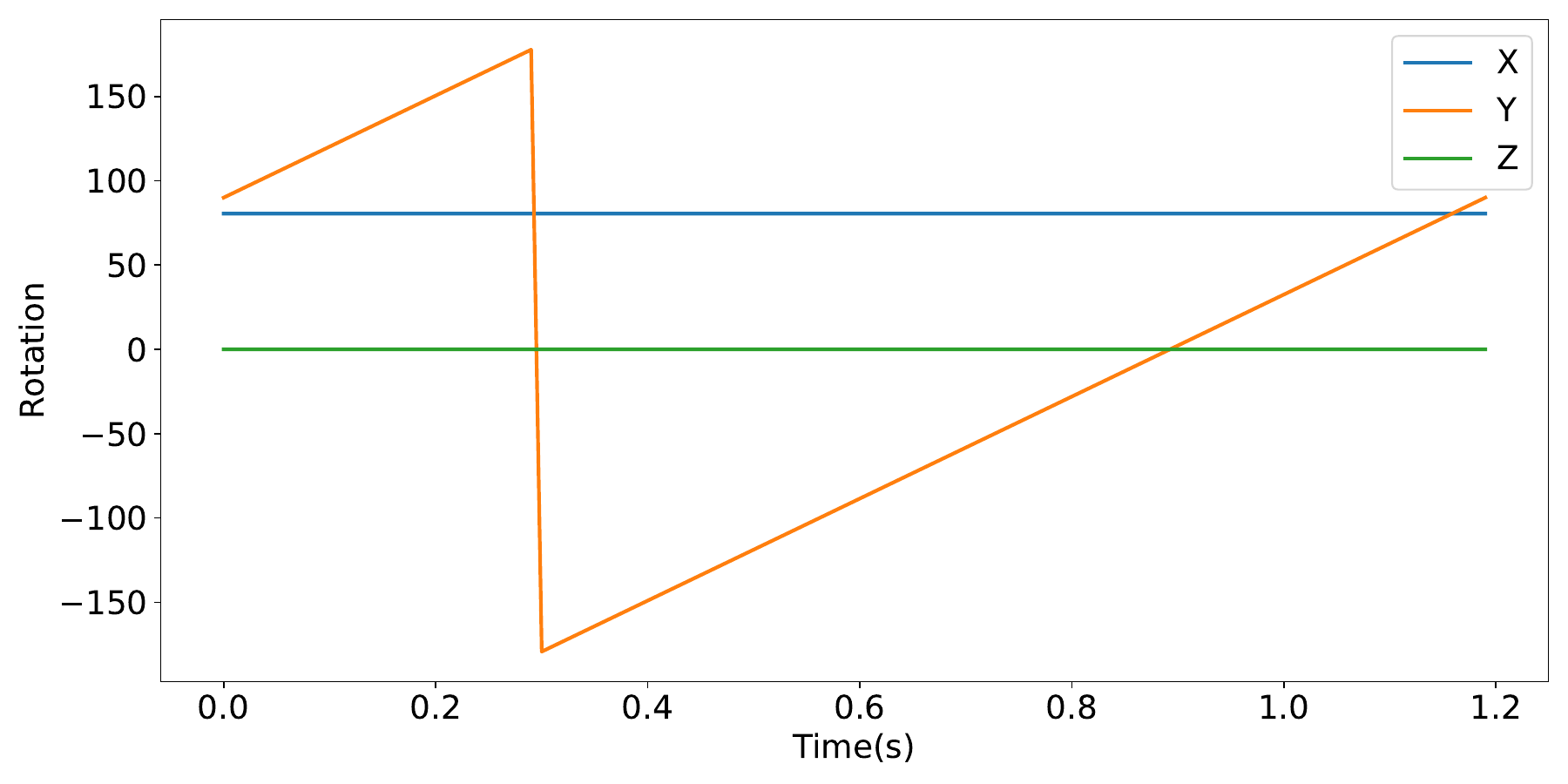}
\vspace{-15pt}
\end{subfigure}
\vspace{-10pt}
\caption{Position and rotation of 3 synthetically generated user movement traces, to accompany our real user traces.}
\label{fig:syn_generation}
\end{figure}

Since the user's movements in real traces are somewhat random, as described in the main text, they might focus on some out-of-scene viewports such as holes where there is no ground truth visual data, affecting the evaluation of our pipeline.
Therefore, we generated 6 synthetic traces.
3 traces of those traces were generated by placing the existing test views from the 3D scenes in sequence.
The remaining 3 synthetic traces follow simple movement patterns such as walking in a circle, spinning on the spot, etc.
Their position and rotation are visualized in \Cref{fig:syn_generation}.

%% file: main.bbl

\begin{thebibliography}{41}


\ifx \showCODEN    \undefined \def \showCODEN     #1{\unskip}     \fi
\ifx \showDOI      \undefined \def \showDOI       #1{#1}\fi
\ifx \showISBNx    \undefined \def \showISBNx     #1{\unskip}     \fi
\ifx \showISBNxiii \undefined \def \showISBNxiii  #1{\unskip}     \fi
\ifx \showISSN     \undefined \def \showISSN      #1{\unskip}     \fi
\ifx \showLCCN     \undefined \def \showLCCN      #1{\unskip}     \fi
\ifx \shownote     \undefined \def \shownote      #1{#1}          \fi
\ifx \showarticletitle \undefined \def \showarticletitle #1{#1}   \fi
\ifx \showURL      \undefined \def \showURL       {\relax}        \fi
\providecommand\bibfield[2]{#2}
\providecommand\bibinfo[2]{#2}
\providecommand\natexlab[1]{#1}
\providecommand\showeprint[2][]{arXiv:#2}

\bibitem[tec(2024)]%
        {tech-report}
 \bibinfo{year}{2024}\natexlab{}.
\newblock \bibinfo{title}{L3GS: Layered 3D Gaussian Splats for Efficient 3D Scene Delivery (technical report)}.
\newblock \bibinfo{howpublished}{\url{https://github.com/mavens-lab/layered_3d_gaussian_splats}}.
\newblock


\bibitem[Barron et~al\mbox{.}(2021)]%
        {barron2021mip}
\bibfield{author}{\bibinfo{person}{Jonathan~T Barron}, \bibinfo{person}{Ben Mildenhall}, \bibinfo{person}{Matthew Tancik}, \bibinfo{person}{Peter Hedman}, \bibinfo{person}{Ricardo Martin-Brualla}, {and} \bibinfo{person}{Pratul~P Srinivasan}.} \bibinfo{year}{2021}\natexlab{}.
\newblock \showarticletitle{Mip-nerf: A multiscale representation for anti-aliasing neural radiance fields}. In \bibinfo{booktitle}{\emph{IEEE/CVF ICCV}}.
\newblock


\bibitem[Barron et~al\mbox{.}(2022)]%
        {barron2022mipnerf360}
\bibfield{author}{\bibinfo{person}{Jonathan~T. Barron}, \bibinfo{person}{Ben Mildenhall}, \bibinfo{person}{Dor Verbin}, \bibinfo{person}{Pratul~P. Srinivasan}, {and} \bibinfo{person}{Peter Hedman}.} \bibinfo{year}{2022}\natexlab{}.
\newblock \showarticletitle{Mip-NeRF 360: Unbounded Anti-Aliased Neural Radiance Fields}.
\newblock \bibinfo{journal}{\emph{CVPR}} (\bibinfo{year}{2022}).
\newblock


\bibitem[Bentaleb et~al\mbox{.}(2019)]%
        {bentaleb2019bandwidth}
\bibfield{author}{\bibinfo{person}{Abdelhak Bentaleb}, \bibinfo{person}{Christian Timmerer}, \bibinfo{person}{Ali~C Begen}, {and} \bibinfo{person}{Roger Zimmermann}.} \bibinfo{year}{2019}\natexlab{}.
\newblock \showarticletitle{Bandwidth prediction in low-latency chunked streaming}. In \bibinfo{booktitle}{\emph{Proceedings of the 29th ACM workshop on network and operating systems support for digital audio and video}}. \bibinfo{pages}{7--13}.
\newblock


\bibitem[Chen et~al\mbox{.}(2024c)]%
        {chen2024nerfhub}
\bibfield{author}{\bibinfo{person}{Bo Chen}, \bibinfo{person}{Zhisheng Yan}, \bibinfo{person}{Bo Han}, {and} \bibinfo{person}{Klara Nahrstedt}.} \bibinfo{year}{2024}\natexlab{c}.
\newblock \showarticletitle{NeRFHub: A Context-Aware NeRF Serving Framework for Mobile Immersive Applications}. In \bibinfo{booktitle}{\emph{Proceedings of the 22nd Annual International Conference on Mobile Systems, Applications and Services}}. \bibinfo{pages}{85--98}.
\newblock


\bibitem[Chen et~al\mbox{.}(2022)]%
        {9729212}
\bibfield{author}{\bibinfo{person}{Cuiqun Chen}, \bibinfo{person}{Mang Ye}, \bibinfo{person}{Meibin Qi}, \bibinfo{person}{Jingjing Wu}, \bibinfo{person}{Yimin Liu}, {and} \bibinfo{person}{Jianguo Jiang}.} \bibinfo{year}{2022}\natexlab{}.
\newblock \showarticletitle{Saliency and Granularity: Discovering Temporal Coherence for Video-Based Person Re-Identification}.
\newblock \bibinfo{journal}{\emph{IEEE Transactions on Circuits and Systems for Video Technology}} \bibinfo{volume}{32}, \bibinfo{number}{9} (\bibinfo{year}{2022}), \bibinfo{pages}{6100--6112}.
\newblock
\urldef\tempurl%
\url{https://doi.org/10.1109/TCSVT.2022.3157130}
\showDOI{\tempurl}


\bibitem[Chen et~al\mbox{.}(2024a)]%
        {chen2024far}
\bibfield{author}{\bibinfo{person}{Yihang Chen}, \bibinfo{person}{Qianyi Wu}, \bibinfo{person}{Mehrtash Harandi}, {and} \bibinfo{person}{Jianfei Cai}.} \bibinfo{year}{2024}\natexlab{a}.
\newblock \showarticletitle{How Far Can We Compress Instant-NGP-Based NeRF?}. In \bibinfo{booktitle}{\emph{Proceedings of the IEEE/CVF Conference on Computer Vision and Pattern Recognition}}. \bibinfo{pages}{20321--20330}.
\newblock


\bibitem[Chen et~al\mbox{.}(2024b)]%
        {chen2024hachashgridassistedcontext}
\bibfield{author}{\bibinfo{person}{Yihang Chen}, \bibinfo{person}{Qianyi Wu}, \bibinfo{person}{Weiyao Lin}, \bibinfo{person}{Mehrtash Harandi}, {and} \bibinfo{person}{Jianfei Cai}.} \bibinfo{year}{2024}\natexlab{b}.
\newblock \bibinfo{title}{HAC: Hash-grid Assisted Context for 3D Gaussian Splatting Compression}.
\newblock
\showeprint[arxiv]{2403.14530}~[cs.CV]
\urldef\tempurl%
\url{https://arxiv.org/abs/2403.14530}
\showURL{%
\tempurl}


\bibitem[Chen et~al\mbox{.}(2023)]%
        {chen2023mobilenerf}
\bibfield{author}{\bibinfo{person}{Zhiqin Chen}, \bibinfo{person}{Thomas Funkhouser}, \bibinfo{person}{Peter Hedman}, {and} \bibinfo{person}{Andrea Tagliasacchi}.} \bibinfo{year}{2023}\natexlab{}.
\newblock \showarticletitle{Mobilenerf: Exploiting the polygon rasterization pipeline for efficient neural field rendering on mobile architectures}. In \bibinfo{booktitle}{\emph{Proceedings of the IEEE/CVF Conference on Computer Vision and Pattern Recognition}}. \bibinfo{pages}{16569--16578}.
\newblock


\bibitem[Fan et~al\mbox{.}(2023)]%
        {fan2023lightgaussian}
\bibfield{author}{\bibinfo{person}{Zhiwen Fan}, \bibinfo{person}{Kevin Wang}, \bibinfo{person}{Kairun Wen}, \bibinfo{person}{Zehao Zhu}, \bibinfo{person}{Dejia Xu}, {and} \bibinfo{person}{Zhangyang Wang}.} \bibinfo{year}{2023}\natexlab{}.
\newblock \showarticletitle{Lightgaussian: Unbounded 3d gaussian compression with 15x reduction and 200+ fps}.
\newblock \bibinfo{journal}{\emph{arXiv preprint arXiv:2311.17245}} (\bibinfo{year}{2023}).
\newblock


\bibitem[Guan et~al\mbox{.}(2023)]%
        {guan2023metastream}
\bibfield{author}{\bibinfo{person}{Yongjie Guan}, \bibinfo{person}{Xueyu Hou}, \bibinfo{person}{Nan Wu}, \bibinfo{person}{Bo Han}, {and} \bibinfo{person}{Tao Han}.} \bibinfo{year}{2023}\natexlab{}.
\newblock \showarticletitle{Metastream: Live volumetric content capture, creation, delivery, and rendering in real time}. In \bibinfo{booktitle}{\emph{Proceedings of the 29th Annual International Conference on Mobile Computing and Networking}}. \bibinfo{pages}{1--15}.
\newblock


\bibitem[Han et~al\mbox{.}(2020)]%
        {han2020vivo}
\bibfield{author}{\bibinfo{person}{Bo Han}, \bibinfo{person}{Yu Liu}, {and} \bibinfo{person}{Feng Qian}.} \bibinfo{year}{2020}\natexlab{}.
\newblock \showarticletitle{ViVo: Visibility-aware mobile volumetric video streaming}. In \bibinfo{booktitle}{\emph{Proceedings of the 26th annual international conference on mobile computing and networking}}. \bibinfo{pages}{1--13}.
\newblock


\bibitem[He et~al\mbox{.}(2018)]%
        {10.1145/3210240.3210323}
\bibfield{author}{\bibinfo{person}{Jian He}, \bibinfo{person}{Mubashir~Adnan Qureshi}, \bibinfo{person}{Lili Qiu}, \bibinfo{person}{Jin Li}, \bibinfo{person}{Feng Li}, {and} \bibinfo{person}{Lei Han}.} \bibinfo{year}{2018}\natexlab{}.
\newblock \showarticletitle{Rubiks: Practical 360-Degree Streaming for Smartphones}. In \bibinfo{booktitle}{\emph{Proceedings of the 16th Annual International Conference on Mobile Systems, Applications, and Services}} (Munich, Germany) \emph{(\bibinfo{series}{MobiSys '18})}. \bibinfo{publisher}{Association for Computing Machinery}, \bibinfo{address}{New York, NY, USA}, \bibinfo{pages}{482–494}.
\newblock
\showISBNx{9781450357203}
\urldef\tempurl%
\url{https://doi.org/10.1145/3210240.3210323}
\showDOI{\tempurl}


\bibitem[Jiang et~al\mbox{.}(2012)]%
        {jiang2012improving}
\bibfield{author}{\bibinfo{person}{Junchen Jiang}, \bibinfo{person}{Vyas Sekar}, {and} \bibinfo{person}{Hui Zhang}.} \bibinfo{year}{2012}\natexlab{}.
\newblock \showarticletitle{Improving fairness, efficiency, and stability in http-based adaptive video streaming with festive}. In \bibinfo{booktitle}{\emph{Proceedings of the 8th international conference on Emerging networking experiments and technologies}}. \bibinfo{pages}{97--108}.
\newblock


\bibitem[Kerbl et~al\mbox{.}(2023)]%
        {kerbl3Dgaussians}
\bibfield{author}{\bibinfo{person}{Bernhard Kerbl}, \bibinfo{person}{Georgios Kopanas}, \bibinfo{person}{Thomas Leimk{\"u}hler}, {and} \bibinfo{person}{George Drettakis}.} \bibinfo{year}{2023}\natexlab{}.
\newblock \showarticletitle{3D Gaussian Splatting for Real-Time Radiance Field Rendering}.
\newblock \bibinfo{journal}{\emph{ACM Transactions on Graphics}} \bibinfo{volume}{42}, \bibinfo{number}{4} (\bibinfo{date}{July} \bibinfo{year}{2023}).
\newblock
\urldef\tempurl%
\url{https://repo-sam.inria.fr/fungraph/3d-gaussian-splatting/}
\showURL{%
\tempurl}


\bibitem[Kerbl et~al\mbox{.}(2024)]%
        {kerbl2024hierarchical}
\bibfield{author}{\bibinfo{person}{Bernhard Kerbl}, \bibinfo{person}{Andreas Meuleman}, \bibinfo{person}{Georgios Kopanas}, \bibinfo{person}{Michael Wimmer}, \bibinfo{person}{Alexandre Lanvin}, {and} \bibinfo{person}{George Drettakis}.} \bibinfo{year}{2024}\natexlab{}.
\newblock \showarticletitle{A hierarchical 3d gaussian representation for real-time rendering of very large datasets}.
\newblock \bibinfo{journal}{\emph{ACM Transactions on Graphics (TOG)}} \bibinfo{volume}{43}, \bibinfo{number}{4} (\bibinfo{year}{2024}), \bibinfo{pages}{1--15}.
\newblock


\bibitem[Knapitsch et~al\mbox{.}(2017)]%
        {knapitsch2017tanksntemple}
\bibfield{author}{\bibinfo{person}{Arno Knapitsch}, \bibinfo{person}{Jaesik Park}, \bibinfo{person}{Qian-Yi Zhou}, {and} \bibinfo{person}{Vladlen Koltun}.} \bibinfo{year}{2017}\natexlab{}.
\newblock \showarticletitle{Tanks and temples: benchmarking large-scale scene reconstruction}.
\newblock \bibinfo{journal}{\emph{ACM Trans. Graph.}} \bibinfo{volume}{36}, \bibinfo{number}{4}, Article \bibinfo{articleno}{78} (\bibinfo{date}{jul} \bibinfo{year}{2017}), \bibinfo{numpages}{13}~pages.
\newblock
\showISSN{0730-0301}
\urldef\tempurl%
\url{https://doi.org/10.1145/3072959.3073599}
\showDOI{\tempurl}


\bibitem[Li et~al\mbox{.}(2023)]%
        {li2023viewport}
\bibfield{author}{\bibinfo{person}{Jie Li}, \bibinfo{person}{Zhixin Li}, \bibinfo{person}{Zhi Liu}, \bibinfo{person}{Pengyuan Zhou}, \bibinfo{person}{Richang Hong}, \bibinfo{person}{Qiyue Li}, {and} \bibinfo{person}{Han Hu}.} \bibinfo{year}{2023}\natexlab{}.
\newblock \showarticletitle{Viewport Prediction for Volumetric Video Streaming by Exploring Video Saliency and Trajectory Information}.
\newblock \bibinfo{journal}{\emph{arXiv preprint arXiv:2311.16462}} (\bibinfo{year}{2023}).
\newblock


\bibitem[Liu and Banerjee(2024)]%
        {liu2024swings}
\bibfield{author}{\bibinfo{person}{Bangya Liu} {and} \bibinfo{person}{Suman Banerjee}.} \bibinfo{year}{2024}\natexlab{}.
\newblock \showarticletitle{Swings: Sliding window Gaussian splatting for volumetric video streaming with arbitrary length}.
\newblock \bibinfo{journal}{\emph{arXiv preprint arXiv:2409.07759}} (\bibinfo{year}{2024}).
\newblock


\bibitem[Liu et~al\mbox{.}(2024)]%
        {liu2024muv2}
\bibfield{author}{\bibinfo{person}{Yu Liu}, \bibinfo{person}{Puqi Zhou}, \bibinfo{person}{Zejun Zhang}, \bibinfo{person}{Anlan Zhang}, \bibinfo{person}{Bo Han}, \bibinfo{person}{Zhenhua Li}, {and} \bibinfo{person}{Feng Qian}.} \bibinfo{year}{2024}\natexlab{}.
\newblock \showarticletitle{MuV2: Scaling up Multi-user Mobile Volumetric Video Streaming via Content Hybridization and Sharing}. In \bibinfo{booktitle}{\emph{Proceedings of the 30th Annual International Conference on Mobile Computing and Networking}}. \bibinfo{pages}{327--341}.
\newblock


\bibitem[Lu et~al\mbox{.}(2024)]%
        {lu2024scaffold}
\bibfield{author}{\bibinfo{person}{Tao Lu}, \bibinfo{person}{Mulin Yu}, \bibinfo{person}{Linning Xu}, \bibinfo{person}{Yuanbo Xiangli}, \bibinfo{person}{Limin Wang}, \bibinfo{person}{Dahua Lin}, {and} \bibinfo{person}{Bo Dai}.} \bibinfo{year}{2024}\natexlab{}.
\newblock \showarticletitle{Scaffold-gs: Structured 3d gaussians for view-adaptive rendering}. In \bibinfo{booktitle}{\emph{Proceedings of the IEEE/CVF Conference on Computer Vision and Pattern Recognition}}. \bibinfo{pages}{20654--20664}.
\newblock


\bibitem[Mildenhall et~al\mbox{.}(2021)]%
        {mildenhall2021nerf}
\bibfield{author}{\bibinfo{person}{Ben Mildenhall}, \bibinfo{person}{Pratul~P Srinivasan}, \bibinfo{person}{Matthew Tancik}, \bibinfo{person}{Jonathan~T Barron}, \bibinfo{person}{Ravi Ramamoorthi}, {and} \bibinfo{person}{Ren Ng}.} \bibinfo{year}{2021}\natexlab{}.
\newblock \showarticletitle{Nerf: Representing scenes as neural radiance fields for view synthesis}.
\newblock \bibinfo{journal}{\emph{Commun. ACM}} \bibinfo{volume}{65}, \bibinfo{number}{1} (\bibinfo{year}{2021}), \bibinfo{pages}{99--106}.
\newblock


\bibitem[M{\"u}ller et~al\mbox{.}(2022)]%
        {muller2022instant}
\bibfield{author}{\bibinfo{person}{Thomas M{\"u}ller}, \bibinfo{person}{Alex Evans}, \bibinfo{person}{Christoph Schied}, {and} \bibinfo{person}{Alexander Keller}.} \bibinfo{year}{2022}\natexlab{}.
\newblock \showarticletitle{Instant neural graphics primitives with a multiresolution hash encoding}.
\newblock \bibinfo{journal}{\emph{ACM Transactions on Graphics}} \bibinfo{volume}{41}, \bibinfo{number}{4} (\bibinfo{year}{2022}), \bibinfo{pages}{1--15}.
\newblock


\bibitem[Narayanan et~al\mbox{.}(2020)]%
        {narayanan2020lumos5g}
\bibfield{author}{\bibinfo{person}{Arvind Narayanan}, \bibinfo{person}{Eman Ramadan}, \bibinfo{person}{Rishabh Mehta}, \bibinfo{person}{Xinyue Hu}, \bibinfo{person}{Qingxu Liu}, \bibinfo{person}{Rostand A.~K. Fezeu}, \bibinfo{person}{Udhaya~Kumar Dayalan}, \bibinfo{person}{Saurabh Verma}, \bibinfo{person}{Peiqi Ji}, \bibinfo{person}{Tao Li}, \bibinfo{person}{Feng Qian}, {and} \bibinfo{person}{Zhi-Li Zhang}.} \bibinfo{year}{2020}\natexlab{}.
\newblock \showarticletitle{Lumos5G: Mapping and Predicting Commercial MmWave 5G Throughput}. In \bibinfo{booktitle}{\emph{Proceedings of the ACM Internet Measurement Conference}} (Virtual Event, USA) \emph{(\bibinfo{series}{IMC '20})}. \bibinfo{publisher}{Association for Computing Machinery}, \bibinfo{address}{New York, NY, USA}, \bibinfo{pages}{176–193}.
\newblock
\showISBNx{9781450381383}
\urldef\tempurl%
\url{https://doi.org/10.1145/3419394.3423629}
\showDOI{\tempurl}


\bibitem[Papantonakis et~al\mbox{.}(2024)]%
        {papantonakis2024reducing}
\bibfield{author}{\bibinfo{person}{Panagiotis Papantonakis}, \bibinfo{person}{Georgios Kopanas}, \bibinfo{person}{Bernhard Kerbl}, \bibinfo{person}{Alexandre Lanvin}, {and} \bibinfo{person}{George Drettakis}.} \bibinfo{year}{2024}\natexlab{}.
\newblock \showarticletitle{Reducing the Memory Footprint of 3D Gaussian Splatting}.
\newblock \bibinfo{journal}{\emph{Proceedings of the ACM on Computer Graphics and Interactive Techniques}} \bibinfo{volume}{7}, \bibinfo{number}{1} (\bibinfo{year}{2024}), \bibinfo{pages}{1--17}.
\newblock


\bibitem[Qian et~al\mbox{.}(2018)]%
        {10.1145/3241539.3241565}
\bibfield{author}{\bibinfo{person}{Feng Qian}, \bibinfo{person}{Bo Han}, \bibinfo{person}{Qingyang Xiao}, {and} \bibinfo{person}{Vijay Gopalakrishnan}.} \bibinfo{year}{2018}\natexlab{}.
\newblock \showarticletitle{Flare: Practical Viewport-Adaptive 360-Degree Video Streaming for Mobile Devices}. In \bibinfo{booktitle}{\emph{Proceedings of the 24th Annual International Conference on Mobile Computing and Networking}} (New Delhi, India) \emph{(\bibinfo{series}{MobiCom '18})}. \bibinfo{publisher}{Association for Computing Machinery}, \bibinfo{address}{New York, NY, USA}, \bibinfo{pages}{99–114}.
\newblock
\showISBNx{9781450359030}
\urldef\tempurl%
\url{https://doi.org/10.1145/3241539.3241565}
\showDOI{\tempurl}


\bibitem[Ren et~al\mbox{.}(2024)]%
        {ren2024octree}
\bibfield{author}{\bibinfo{person}{Kerui Ren}, \bibinfo{person}{Lihan Jiang}, \bibinfo{person}{Tao Lu}, \bibinfo{person}{Mulin Yu}, \bibinfo{person}{Linning Xu}, \bibinfo{person}{Zhangkai Ni}, {and} \bibinfo{person}{Bo Dai}.} \bibinfo{year}{2024}\natexlab{}.
\newblock \showarticletitle{Octree-gs: Towards consistent real-time rendering with lod-structured 3d gaussians}.
\newblock \bibinfo{journal}{\emph{arXiv preprint arXiv:2403.17898}} (\bibinfo{year}{2024}).
\newblock


\bibitem[Schwarz et~al\mbox{.}(2007)]%
        {schwarz2007overview}
\bibfield{author}{\bibinfo{person}{Heiko Schwarz}, \bibinfo{person}{Detlev Marpe}, {and} \bibinfo{person}{Thomas Wiegand}.} \bibinfo{year}{2007}\natexlab{}.
\newblock \showarticletitle{Overview of the scalable video coding extension of the H. 264/AVC standard}.
\newblock \bibinfo{journal}{\emph{IEEE Transactions on circuits and systems for video technology}} \bibinfo{volume}{17}, \bibinfo{number}{9} (\bibinfo{year}{2007}), \bibinfo{pages}{1103--1120}.
\newblock


\bibitem[Shi et~al\mbox{.}(2024)]%
        {shi2024lapisgslayeredprogressive3d}
\bibfield{author}{\bibinfo{person}{Yuang Shi}, \bibinfo{person}{Simone Gasparini}, \bibinfo{person}{Géraldine Morin}, {and} \bibinfo{person}{Wei~Tsang Ooi}.} \bibinfo{year}{2024}\natexlab{}.
\newblock \bibinfo{title}{LapisGS: Layered Progressive 3D Gaussian Splatting for Adaptive Streaming}.
\newblock
\showeprint[arxiv]{2408.14823}~[cs.CV]
\urldef\tempurl%
\url{https://arxiv.org/abs/2408.14823}
\showURL{%
\tempurl}


\bibitem[Shuai et~al\mbox{.}(2024)]%
        {shuai2024LoG}
\bibfield{author}{\bibinfo{person}{Qing Shuai}, \bibinfo{person}{Haoyu Guo}, \bibinfo{person}{Zhen Xu}, \bibinfo{person}{Haotong Lin}, \bibinfo{person}{Sida Peng}, \bibinfo{person}{Hujun Bao}, {and} \bibinfo{person}{Xiaowei Zhou}.} \bibinfo{year}{2024}\natexlab{}.
\newblock \showarticletitle{Real-Time View Synthesis for Large Scenes with Millions of Square Meters}.
\newblock


\bibitem[Sitzmann et~al\mbox{.}(2018)]%
        {sitzmann2018saliency}
\bibfield{author}{\bibinfo{person}{Vincent Sitzmann}, \bibinfo{person}{Ana Serrano}, \bibinfo{person}{Amy Pavel}, \bibinfo{person}{Maneesh Agrawala}, \bibinfo{person}{Diego Gutierrez}, \bibinfo{person}{Belen Masia}, {and} \bibinfo{person}{Gordon Wetzstein}.} \bibinfo{year}{2018}\natexlab{}.
\newblock \showarticletitle{Saliency in VR: How do people explore virtual environments?}
\newblock \bibinfo{journal}{\emph{IEEE transactions on visualization and computer graphics}} \bibinfo{volume}{24}, \bibinfo{number}{4} (\bibinfo{year}{2018}), \bibinfo{pages}{1633--1642}.
\newblock


\bibitem[Stein et~al\mbox{.}(2022)]%
        {stein2022eye}
\bibfield{author}{\bibinfo{person}{Niklas Stein}, \bibinfo{person}{Gianni Bremer}, {and} \bibinfo{person}{Markus Lappe}.} \bibinfo{year}{2022}\natexlab{}.
\newblock \showarticletitle{Eye tracking-based lstm for locomotion prediction in vr}. In \bibinfo{booktitle}{\emph{2022 IEEE conference on virtual reality and 3D user interfaces (VR)}}. IEEE, \bibinfo{pages}{493--503}.
\newblock


\bibitem[Sun et~al\mbox{.}(2024)]%
        {sun20243dgstream}
\bibfield{author}{\bibinfo{person}{Jiakai Sun}, \bibinfo{person}{Han Jiao}, \bibinfo{person}{Guangyuan Li}, \bibinfo{person}{Zhanjie Zhang}, \bibinfo{person}{Lei Zhao}, {and} \bibinfo{person}{Wei Xing}.} \bibinfo{year}{2024}\natexlab{}.
\newblock \showarticletitle{3dgstream: On-the-fly training of 3d gaussians for efficient streaming of photo-realistic free-viewpoint videos}. In \bibinfo{booktitle}{\emph{Proceedings of the IEEE/CVF Conference on Computer Vision and Pattern Recognition}}. \bibinfo{pages}{20675--20685}.
\newblock


\bibitem[Sun et~al\mbox{.}(2025)]%
        {sun2025lts}
\bibfield{author}{\bibinfo{person}{Yuan-Chun Sun}, \bibinfo{person}{Yuang Shi}, \bibinfo{person}{Cheng-Tse Lee}, \bibinfo{person}{Mufeng Zhu}, \bibinfo{person}{Wei~Tsang Ooi}, \bibinfo{person}{Yao Liu}, \bibinfo{person}{Chun-Ying Huang}, {and} \bibinfo{person}{Cheng-Hsin Hsu}.} \bibinfo{year}{2025}\natexlab{}.
\newblock \showarticletitle{{LTS}: A {DASH} Streaming System for Dynamic Multi-Layer {3D Gaussian} Splatting Scenes}. In \bibinfo{booktitle}{\emph{The 16th ACM Multimedia Systems Conference, MMSys 2025, 2025}}. \bibinfo{publisher}{{ACM}}.
\newblock


\bibitem[Wang et~al\mbox{.}(2024)]%
        {wang2024v}
\bibfield{author}{\bibinfo{person}{Penghao Wang}, \bibinfo{person}{Zhirui Zhang}, \bibinfo{person}{Liao Wang}, \bibinfo{person}{Kaixin Yao}, \bibinfo{person}{Siyuan Xie}, \bibinfo{person}{Jingyi Yu}, \bibinfo{person}{Minye Wu}, {and} \bibinfo{person}{Lan Xu}.} \bibinfo{year}{2024}\natexlab{}.
\newblock \showarticletitle{V\^{} 3: Viewing Volumetric Videos on Mobiles via Streamable 2D Dynamic Gaussians}.
\newblock \bibinfo{journal}{\emph{ACM Transactions on Graphics (TOG)}} \bibinfo{volume}{43}, \bibinfo{number}{6} (\bibinfo{year}{2024}), \bibinfo{pages}{1--13}.
\newblock


\bibitem[Wang et~al\mbox{.}(2018)]%
        {wang2018revisiting}
\bibfield{author}{\bibinfo{person}{Wenguan Wang}, \bibinfo{person}{Jianbing Shen}, \bibinfo{person}{Fang Guo}, \bibinfo{person}{Ming-Ming Cheng}, {and} \bibinfo{person}{Ali Borji}.} \bibinfo{year}{2018}\natexlab{}.
\newblock \showarticletitle{Revisiting video saliency: A large-scale benchmark and a new model}. In \bibinfo{booktitle}{\emph{Proceedings of the IEEE Conference on computer vision and pattern recognition}}. \bibinfo{pages}{4894--4903}.
\newblock


\bibitem[Xu et~al\mbox{.}(2019)]%
        {10.1145/3359989.3365413}
\bibfield{author}{\bibinfo{person}{Tan Xu}, \bibinfo{person}{Bo Han}, {and} \bibinfo{person}{Feng Qian}.} \bibinfo{year}{2019}\natexlab{}.
\newblock \showarticletitle{Analyzing viewport prediction under different VR interactions}. In \bibinfo{booktitle}{\emph{Proceedings of the 15th International Conference on Emerging Networking Experiments And Technologies}} (Orlando, Florida) \emph{(\bibinfo{series}{CoNEXT '19})}. \bibinfo{publisher}{Association for Computing Machinery}, \bibinfo{address}{New York, NY, USA}, \bibinfo{pages}{165–171}.
\newblock
\showISBNx{9781450369985}
\urldef\tempurl%
\url{https://doi.org/10.1145/3359989.3365413}
\showDOI{\tempurl}


\bibitem[Ye et~al\mbox{.}(2024a)]%
        {ye2024gaussiangroupingsegmentedit}
\bibfield{author}{\bibinfo{person}{Mingqiao Ye}, \bibinfo{person}{Martin Danelljan}, \bibinfo{person}{Fisher Yu}, {and} \bibinfo{person}{Lei Ke}.} \bibinfo{year}{2024}\natexlab{a}.
\newblock \bibinfo{title}{Gaussian Grouping: Segment and Edit Anything in 3D Scenes}.
\newblock
\showeprint[arxiv]{2312.00732}~[cs.CV]
\urldef\tempurl%
\url{https://arxiv.org/abs/2312.00732}
\showURL{%
\tempurl}


\bibitem[Ye et~al\mbox{.}(2024b)]%
        {ye2024dissecting}
\bibfield{author}{\bibinfo{person}{Wei Ye}, \bibinfo{person}{Xinyue Hu}, \bibinfo{person}{Steven Sleder}, \bibinfo{person}{Anlan Zhang}, \bibinfo{person}{Udhaya~Kumar Dayalan}, \bibinfo{person}{Ahmad Hassan}, \bibinfo{person}{Rostand~AK Fezeu}, \bibinfo{person}{Akshay Jajoo}, \bibinfo{person}{Myungjin Lee}, \bibinfo{person}{Eman Ramadan}, {et~al\mbox{.}}} \bibinfo{year}{2024}\natexlab{b}.
\newblock \showarticletitle{Dissecting Carrier Aggregation in 5G Networks: Measurement, QoE Implications and Prediction}. In \bibinfo{booktitle}{\emph{Proceedings of the ACM SIGCOMM 2024 Conference}}. \bibinfo{pages}{340--357}.
\newblock


\bibitem[Yin et~al\mbox{.}(2015)]%
        {yin2015mpc}
\bibfield{author}{\bibinfo{person}{Xiaoqi Yin}, \bibinfo{person}{Abhishek Jindal}, \bibinfo{person}{Vyas Sekar}, {and} \bibinfo{person}{Bruno Sinopoli}.} \bibinfo{year}{2015}\natexlab{}.
\newblock \showarticletitle{A Control-Theoretic Approach for Dynamic Adaptive Video Streaming over HTTP}. In \bibinfo{booktitle}{\emph{Proceedings of the 2015 ACM Conference on Special Interest Group on Data Communication}} (London, United Kingdom) \emph{(\bibinfo{series}{SIGCOMM '15})}. \bibinfo{publisher}{Association for Computing Machinery}, \bibinfo{address}{New York, NY, USA}, \bibinfo{pages}{325–338}.
\newblock
\showISBNx{9781450335423}
\urldef\tempurl%
\url{https://doi.org/10.1145/2785956.2787486}
\showDOI{\tempurl}


\bibitem[Yu et~al\mbox{.}(2024)]%
        {yu2024mip}
\bibfield{author}{\bibinfo{person}{Zehao Yu}, \bibinfo{person}{Anpei Chen}, \bibinfo{person}{Binbin Huang}, \bibinfo{person}{Torsten Sattler}, {and} \bibinfo{person}{Andreas Geiger}.} \bibinfo{year}{2024}\natexlab{}.
\newblock \showarticletitle{Mip-splatting: Alias-free 3d gaussian splatting}. In \bibinfo{booktitle}{\emph{Proceedings of the IEEE/CVF Conference on Computer Vision and Pattern Recognition}}. \bibinfo{pages}{19447--19456}.
\newblock


\end{thebibliography}
